\documentclass[aps,reprint,groupedaddress, amsmath,amssymb]{revtex4-2}

\usepackage{extarrows}
\usepackage{graphicx,titlesec,amsthm}
\usepackage[colorlinks,urlcolor=blue,citecolor=blue]{hyperref}

\newcommand{\ee}{\mathrm{e}}
\newcommand{\dd}{\mathrm{d}}
\newcommand{\ii}{\mathrm{i}}
\newcommand{\sn}{\mathrm{sn}}
\newcommand{\dn}{\mathrm{dn}}
\newcommand{\cn}{\mathrm{cn}}

\newcommand{\scd}{\mathrm{scd}}
\newcommand{\off}{\mathrm{off}}
\newcommand{\diag}{\mathrm{diag}}

\newtheorem{define}{Definition}

\begin{document}
\title{Elliptic-rogue waves and modulational instability 
in nonlinear soliton equations}

\author{Liming Ling$^{1}$}
\email{linglm@scut.edu.cn}
\author{Xuan Sun$^{1}$} 
\email{masunx@mail.scut.edu.cn}

\affiliation{$^{1}$School of Mathematics, South China University of Technology, Guangzhou 510640, China}

\date{\today}
 
\begin{abstract}
We present elliptic-rogue wave solutions for integrable nonlinear soliton equations in theta functions. 
Unlike solutions generated on the plane wave background, these solutions depict rogue waves emerging on elliptic function backgrounds.
By refining the modified squared wave function method in tandem with the Darboux-B\"acklund transformation, we establish a quantitative correspondence between elliptic-rogue waves and the modulational instability. 
This connection reveals that the modulational instability of elliptic function solutions triggers rational-form solutions displaying elliptic-rogue waves, whereas the modulational stability of elliptic function solutions results in the rational-form solutions exhibiting the elliptic-solitons or elliptic-breathers.
Moreover, this approach enables the derivation of higher-order elliptic-rogue waves, offering a versatile framework for constructing elliptic-rogue waves and exploring modulational stability in other integrable equations.
\end{abstract}
 
\maketitle

\section{Introduction}

The nonlinear Schr\"odinger (NLS) equation, the modified Korteweg–de Vries (mKdV) equation, and the sine-Gordon (SG) equation are renowned models for the dynamics of waves \cite{Dodd-1982-solitons,Newell-1985-solitons}. 
They all belong to the class of integrable nonlinear soliton (INS) equations, distinguished by their soliton solutions \cite{AblowitzBC-22}.	
This class represents a distinct category within nonlinear wave equations and finds significant utility in studying various physical phenomena \cite{Agrawal-2000-nonlinear,Boyd-2008-nonlinear,Shabat-1972-exact}, including fluids, plasmas, and optical systems.

Rogue waves (RWs), characterized by their sudden appearance and disappearance without a trace \cite{AkhmedievaAT-09,AkhmedievaSA-09}, have been observed in various fields of physics, including oceanography \cite{AkhmedievaAT-09,AkhmedievaSA-09,ChabchoubHA-11,KristianHP-08}, quantum mechanics \cite{KonotopS-02,BludovKA-09,SolliRKJ-07}, optics \cite{DudleyDEG-14,AlexanderJSHM-11,ZaviyalovFIL-12}, and plasma physics \cite{StenfloM-10,AwadyM-11}.
These diverse experimental observations of RWs have spurred the exploration of RW solutions for INS equations such as the NLS equation \cite{AkhmedievAS-09,DubardGKM-10,AnkiewiczKA-11,BaronioDCW-12,KedzioraAA-12,ChowduryKA-17,GuoLL-12}, Hirota equation \cite{AnkiewiczSA-10}, coupled NLS equation \cite{GuoL-11}, and derivative NLS equation \cite{GuoLL-13}, among others.
So far, the research on RWs on plane wave backgrounds has reached a relatively advanced stage.

Nowadays, much attention is dedicated to constructing and analyzing solutions on complicated wave backgrounds, particularly those arising from elliptic function backgrounds in photonic crystal fibers and nonlinear metamaterials \cite{Russell-2003,LapineSK-2014}.
A multitude of elliptic-localized wave solutions \cite{Gesztesy2003-soliton,GesztesyH-99,CaliniS-17,KedzioraAA-14,Wright-16,ChenP-18,ChenPW-19,Chen-19-mKdV,ChenP-21,LiG-20-SG,PelinovskyW-20-SG,ChenPW-20,AlejoMP-17-SG,LingS-22-SG,LingS-22-mKdV,LingS-21-mKdV-stability,FengLT-20} of the INS equations has been constructed by developing the integrable system methods \cite{AnkiewiczSA-10,BaronioDCW-12,KedzioraAA-12,ChowduryKA-17,GuoLL-12,GuoL-11,BettelheimSM-22,KrajenbrinkL-21,GuoLL-13,BilmanLM-20-extreme,DongLZ-22,MallickMS-22,Ablowitz1974-AKNS}. 
These elliptic-localized wave solutions exhibit the dynamic behaviors of solitons, breathers, and RWs on elliptic function backgrounds called elliptic-solitons, elliptic-breathers, and elliptic-rogue waves (eRWs), respectively.
Moreover, elliptic-localized waves have been observed in experiments within the realms of nonlinear optics and hydrodynamics \cite{XuCPK-20}.

Naturally, we are inclined to explore the exciting conditions of the above-mentioned elliptic-localized waves. 
On the plane wave backgrounds, the modulational instability (MI) offers valuable insights into RWs, encompassing the existence of RWs \cite{BaronioCDLOW-14,BaronioCGWC-15}, 
the exciting conditions of RWs \cite{ZhaoL-16,LingZY-17,LingZ-19}, 
and the correspondence principle for RWs \cite{ChenBPHBGA-2022}. 
These insights offer a profound understanding of the dynamic behaviors of RWs \cite{AlejoM-13,ZakharovO-09,ErcolaniFM-90,DudleyGDKA-09,ErkintaloHKFADG-11,TaiHT-86,AblowitzC-21,OnoratoRBMA-13,BaronioCDLOW-14,BaronioCGWC-15,ZhaoL-16,LingZY-17,LingZ-19,ChenBPHBGA-2022}.
However, studying the MI of elliptic function solutions remains a challenge.
While spectral and orbital stability analyses of elliptic function solutions for the NLS equation and the mKdV equation have been conducted \cite{FengLT-20,AlejoMP-17,LingS-21-mKdV-stability,DeconinckU-2020orbital,DeconinckS-2017}, systematic MI analyses of elliptic function solutions for integrable nonlinear soliton (INS) equations remain an open question.
Numerical methods have been employed to assess the stability outcomes of the NLS equation \cite{ChenPW-20}, but a comprehensive MI analysis of elliptic function solutions for INS equations is yet to be tackled.
Therefore, our primary objective is to quantitatively reveal the correspondence between eRWs and MI, and to deduce general higher-order eRWs expressed in a rational form, which have never been reported.

In this article, we propose a method based on the Darboux-B\"acklund transformation and theta functions to construct rational-elliptic-localized wave solutions for INS equations.
Integrating this approach with the modified squared wave function (MSW) method establishes a quantitative correspondence between eRWs and MI, and correlates modulational stability (MS) with elliptic-solitons/breathers.
This method also enables the construction of higher-order eRW solutions for INS equations.
To illustrate these results, we apply our approach to three INS equations: the NLS equation, the mKdV equation, and the SG equation.

The structure of this paper is organized as follows.
In Sec. \ref{section:INSE}, we derive an infinite number of INS equations from the classical Ablowitz-Kaup-Newell-Segur (AKNS) system and present elliptic function solutions expressed in terms of theta functions for these equations.
In Sec. \ref{section:MI}, we dedicate to studying the MI and baseband MI of these elliptic function solutions. 
In Sec. \ref{section:eRWs}, we construct rational-elliptic-localized wave solutions. 
In Sec. \ref{section:correspondence}, we provide the correspondence between the MI analysis and eRWs, revealing the connection between MI and eRWs, as well as the connection between MS and elliptic-solitons/breathers.
In Sec. \ref{section:higher-order}, we are dedicated to constructing multi-higher-order eRWs in theta functions.
In Sec. \ref{section:conclution}, we give conclusions and perspective.
The appendixes provide useful formulas and detailed proofs for certain equations.

\section{Integrable nonlinear soliton (INS) equations and their elliptic function solutions}\label{section:INSE}
The INS equations include well-known equations such as the NLS equation, the mKdV equation, the SG equation, and others, which predominantly model wave dynamics in various physical systems like fluids, plasmas, and optical systems \cite{Dodd-1982-solitons,Newell-1985-solitons,AblowitzBC-22}. 
The classical AKNS system \cite{Ablowitz1974-AKNS} could derive infinite numbers of INS equations by the AKNS spectral problem $\Phi_x=\mathbf{U}\Phi$, where $\Phi\equiv\Phi(x,\mathbf{t};\lambda)$ is called the wave function with variables $x\in \mathbb{R}$, $\mathbf{t}=(\cdots,t_{-1},t_1,\cdots)\in \mathbb{R}^{\infty}$ and the spectral parameter $\lambda$; 
$\mathbf{U}\equiv \mathbf{U}(\lambda;\mathbf{Q})$ is defined by $\lambda$ and the anti-diagonal matrix $\mathbf{Q}$. 
Considering the positive power flow, we set a $2\times 2$ matrix function $\Phi$ as 
\begin{equation}\label{eq:Phi-m}
	\Phi=m\exp\left(-\ii\lambda \sigma_3 \left(x+\sum_{n=1}^{\infty}\lambda^n t_n\right) \right),
\end{equation}  
where $\sigma_3:=\diag\left(1,-1\right)$ is the third Pauli matrix; $m\equiv m(\lambda;x,\mathbf{t})$ is a holomorphic matrix function smoothly depending on $x$ and $\mathbf{t}$. 
As $\lambda\rightarrow \infty$, the matrix function $m$ could be expressed as
\begin{equation}\label{eq:m-expand-infty}
	m=\mathbb{I}_2+m_1(x,\mathbf{t})\lambda^{-1}+m_2(x,\mathbf{t})\lambda^{-2}+\mathcal{O}\big(\lambda^{-3}\big).
\end{equation}
Taking the derivative of variables $x$ and $t_n$ and letting $\Phi_{x}\Phi^{-1}
=\mathbf{U}(\lambda;\mathbf{Q})+\mathcal{O}(\lambda^{-1})$, $\Phi_{t_n}\Phi^{-1}
=\mathbf{V}_{n}(\lambda;\mathbf{Q})+\mathcal{O}(\lambda^{-1})$,
we obtain that matrices $\mathbf{U}(\lambda;\mathbf{Q})$ and $\mathbf{V}_n(\lambda;\mathbf{Q})$ are 
\begin{equation}\label{eq:U-V-n}
	\begin{split}
		\mathbf{U}(\lambda;\mathbf{Q})
		:=&-\ii\left( \lambda  m\sigma_3m^{-1}\right)_{+}, \\
		\mathbf{V}_n(\lambda;\mathbf{Q})
		:=&-\ii\left( \lambda^{n+1} m\sigma_3m^{-1}\right)_{+},
	\end{split}
\end{equation}
where $(\cdot)_{+}$ defines the regular part of the spectral parameter $\lambda$. 
Combined with Eqs. \eqref{eq:m-expand-infty} and \eqref{eq:U-V-n}, the $x$-part of the Lax pair could be written as
\begin{equation}\label{eq:Lax-x-part}
	\begin{split}	
	\Phi_x=&\mathbf{U}(\lambda;\mathbf{Q}) \Phi, \qquad \mathbf{U}(\lambda;\mathbf{Q})=-\ii \lambda \sigma_3 +\ii \mathbf{Q}, \\
	\mathbf{Q}=&\left[\sigma_3,m_1(x,\mathbf{t})\right]
	=\begin{bmatrix}
		0 &  q \\  r & 0
	\end{bmatrix},
	\end{split}
\end{equation}
where $q\equiv q(x,\mathbf{t})$, $r\equiv r(x,\mathbf{t})$, and $\left[\mathbf{A},\mathbf{B}\right]$ is defined as the commutator $\left[\mathbf{A},\mathbf{B}\right]=\mathbf{AB}-\mathbf{BA}$. Based on them, we construct the matrix $\mathbf{V}_n(\lambda;\mathbf{Q})$. Set $\Psi\equiv\Psi(\lambda;x,\mathbf{t}):=m\sigma_3 m^{-1}$ with $\Psi^2=\mathbb{I}$. Together with Eqs. \eqref{eq:m-expand-infty} and \eqref{eq:U-V-n}, the matrix function $\Psi$ could be represented as a summation form and the matrix $\mathbf{V}_n(\lambda;\mathbf{Q})$ could be rewritten by matrix functions $\Psi_i\equiv \Psi_i(x,\mathbf{t})$ as follows:
\begin{equation}\label{eq:Theta-expand-lambda-infty}
	\begin{split}
		\Psi=\sum_{i=0}^{\infty}\Psi_i\lambda^{-i}, \quad	\mathbf{V}_n(\lambda;\mathbf{Q})=-\ii \sum_{i=0}^{n+1}\Psi_i\lambda^{n+1-i}.
	\end{split}
\end{equation} 
Based on the stationary zero curvature equation
$\Psi_x=\left[\mathbf{U}(\lambda;\mathbf{Q}), \Psi\right]$,
$\Psi_{t_n}=\left[\mathbf{V}_n(\lambda;\mathbf{Q}), \Psi\right]$, matrices $\Psi_i$ in Eq. \eqref{eq:Theta-expand-lambda-infty} are expressed as follows: $\Psi_0=\sigma_3$, $\Psi_{1}=-\mathbf{Q}$, and
\begin{equation}\label{eq:Theta-i-expression}
	\begin{split}
		\Psi_{i+1}^{\off}
		=&\ \frac{\sigma_3}{2}\left(\ii \Psi_{i,x}^{\off}+\left[\mathbf{Q},\Psi_{i}^{\diag}\right]\right),\\
		\Psi_{i+1}^{\diag}
		=&\ -\frac{\sigma_3}{2}\sum_{j=1}^{i}\left(\Psi_{j}\Psi_{i+1-j}\right)^{\diag}, 
	\end{split}
\end{equation}
$i=1,2,\cdots$.
Plugging matrices $\Psi_i$ into Eq.  \eqref{eq:Theta-expand-lambda-infty}, we get expressions for matrices $\mathbf{V}_n(\lambda;\mathbf{Q})$. 
In such a case, the compatibility conditions of ordinary differential equations $\Phi_{x}=\mathbf{U}(\lambda;\mathbf{Q})\Phi$ and $\Phi_{t_n}=\mathbf{V}_n(\lambda;\mathbf{Q})\Phi$ could deduce the related INS equation \cite{Wright-19} under different symmetries. 
Moreover, by combining a linear combination of the aforementioned ordinary differential equations with $\hat{t}_{n}=\sum_{i=1}^{n}a_{i}t_i$, we can derive various INS equations. 

\subsection{INS equations and their Lax pairs}\label{subsection:NLS-Lax}
Consider the second-order positive power flow and suppose matrices $\mathbf{U}(\lambda;\mathbf{Q})$ and $\mathbf{V}_1(\lambda;\mathbf{Q})$ satisfying the su(2)-symmetry
\begin{equation}\label{eq:symmetry-su}	  \mathbf{U}^{\dagger}(\lambda^*;\mathbf{Q})=-\mathbf{U}(\lambda,\mathbf{Q}), \,\,\, \mathbf{V}_1^{\dagger}(\lambda^*;\mathbf{Q})=-\mathbf{V}_1(\lambda,\mathbf{Q}),
\end{equation}
with $r=q^*$. Setting $t=t_1$, the Lax pair could be written as
\begin{equation}\label{eq:NLS-Lax-pair}
	\begin{split}
		&\Phi_x
		=\ \mathbf{U}(\lambda;\mathbf{Q})\Phi,\qquad
		\Phi_{t}
		=\ \mathbf{V}_1(\lambda;\mathbf{Q})\Phi, \\	&\mathbf{V}_1(\lambda;\mathbf{Q})
		=\lambda\mathbf{U}(\lambda;\mathbf{Q})+\frac{ \sigma_3}{2}\left(\ii\mathbf{Q}^2-\mathbf{Q}_{x}\right),
	\end{split}
\end{equation}
with $\lambda \in (\mathbb{C}\cup \{\infty\})\backslash \mathbb{R}$ and matrix $\mathbf{U}(\lambda;\mathbf{Q})$ defined in Eq. \eqref{eq:Lax-x-part}. Based on the compatibility conditions $\Phi_{xt}(x,t;\lambda)=\Phi_{tx}(x,t;\lambda)$, we deduce the focusing NLS equation as
\begin{equation}\label{eq:NLS-equation}\tag{NLS}
	\ii q_{t}+\frac{1}{2} q_{xx}+|q|^2q=0.
\end{equation}

Consider the third-order positive power flow and suppose matrices $\mathbf{U}(\lambda;\mathbf{Q})$ and $\mathbf{V}_2(\lambda;\mathbf{Q})$ satisfy the su(2)-symmetry \eqref{eq:symmetry-su} and the twist-symmetry 
\begin{equation}\label{eq:symmetry-twist}	\mathbf{U}^{\top}\!(-\lambda;\mathbf{Q})\!=\!-\mathbf{U}(\lambda,\mathbf{Q}),
\mathbf{V}_2^{\top}\!(-\lambda;\mathbf{Q})\!=\!-\mathbf{V}_2(\lambda,\mathbf{Q}),
\end{equation}
which implies $\mathbf{Q}^{\top}=-\mathbf{Q}$ and $q\in \ii \mathbb{R}$.
Without loss of generality, we set $\ii q=u(x,\mathbf{t})\in \mathbb{R}$ and $t=t_2/4$ and then obtain the Lax pair by the third-order flow as
\begin{equation}\label{eq:mKdV-Lax-pair}
	\begin{split}
		&\Phi_x=\mathbf{U}(\lambda;\mathbf{Q})\Phi, \qquad 
		\Phi_{t}= \mathbf{V}_2(\lambda;\mathbf{Q})\Phi,\\
		&\mathbf{V}_2(\lambda;\mathbf{Q})	=4\lambda\mathbf{V}_1(\lambda;\mathbf{Q}) -\ii\left(\mathbf{Q}_{xx}+2 \mathbf{Q}^3\right),
	\end{split}
\end{equation}
with $\lambda \in (\mathbb{C}\cup \{\infty\})\backslash \mathbb{R}$ and matrices $\mathbf{U}(\lambda;\mathbf{Q})$ and $\mathbf{V}_1(\lambda;\mathbf{Q})$ are defined in Eqs. \eqref{eq:Lax-x-part} and \eqref{eq:NLS-Lax-pair}. By the compatibility conditions and setting $t=t_2/4$, we obtain the focusing mKdV equation:
\begin{equation}\label{eq:mKdV-equation}\tag{mKdV}
	u_{t}+u_{xxx}+6u^2u_x=0. 
\end{equation}

For the SG equation, we consider the negative power flow and expand the matrix function $\Phi$ with respect to the spectral parameter $\lambda$ at the zero point ($\lambda\rightarrow 0$)
\begin{equation}\nonumber
	\Phi=\hat{m}\exp\left(-\ii \sum_{i=1}^{\infty}\frac{t_{-i}}{\lambda^i}\sigma_3\right),
\end{equation}
where $\hat{m}=\hat{m}_0(x,\mathbf{t})+\hat{m}_1(x,\mathbf{t})\lambda+\hat{m}_2(x,\mathbf{t})\lambda^2+\mathcal{O}(\lambda^3)$,
and $\hat{m}\equiv \hat{m}(\lambda;x,\mathbf{t})$ is a holomorphic matrix function in a punctured neighborhood of the infinity on the Riemann sphere.
Taking the derivative of variables $t_{-i}$, we obtain $\Phi_{t_{-i}}\Phi^{-1}=\hat{m}_{t_{-i}}\hat{m}^{-1}-\ii\lambda^{-i} \hat{m}\sigma_3\hat{m}^{-1}=\mathbf{V}_{-i}(\lambda;\mathbf{Q})+\mathcal{O}(1)$.
Set $\hat{\Psi}\equiv\hat{\Psi}(\lambda;x,\mathbf{t}):=\hat{m}\sigma_3\hat{m}^{-1}$, which deduces $\det(\hat{\Psi})=-1$ and	$\mathrm{Tr}(\hat{\Psi})=0$. 
Matrix functions $\hat{\Psi}$ and $\mathbf{V}_{-n}(\lambda;\mathbf{Q})$ could be rewritten as
\begin{equation}\label{eq:Theta-hat-expand-lambda-infty}
		\hat{\Psi} \!
		= \! \hat{\Psi}_0 \!	+ \! \hat{\Psi}_1\lambda \! +\! \mathcal{O}(\lambda^{2}), \,\, \mathbf{V}_{-n}(\lambda;\mathbf{Q})\!
		= \!-\ii  	\sum_{i=0}^{n-1}\hat{\Psi}_i\lambda^{i-n},
\end{equation} 
where $\hat{\Psi}_i\equiv \hat{\Psi}_i(x,\mathbf{t})$.
Plugging the matrix $\Phi$ in Eq. \eqref{eq:Phi-m} into the Lax pair $\Phi_{t_{-i}}=\mathbf{V}_{-i}(\lambda;\mathbf{Q})\Phi$ and collecting coefficients of the spectral parameter $\lambda$, we derive matrix functions $\hat{\Psi}_{i}$, $i=0,1,\cdots$, expressed in terms of matrices $m_i(x,\mathbf{t})$, $i=0,1,\cdots$. 
Combined with the compatibility conditions, infinite numbers of INS equations are derived under the negative power flows of the AKNS system. 

Consider the $t_{-1}$-part of the Lax pair $\Phi_{t_{-1}}=\mathbf{V}_{-1}(\lambda;\mathbf{Q})\Phi$. Combined with Eq. \eqref{eq:Theta-hat-expand-lambda-infty}, the matrix $\hat{\Psi}_{0}$ is expressed as $\hat{\Psi}_{0}=4\ii m_{1,t_{-1}}(x,\mathbf{t})$. Since $2\sigma_3 m_1^{\mathrm{off}}(x,\mathbf{t})=\mathbf{Q}$, we obtain $\hat{\Psi}_{0}^{\mathrm{off}}=2\ii\sigma_3\mathbf{Q}_{t_{-1}}$. Combining equations $\det(\hat{\Psi})=-1$ and	$\mathrm{Tr}(\hat{\Psi})=0$ with Eq. \eqref{eq:Theta-hat-expand-lambda-infty}, we obtain $\det(\hat{\Psi}_0)=-1$ and	$\mathrm{Tr}(\hat{\Psi}_0)=0$, which implies
\begin{equation}\nonumber
	\hat{\Psi}_0
	=
	\begin{bmatrix}
		- \sqrt{1+4r_{t_{-1}}q_{t_{-1}}}
		&2 \ii q_{t_{-1}}\\
		-2\ii r_{t_{-1}} 
		& \sqrt{1+4r_{t_{-1}}q_{t_{-1}}}
	\end{bmatrix}.
\end{equation}
If matrices $\mathbf{U}(\lambda;\mathbf{Q})$ and $\mathbf{V}_{-1}(\lambda;\mathbf{Q})$ satisfy the su(2)-symmetry \eqref{eq:symmetry-su} and twist-symmetry \eqref{eq:symmetry-twist}, i.e., $\mathbf{Q}^{\top}=-\mathbf{Q}$, $q\in \ii \mathbb{R}$, we can set $q=-\ii v_x/2$ and choose $t=4t_{-1}$ without loss of generality. 
The Lax pair, derived from the negative power flow and satisfying su(2)-symmetry \eqref{eq:symmetry-su} and twist-symmetry \eqref{eq:symmetry-twist}, can be expressed as
\begin{equation}\label{eq:SG-Lax-pair}
	\begin{split}
		&\Phi_x
		= \mathbf{U}(\lambda;\mathbf{Q})\Phi, \quad	
		\Phi_t
		=\mathbf{V}_{-1}(\lambda;\mathbf{Q})\Phi,\\
		&\mathbf{V}_{-1}(\lambda;\mathbf{Q})=\frac{\hat{\Psi}_0}{4}
		=-\frac{\ii}{4\lambda}
		\begin{bmatrix}
			-  \cos(\hat{v}) & \sin(\hat{v}) \\ 
			\sin(\hat{v})  & \cos(\hat{v})
		\end{bmatrix},
	\end{split}
\end{equation}
where $\hat{v}=\arcsin\left(v_{xt}\right)$. 
By the compatibility conditions, we obtain $\hat{v}=v$ and the SG equation
\begin{equation}\label{eq:SG-equation}\tag{SG}
	v_{xt}=\sin(v).
\end{equation}
In this way, we could also obtain infinite numbers of INS equations based on the negative power flows of the AKNS system. 
The solutions of these INS equations could be expressed in terms of the $(1,2)$-element of the anti-diagonal matrix $\mathbf{Q}$, denoted as $\mathbf{Q}_{12}$, which corresponds to $\mathbf{Q}_{12}=q=-\ii u=-\ii v_x/2$.
We will express solutions using $\mathbf{Q}_{12}$ instead of $q$, $u$, and $v_x$.

\subsection{Elliptic function solutions of INS equations}\label{subsection:elliptic-solution}
The MSW method \cite{Wright-97,Shin-12} and the algebro-geometric approach \cite{BelokolosBEI-94,Kamchatnov-90,Kamchatnov-97} stand out as two highly effective methods for constructing elliptic function solutions of INS equations \cite{FengLT-20,LingS-21-mKdV-stability,LingS-22-mKdV,LingS-22-SG}. 
In this subsection, we utilize the MSW method and algebro-geometric approaches to present details for deriving elliptic function solutions of INS equations. Then, we express these solutions in terms of theta functions.
We will apply this method to the \eqref{eq:NLS-equation} equation, the \eqref{eq:mKdV-equation} equation, and the \eqref{eq:SG-equation} equation.
By doing so, we will obtain uniform expressions for elliptic function solutions.

For the genus-g case, set
\begin{equation}\label{eq:L-expand-Theta}
	\begin{split}
		\mathbf{L}:
		=&\left(\sum_{n=0}^{g+1}\alpha_n \lambda^{n}\Psi\right)_{+}
		=\sum_{n=0}^{g+1}\sum_{j=0}^{g+1-n}\alpha_{n+j}\Psi_{j}\lambda^n,
	\end{split}
\end{equation}
where $\mathbf{L}\equiv\mathbf{L}(\lambda;x,\mathbf{t})$ and functions $\Psi$, $\Psi_j$, $j=0,1,2,\cdots$ are defined in Eqs. \eqref{eq:Theta-expand-lambda-infty} and \eqref{eq:Theta-i-expression}. 
Let the matrix function $\mathbf{L}$ also satisfy the stationary zero curvature equations
\begin{equation}\label{eq:zero-curvature-equation-L}
	\mathbf{L}_x=\left[\mathbf{U}(\lambda,\mathbf{Q}),\mathbf{L}\right], \quad 
	\mathbf{L}_{\hat{t}_i}=\left[\hat{\mathbf{V}}_i(\lambda,\mathbf{Q}),\mathbf{L}\right],
\end{equation}
where the matrix $\hat{\mathbf{V}}_i(\lambda;\mathbf{Q})=\sum_{j=1}^{i}a_{j}\mathbf{V}_j(\lambda;\mathbf{Q})$, $a_j\in \mathbb{R}$. 
Since $\Psi_x=\left[\mathbf{U}(\lambda;\mathbf{Q}), \Psi\right]$ and $\Psi_{t_i}=\left[\mathbf{V}_i(\lambda;\mathbf{Q}), \Psi\right]$, parameters $\alpha_j$ satisfy $\alpha_{j,x}=0$ and $\alpha_{j,\hat{t}_i}=0$, $j=0,1,\cdots,g+1$. 
When $i>0$, the following equations hold 
\begin{equation}\label{eq:zero-equation-case}
	\ii\!\! \!\sum_{j=0}^{g+1-n} \!\! \! \alpha_{j+n} \Psi_{j,t_{i}} \!=\! \sum_{l=0}^{n}\!\left[\sum_{j=0}^{i} \! a_j\Psi_{j+1-n+l},\!\! \sum_{j=0}^{g+1-l}\!\!\!\alpha_{j+l} \Psi_{j}\right]\!\!.\!\!\!
\end{equation}
By the matrix function $\mathbf{L}$ in Eq. \eqref{eq:L-expand-Theta}, we set 
\begin{equation}\label{eq:define-P}
	\mathbf{L}
	\!=\!\begin{bmatrix}
		L_{11} & L_{12} \\
		L_{21} & -L_{11}
	\end{bmatrix}\!, \, \det\left(\mathbf{L}\right)\!=\!-L_{11}^2-L_{12}L_{21}\!=\!P(\lambda),
\end{equation}
where $P(\lambda)= \prod_{i=1}^{2g+2}(\lambda-\lambda_i)=\sum_{i=0}^{2g+2}s_i\lambda^i$, $s_{2g+2}=1$. 
By comparing the exact expression of the matrix $\mathbf{L}$ in Eq. \eqref{eq:L-expand-Theta} with the function $P(\lambda)$, we can derive certain equations crucial to the construction of elliptic function solutions for INS equations.
In this context, we consider the genus-1 case as an example. 

Under the genus-1 case with $\alpha_2=1$ in Eq. \eqref{eq:L-expand-Theta} and $i=1$ in Eq. \eqref{eq:zero-curvature-equation-L}, and combined with Eqs. \eqref{eq:L-expand-Theta}-\eqref{eq:zero-equation-case}, the matrix function $\mathbf{L}$ is
\begin{equation}\label{eq:exact-L-expression}
	\begin{split}
		\mathbf{L}=&-\ii \sigma_3\lambda^2+\ii( \mathbf{Q}- \alpha_1\sigma_3)\lambda+\ii\sigma_3\left(\ii\mathbf{Q}_{x}+\mathbf{Q}^2\right)/2\\
		&+\ii \alpha_1 \mathbf{Q}-\ii \alpha_0\sigma_3,
	\end{split}
\end{equation}
and functions $L_{ij}\equiv L_{ij}(\lambda;x,\mathbf{t})$ are
\begin{equation}\label{eq:elements-L}
	\begin{split}
		L_{11}(\lambda;x,\mathbf{t})=&-\ii\lambda^2-\ii \alpha_1\lambda-\ii(\alpha_0-\left|q\right|^2/2),\\
		L_{12}(\lambda;x,\mathbf{t})=&-L_{21}^*(\lambda^*;x,\mathbf{t})=\ii q\left(\lambda-\mu\right), 
	\end{split}
\end{equation}
where $\mu=-\ii(\ln q)_x/2 -\alpha_1$.
Since $-\mathbf{L}^{\dagger}(\lambda^*;x,\mathbf{t})=\mathbf{L}(\lambda;x,\mathbf{t})$ gained by the existence and uniqueness of the ordinary differential equation, we obtain $P(\lambda)=P^*(\lambda^*)$. 
As a result, the parameter $\lambda_i$, representing the zeros of the polynomial $P(\lambda)$ in Eq. \eqref{eq:define-P}, satisfies the relationship $\lambda_{j+2}=\lambda_j^*$, $j=1,2$.

Together with Eqs. \eqref{eq:define-P} and \eqref{eq:elements-L}, we obtain
\begin{equation}\label{eq:elements-s}
	\begin{split}
		&2\alpha_1=s_3, \qquad 
		\alpha_1^2+2\alpha_0=s_2, \\
		&\alpha_1(2\alpha_0-|q|^2)-|q|^2(\mu 	+\mu^*)=s_1,\\
		&(2\alpha_0-|q|^2)^2+4|q|^2\mu \mu^*=4s_0.
	\end{split}
\end{equation} 
For the NLS equation, we consider the case with parameters $a_1=1$, $i=1$, and $g=1$ in Eqs. \eqref{eq:L-expand-Theta} and \eqref{eq:zero-curvature-equation-L}. By Eq. \eqref{eq:zero-equation-case}, we obtain $\alpha_1 \Psi_{1,t_{1}}
=-\ii \left[\Psi_{2}, \alpha_1 \Psi_0\right]-\ii \left[\Psi_{1}, \alpha_0 \Psi_0\right]$. 
Combining with functions $\Psi_i$, $i=0,1,2$, shown in Eq. \eqref{eq:Theta-i-expression}, we get $\mathbf{Q}_t-2\ii \alpha_0 \sigma_3 \mathbf{Q}+\alpha_1\mathbf{Q}_x=0$ and $\alpha_0=0$. 
Together with Eq. \eqref{eq:elements-s} and the definition of $\mu$, we obtain $2(\mu-\mu^*)=-\ii |q|^2_x/|q|^2$ and $|q|^4(\mu-\mu^*)^2=[\alpha_1(2\alpha_0-|q|^2)-s_1]^2-|q|^2[4s_0-(2\alpha_0-|q|^2)^2]$, which deduce the equation $-|q|^4_x=4|q|^6+|q|^4(s_3^2-2(4s_2-s_3^2))+|q|^2(4s_1s_3-16s_0+(4s_2-s_3^2)(2s_3^2+1)/4)+4s_1^2+(4s_2-s_3^2)(s_3^2-16s_1s_3)/16=4(|q|^2+d_1^2)(|q|^2-d_2^2)(|q|^2-d_3^2)$ with three real-valued parameters $d_1<d_2<d_3$. 
Upon solving it, we obtain the elliptic function solution $|q|^2(x,t)=k^2\alpha^2\left(\sn^2(K_l)-\sn^2\left(\alpha \left(x+s_1t/2\right)\right)\right)$, where parameters $d_1$, $d_2$, and $d_3$ \cite{LingS-21-mKdV-stability} are well-defined as follows:
\begin{equation}\label{eq:define-d-1-2-3-Kl}
	d_1=-\dn(K_l), \quad 
	d_2=\ii k\cn(K_l), \quad 
	d_3=k\sn(K_l),
\end{equation} 
with $d_{1,2,3}\in \mathbb{R}$, $d_1\le 0\le d_2\le d_3$, $K_l=K+4\ii lK$, $k\in (0,1)$, $l\in \left[0,-\ii \tau/4\right]$ and $\alpha\in \mathbb{R}$.
The values of parameters $l,k$ and $\alpha$ depend on parameters $s_i$, $i=1,2,3$. In combination with the Lax pair, the solution $q$ can be expressed as follows:
\begin{equation}\label{eq:NLS-solution-seed}
	q=-\ii \sqrt{v\left(\xi\right)}\exp \!\left[\ii s_2 t \!- \!\frac{\ii s_1(x+s_1t)}{2} \!- \!\int_{0}^{\xi}\frac{\ii m\dd s}{v(s)}\right]\! ,
\end{equation}
where $m=\sqrt{d_1d_2d_3}$ and $\xi=x+s_1t/2 $.
The detailed calculation process is given in the article \cite{FengLT-20}.

As previously mentioned, utilizing the Lax pair of the mKdV equation discussed in Sec. \ref{subsection:NLS-Lax}, we consider the case with parameters $a_2=4$, $a_1=0$, $i=2$, and $g=1$ in Eqs. \eqref{eq:L-expand-Theta} and \eqref{eq:zero-curvature-equation-L}. 
Consequently, we get $\alpha_1=0$ and express $\mathbf{L}$ as shown in Eq. \eqref{eq:exact-L-expression}. Furthermore, the derived equation is $\mathbf{Q}_t+4\alpha_2\mathbf{Q}_x=0$. 
Taking these considerations into account, the elliptic function solutions of the mKdV equation can be expressed as follows:
\begin{equation}\label{eq:mKdV-solution-seed}
	u=\alpha k \cn(\alpha (x-st)), \qquad 
	u=\alpha \dn(\alpha (x-st)),
\end{equation}
where $s=\alpha^2(d_3^2+d_2^2-d_1^2)$ with $l=0$ for $\cn$-type solutions and $l=-\ii \tau/4$ for $\dn$-type solutions.
The detailed calculation process for obtaining elliptic function solutions of the mKdV equation is given in the article \cite{LingS-21-mKdV-stability}.

To consider the matrix $\mathbf{L}$ of the SG equation, we focus on the negative power of the spectral parameter $\lambda$. Combined with Eq. \eqref{eq:zero-curvature-equation-L}, the equation is 
\begin{equation}\label{eq:zero-equation-case-1}
	\ii \!\sum_{j=0}^{g+1-n}\!\alpha_{n+j}\! \Psi_{j,\hat{t}_{-i}}= b\sum_{l=0}^{n-1} \!\left[\hat{\Psi}_{i+l-n}, \sum_{j=0}^{g+1-l} \!\alpha_{j+l} \Psi_{j}\right], 
\end{equation}
where $n=0,1,\cdots,g+1$, $\hat{\Psi}_{i}$ is defined in Eq. \eqref{eq:Theta-hat-expand-lambda-infty}. Similarly, when $g=1$, $b=1/4$, and $i=1$ in Eqs. \eqref{eq:L-expand-Theta} and \eqref{eq:zero-equation-case-1}, 
elliptic function solutions of the SG equation \eqref{eq:SG-equation} could be written as
\begin{subequations}\label{eq:SG-solution-seed}
	\begin{align}
		v_x=&2\alpha \dn(\alpha (x-st)), \quad l=-\ii \tau/4, \label{eq:SG-solution-seed-dn}\\
		v_x=&2\alpha k\cn(\alpha (x-st)), \quad l= 0, \label{eq:SG-solution-seed-cn}
	\end{align}
\end{subequations}
where $s=(-1)^{n}\left(\alpha k\right)^{-2}d_3^2$.

For ease of representation, we introduce a coordinate transformation:
\begin{equation}\label{eq:coordinate-transformation}
	(x,t)\xlongequal[\eta=t]{\xi=x-st}(\xi ,\eta),
\end{equation}
where $s$ is the background velocity of solutions. By integral formulas \eqref{eq:transformation-elliptic-theta}, we express elliptic function solutions \eqref{eq:NLS-solution-seed}, \eqref{eq:mKdV-solution-seed}, and \eqref{eq:SG-solution-seed} in theta functions.

Elliptic function solutions of INS equations could be expressed in terms of theta functions as follows:
\begin{equation}\label{eq:solution-seed}
	\mathbf{Q}_{12}=\ii \gamma \frac{\vartheta_2(\hat{\alpha}\xi+2\ii l)}{\vartheta_4(\hat{\alpha}\xi)}\ee^{\ii (\omega\xi+\kappa \eta)},\quad \gamma=\frac{\alpha\vartheta_2\vartheta_4}{\vartheta_3\vartheta_3(2\ii l)},	
\end{equation}
where $\omega= \ii \alpha Z(K_l)$ and $\hat{\alpha}=\alpha/(2K)$; $\xi$ and $\eta$ are defined in Eq. \eqref{eq:coordinate-transformation}; parameters $K_l$ and $d_{1,2,3}$ are defined in Eq. \eqref{eq:define-d-1-2-3-Kl}; and $\vartheta_i(x)$ with $\vartheta_i\equiv\vartheta_i(0)$, $i=1,2,3,4$ are called theta functions defined in Def. \ref{define:theta}.
Parameters $s$ and $\kappa$ are defined by different equations as follows:
\begin{itemize}
	\item For the \eqref{eq:NLS-equation} equation, parameters are defined as $s=0$ and $\kappa=\alpha^2(d_3^2+d_2^2-d_1^2)/2 $;
	\item For the \eqref{eq:mKdV-equation} equation, parameters are defined as $s=\alpha^2(d_3^2+d_2^2-d_1^2)$ and $\kappa=0$;
	\item For the \eqref{eq:SG-equation} equation, parameters are defined as $s=(-1)^{n}\left(\alpha k\right)^{-2}d_3^2$ and $\kappa=0$.
\end{itemize}

\begin{figure}[h]
	\centering
	\includegraphics[width=0.9\linewidth]{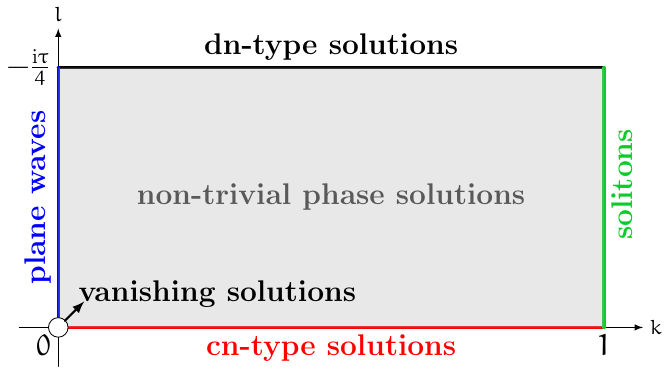}
	\caption{The sketch map for the elliptic function solution \eqref{eq:solution-seed} of INS equations with different values of parameters $l\in [0,-\ii \tau/4]$ and $k\in (0,1)$. }
	\label{fig:solution-k-l}
\end{figure}

In this paper, we just consider the region of the parameter $l$ and the modulus $k$ as $l\in[0,-\ii \tau/4]$ and $k\in(0,1)$, respectively.
When $l=0$, the solution \eqref{eq:solution-seed} could be written as $\alpha k \cn(\alpha \xi)\ee^{\ii \kappa \eta}$, where $\cn(\cdot)$ is defined in Def. \ref{define:elliptic-function}. 
The solution is referred to as the $\cn$-type solution, representing a trivial phase solution. 
When $l=-\ii \tau/4$, the solution \eqref{eq:solution-seed}  transforms into $\alpha \dn(\alpha \xi)\ee^{\ii \kappa \eta}$, known as the $\dn$-type solution, which is also a trivial phase solution. 
For $l\in \left(0,-\ii \tau/4\right)$, solutions \eqref{eq:solution-seed} fall into the category of non-trivial phase solutions.
We can confine our consideration to the modulus $k\in (0,1)$.
This limitation arises from the fact that elliptic functions with $k$ in the interval $(1,\infty)$ can be transformed into the case $(0,1)$ through reciprocal modulus transformation formulas \cite[p.38]{ByrdF-54}. 
As the modulus $k\rightarrow 0$, solutions tend to degenerate into constants, while as $k\rightarrow 1$, solutions tend to degenerate into solitons. 
Based on these characteristics, different types of solutions can be distinguished by varying the values of parameters $l$ and $k$, as depicted in the sketch map shown in Fig.\ref{fig:solution-k-l}. 

\subsection{The fundamental solutions of related Lax pairs}
Under the coordinate transformation \eqref{eq:coordinate-transformation}, we can construct the fundamental solutions of the related Lax pair.
Let $\pm y$ denote the two eigenvalues of the matrix $\mathbf{L}\equiv\mathbf{L}(\lambda;\xi,\eta)$.
By analyzing eigenvectors of the matrix $\mathbf{L}$ corresponding to the eigenvalues $\pm y$, we can set $\phi_{2i}\equiv\phi_{2i}(\xi,\eta)=r_i(\xi,\eta;\lambda)\phi_{1i}(\xi,\eta)\equiv r_i\phi_{1i}$, $r_{1,2}=-(L_{11} \pm y)/L_{12}=L_{21}/(L_{11}\mp y)$, $i=1,2$. 
Functions $\phi_{ij}$ and $L_{ij}$, $i,j=1,2$, represent the $(i,j)$-elements of matrices $\Phi\equiv\Phi(\xi,\eta;\lambda)$ and $\mathbf{L}$ respectively. 
The fundamental solution of the Lax pair could be expressed as 
\begin{equation}\label{eq:Phi-elliptic}
	\Phi=
	\begin{bmatrix}
		\sqrt{\varrho(\xi)-\beta_1}\ee^{\theta_1 }& 
		\sqrt{\varrho(\xi)-\beta_2}\ee^{\theta_2} \\
		r_1\sqrt{\varrho(\xi)-\beta_1}\ee^{\theta_1} &
		r_2\sqrt{\varrho(\xi)-\beta_2}\ee^{\theta_2}
	\end{bmatrix},
\end{equation}
where $\varrho(\xi)=k^2\alpha^2\left(\sn^2(K_l)-\sn^2(\alpha \xi)\right)$, $\beta_{1,2}=2\lambda^2+\alpha^2(d_3^2+d_2^2-d_1^2)/2\mp2y$,
\begin{equation}\label{eq:define-theta-1-2}
	\begin{split}
		r_{1,2}=&\ii \sqrt{\frac{\mathbf{Q}_{12}^*(\lambda-\mu^*)(\varrho(\xi)-\beta_{2,1})}{\mathbf{Q}_{12}(\lambda-\mu)(\varrho(\xi)-\beta_{1,2})}},\\
		\theta_{1,2} =  &\int_{0}^{\xi }\frac{\ii \lambda \beta_{1,2}\dd  x }{\varrho(x)-\beta_{1,2}} \! + \!\frac{\ii \lambda}{2}\xi  +  \ii V_{1,2}  \eta,
	\end{split}
\end{equation}
and functions $V_{1,2}\equiv V_{1,2}(z)$ are dependent on equations as follows: 
\begin{equation}\label{eq:define-V}
	\begin{split}
		 V_{1,2}(z)=&\kappa/2 \pm y(z), \qquad \text{NLS};\\
		 V_{1,2}(z)=&\pm 4 \lambda(z) y(z), \qquad \text{mKdV};\\
		 V_{1,2}(z)=&\pm s y(z)/\lambda(z) ,\qquad \text{SG}.
	\end{split}
\end{equation}
The detailed process can be obtained in \cite{FengLT-20,LingS-21-mKdV-stability,LingS-22-mKdV,LingS-22-SG}.

Next, our objective is to introduce a uniform parameter $z$ and utilize theta functions to represent the solution \eqref{eq:Phi-elliptic}. 
Upon reviewing the Dubrovin-type equation for $\mu$ and setting $\lambda=\mu$, we get $(L_{12} )_{\xi}=-\ii q\mu_{\xi}$ and $(L_{12})_{\xi}=-2\ii q_{\xi} L_{11}$ by Eq. \eqref{eq:zero-curvature-equation-L}, which implies $\mu_{\xi}=2 L_{11}$. 
Since $\pm y$ are eigenvalues of the matrix $\mathbf{L}$, we obtain $\mu_{\xi}=2 L_{11}=-2\ii y$. 
Functions $\lambda$ and $y$ are expressed by the uniform parameter $z$:
\begin{equation}\label{eq:y_mu_lambda}
	\lambda(z)\!=\!\mu\left(\frac{z_l}{\alpha}\right),\,\, y(z)=\frac{\alpha}{4K}\frac{\dd }{\dd z}\mu\left(\frac{z_l}{\alpha}\right)=\frac{\alpha}{4K}\frac{\dd \lambda(z)}{\dd z},
\end{equation}
with $z_l=2\ii (z-l)K$.
Since $\mu=-\ii (\ln (q))_{x}-\alpha_1$, we get
\begin{equation}\label{eq:lambda-y}
	\begin{split}
		\lambda(z)=&\frac{\ii 	\alpha}{2}\left(\frac{\scd(z_l)-\scd(K_l)}{\sn^2(K_l)-\sn^2(z_l)}\right),\\
		y(z)=&\alpha^2 k^2 	\big(\sn^2(z_l)-\sn^2(-z_l-K_l+\ii K')\big)/4,
	\end{split}
\end{equation}
with $\scd(\cdot):=\sn(\cdot)\cn(\cdot)\dn(\cdot)$ and $K_l$ defined in Eq. \eqref{eq:define-d-1-2-3-Kl}. 
It solves the algebraic curve $y^2=-\lambda^4-s_3\lambda^3-s_2\lambda^2-s_1\lambda-s_0$ automatically. The conformal map $\lambda(z)$ \cite{FengLT-20,LingS-21-mKdV-stability,LingS-22-mKdV,LingS-22-SG} maps the rectangular area
\begin{equation}\label{eq:set-S}
	S:=\{z\in \mathbb{C}|\  |\Re(z)-l|\le -\ii \tau/2, |\Im(z)|\le 1/4 \},
\end{equation}
into the whole complex plane $\mathbb{C}\cup \{\infty\}$ with two cuts. 
Therefore, we turn to study $z\in S$ instead of $\lambda\in \mathbb{C}\cup \{\infty\}$ in the following. 

Combined with Eq. \eqref{eq:trans-elliptic-theta}, the corresponding fundamental solutions \eqref{eq:Phi-elliptic} of the related Lax pair could be expressed in theta functions:
\begin{equation}\label{eq:Phi-solution}
	\Phi
	=\frac{\gamma \vartheta_3(2\ii l)}{\vartheta_4(\hat{\alpha}\xi)}\Upsilon
	\begin{bmatrix}
		\frac{\vartheta_1(\ii(z-l) -\hat{\alpha}\xi )}{\vartheta_4(\ii(z-l))} &
		\frac{\vartheta_3(\ii (z+l)+\hat{\alpha}\xi )}{\vartheta_2(\ii(z+l))} \\
		\frac{\ii\vartheta_3(\ii (z+l)-\hat{\alpha}\xi )}{\vartheta_2(\ii (z+l))}
		&\frac{\ii\vartheta_1(\ii (z-l)+\hat{\alpha}\xi )}{\vartheta_4(\ii( z-l))}
	\end{bmatrix}\mathbf{E}, 
\end{equation}	
where $\mathbf{E}=\diag\left(\ee^{\ii W_1(z)\xi+\ii V_1(z)\eta},\ee^{\ii W_2(z)\xi+\ii V_2(z)\eta}\right)$,
\begin{equation}\label{eq:E1-E2-W}
	\begin{split}
		W_1(z):=&\lambda-\ii \alpha Z(z_l), \quad \Upsilon=\diag(1,\ee^{-\ii (\omega \xi+ \kappa \eta)}), \\ W_2(z):=&\lambda+\hat{\alpha}\pi-\ii \alpha Z(\ii K'- K_l-z_l).
	\end{split}
\end{equation} 
Functions $\lambda\equiv \lambda(z)$ and $y\equiv y(z)$ are defined in Eq. \eqref{eq:lambda-y}. $K\equiv K(k)$ and $K'\equiv K(k')$ are both the first complete elliptic integrals defined in Def. \ref{define:complete-elliptic-integrals}; 
$\vartheta_i(z)$, $i=1,2,3,4$, are theta functions defined in Def. \ref{define:theta}; 
$Z(x)\equiv Z(x,k)$ is the Zeta function defined in Def. \ref{define:Zeta}; 
parameters $\kappa$ and $s$ are defined in Eq. \eqref{eq:solution-seed};
and functions $V_{1,2}(z)$ are defined in Eq. \eqref{eq:define-V}.
It should be noticed that functions $W_1(z)$ and $W_2(z)$ are not mutually independent. 
Combined Eqs. \eqref{eq:lambda-y} and \eqref{eq:E1-E2-W} with formulas listed in Appendix \ref{appendix:Elliptic functions}, functions $W_{1,2}(z)$ satisfy
\begin{equation}\label{eq:W-1-2}
	W_2(z)-\omega+W_1(z)=0,
\end{equation}
where $\omega$ is defined in Eq. \eqref{eq:solution-seed}, which implies $W'_2(z)=-W_1'(z)$, shown in Appendix \ref{appendix:Lemma}. Similarly, based on the definition of functions $V_{1,2}(z)$ in Eq. \eqref{eq:define-V}, we get $V'_2(z)=-V'_1(z)$.

Denote the branch points as
\begin{equation}\label{eq:define-branch}
	\mathcal{B}:=\left\{z_{1,2}=- \ii /4  \mp\ii \tau /4,\quad z_{3,4}=\ii /4  \mp\ii \tau /4\right\}.
\end{equation}
Substituting $z_i\in \mathcal{B}$, we can directly deduce $y(z_i)=0$. 
Combining this with fundamental solutions expressed by the spectral parameter $\lambda$ and the function $y$ in Eq. \eqref{eq:Phi-elliptic}, we obtain $\beta_1=\beta_2$, $\theta_1=\theta_2$, and $r_1=r_2$.
Thus, when $z_i\in \mathcal{B}$, two column vectors of the fundamental solution $\Phi(\xi,\eta;\lambda_i)$ in Eq. \eqref{eq:Phi-solution} exhibit a linear correlation, which means $\det(\Phi(\xi,\eta;\lambda_i))=0$, $\lambda_i=\lambda(z_i)$.

\section{The MI analysis}\label{section:MI}
The MI analysis involves linearizing INS equations and calculating eigenvalues $\Omega$ and eigenvectors $\mathbf{p}$ of the related linear operator $\mathcal{L}$. Perturbed functions of elliptic function solutions \eqref{eq:solution-seed} for INS equations have the form
\begin{equation}\label{eq:perturbed-solutions}
	p(\xi ,\eta)=\ee^{\ii  (\omega \xi+\kappa \eta)
	}\left(p_1(\xi)\ee^{\ii \Omega \eta}+p_2^*(\xi)\ee^{-\ii \Omega^* \eta}\right).
\end{equation}
Substituting the solution $\mathbf{Q}_{12}+\epsilon p(\xi,\eta)$, $|\epsilon| \ll 1$ into INS equations, we obtain that the linearized INS equation could be transformed into
\begin{equation}\label{eq:soliton-linearized}
	\mathcal{L}
	\mathbf{p}=\Omega
	\mathbf{p}, \qquad  \mathbf{p}=\begin{bmatrix}
		p_1(\xi) & p_2(\xi)
	\end{bmatrix}^{\top},
\end{equation}
where $\mathbf{p}$ is the eigenvector corresponding to the eigenvalue $\Omega\in \mathbb{C}$ of the linear operator $\mathcal{L}$.
Therefore, the MS/MI problems revolve around studying the eigenvalue $\Omega$ of Eq. \eqref{eq:soliton-linearized} for the bounded function $\mathbf{p}$.

Define functions $I_{1,2}(z)$, $\Gamma_i$, and $\Lambda_i$ as follows:
\begin{equation}\label{eq:define-Gamma-Lambda}
	\begin{split}
		&I_{1,2}(z):=2W_{1,2}(z)\pm\hat{\alpha}\pi - \omega, \,\, 
		\Lambda_{i}:=I'_{1}(z_i)/(4\ii K),\\
		&\Gamma_{i}:=\Omega'_{1}(z_i)/(4\ii K), \quad z_i\in \mathcal{B}.
	\end{split}
\end{equation}
Introduce a subset of the set $S$ in Eq. \eqref{eq:set-S} as
\begin{equation}\label{eq:define-set-Sb}
	S_b :=\left\{z\in S\left| \Im(I_j(z))=0, \quad j=1,2\right.\right\}.
\end{equation}
Since $W_1(z)=\omega-W_2(z)$, $\omega\in \mathbb{R}$, as stated in Eq. \eqref{eq:W-1-2}, we know $\Im(I_1(z))=-\Im(I_2(z))$.
Then, to determine the set of $S_b$, it is equivalent to getting the set about $z$ such that $\Im(I_1(z)-I_2(z))=0$.
By Eq. \eqref{eq:derivative-zeta}, the derivative of the function $I_1(z)-I_2(z)$ is $(I_1(z)-I_2(z))'
=4\alpha K \left(\dn^2(z_l)+\dn^2(z_l+K_l-\ii K')-2E/K\right)$.
In the view of \cite{LingS-21-mKdV-stability}, we could obtain that when $l=0$ and $l=-\ii \tau/4$, 
the set $S_b$ consists of the real line and two curves. 
For $l=-\ii \tau/4$, two curves are parallel to the real line.
For $l=0$, if $2E/K \ge 1$, two curves in set $S_b$ are intersecting with the real axis; 
if $2E/K<1$, two curves in set $S_b$ are intersecting with the imaginary axis.
For the case $l\in (0,-\ii \tau/4)$ described in \cite{FengLT-20}, we could also obtain the similar results, by using a similar approach as in the case of $l=0$, which we are unable to present here.

By symmetric properties of the Hamiltonian systems \cite{MarianaK-08}, eigenvalues $\Omega(z)$ for INS equations derived by the AKNS system are symmetric about the imaginary axis.
This symmetry implies that both $\Omega(z)$ and $-\Omega^*(z)$ are eigenvalues of the operator \eqref{eq:soliton-linearized}.
Thus, if exists $z\in S_b$ such that $\Omega(z)\in \mathbb{C}\backslash \mathbb{R}$, the elliptic function solution is modulationally unstable; if $\Omega(z)\in \mathbb{R}$, for any $z\in S_b$, the elliptic function solution is modulationally stable.
If $\Omega_{1,2}(z)=0$ and $I_{1,2}(z)=0$, the corresponding consequences indicate baseband MI or baseband MS \cite{BaronioCDLOW-14,BaronioCGWC-15}.

In summary, we obtain the following results about the stability of elliptic function solutions.
Set two linearly independent solutions $\mathbf{p}_j$ as 	\begin{equation}\label{eq:eigenvectors}
	\mathbf{p}_j=\begin{bmatrix}
		\Phi_{1j}^2\ee^{-\ii  \omega \xi-2 \ii V_j\eta}\\[3pt]
		\Phi_{2j}^2\ee^{\ii  \omega\xi-2\ii V_j\eta}
	\end{bmatrix}
	=\begin{bmatrix}
		\hat{p}_{1j}(\xi)\ee^{\ii I_j\xi} \\
		\hat{p}_{2j}(\xi)\ee^{\ii I_j\xi}
	\end{bmatrix},\,\,\, j=1,2,
\end{equation}	
where $I_j\equiv I_j(z), z\in S_b$, defined in Eq. \eqref{eq:define-Gamma-Lambda}; $\hat{p}_{ij}(\xi)$ are $2K$ periodic functions; 
and $\Phi_{ij}$ is the $(i,j)$-element of $\Phi$ defined in \eqref{eq:Phi-solution}.
Together with Eqs. \eqref{eq:soliton-linearized} and \eqref{eq:eigenvectors}, it becomes apparent that the eigenvalues corresponding to eigenvectors $\mathbf{p}_j$ are represented as $\Omega_j\equiv \Omega_j(z)=2V_j-\kappa$. 
By Eqs. \eqref{eq:Phi-solution}, \eqref{eq:define-set-Sb} and \eqref{eq:eigenvectors}, it is evident that eigenvectors $\mathbf{p}_j$ remain bounded only when $z\in S_b$.
Consequently, studying the MS/MI of elliptic function solutions for INS equations is transformed into examining the values of $\Omega_j(z)$ for $z\in S_b$.
When $\Omega_{1,2}(z)\in \mathbb{C}\backslash \mathbb{R}, z\in S_b\backslash \mathcal{B}$, elliptic function solutions are modulationally unstable; when $\Omega_{1,2}(z)\in \mathbb{R}, z\in S_b \backslash \mathcal{B}$, elliptic function solutions are modulationally stable. 
	
Consider a special case $\Im(I_{1,2}(z))\equiv 0$, where $z\in S$. 
The tangent vector field of the level curves $\Im(I_j(z))\equiv \mathrm{const}$ can then be represented as
\begin{equation}\nonumber
	\left(-\frac{\dd \Im(I_j)}{\dd \Im(z)},\frac{\dd \Im(I_j)}{\dd \Re(z) }\right)
	=\left(-\Re\left(\frac{\dd I_j}{\dd z}\right),\Im\left(\frac{\dd I_j}{\dd z}\right)\right),
\end{equation}
where $I_j\equiv I_j(z)$.
The derivative of $I_j(z)$ along the line $\Im(I_j(z))=0$ could be expressed as
\begin{equation}\nonumber
	\begin{split}
		&\left(\frac{\dd I_j}{\dd \Re(z)},\frac{\dd I_j}{\dd \Im(z)}\right)\cdot \left(-\frac{\dd \Im(I_j)}{\dd \Im(z) },\frac{\dd \Im(I_j)}{\dd \Re(z) }\right)\\
		=&-\left(\Re \left(\frac{\dd I_j}{\dd z}\right) \right)^2-\left(\Im \left(\frac{\dd I_j}{\dd z}\right) \right)^2 \le 0,
	\end{split}
\end{equation}
which implies that functions $I_j(z)$ along the curve $\Im(I_j(z))\equiv 0$ is monotonicity.
Plugging $z_i\in \mathcal{B}\subset S_b$, we get $I_j(z_i)=0$. Based on the monotonicity, we could obtain that if and only if $z\in \mathcal{B}\subset S_b$, the equation $I_j(z)=0$ holds.
Since $I_{1,2}(z_i)=0$ and $\Omega_{1,2}(z_i)=0$, $z_i\in \mathcal{B}$, we cannot directly justify whether solutions are stable or not. 
Then, we turn to consider the limitation 
\begin{equation}\label{eq:limitation}
	\!\!\!\! \!\! \lim_{z\rightarrow z_i\in \mathcal{B}} \!\!\frac{\Omega_{1,2}(z)}{I_{1,2}(z)}
	\!=\!\frac{\Omega'_{1,2}(z_i)(z-z_i)\!+\!\mathcal{O}((z-z_i)^2)}{I'_{1,2}(z_i)(z-z_i)\!+\!\mathcal{O}((z-z_i)^2)}
	\!=\!\frac{\Gamma_{i}}{\Lambda_{i}},\!\!\!\!\!
\end{equation}	
since $\Omega'_2(z)=-\Omega'_1(z)$ and $I'_{1}(z)=-I'_{2}(z)$.
By Eq. \eqref{eq:limitation}, the perturbed function $p(\xi ,\eta)$ \eqref{eq:perturbed-solutions} at branch points could be expressed as $p(\xi,\eta)=(\hat{p}_1(\xi)\ee^{\ii I_j(z_i)(\xi+ \Gamma_{i}\eta/\Lambda_i )}+\hat{p}_2^*(\xi)\ee^{-\ii I_j(z_i)(\xi+ \Gamma_{i}^*\eta/\Lambda_i^* )})\ee^{\ii (\omega \xi+\kappa \eta)}$, $z_i\in \mathcal{B}$.
Thus, when $\Gamma_{i}/\Lambda_{i}\in \mathbb{R}$, the perturbed function $p(\xi ,\eta)$ consistently maintains its diminutive magnitude as $\eta$ increases, indicating a stable dynamic behavior. 
In cases when $\Gamma_i/\Lambda_i\in \mathbb{C}\backslash\mathbb{R}$, $p(\xi,\eta)$ is not always sufficiently small,
which reveals that elliptic function solutions are baseband modulationally unstable \cite{BaronioCDLOW-14,BaronioCGWC-15}.
Conversely, when $\Gamma_i/\Lambda_i\in \mathbb{R}$, the small perturbed function $p(\xi,\eta)$ would float within a small value and not increase or decrease over time, which corresponds to the baseband MS.
In summary, we obtain: 
\begin{itemize}
	\item[(1)] If exists $z_i\in S_b$ defined in Eq. \eqref{eq:define-set-Sb}, such that $\Omega_j(z_i)\in \mathbb{C}\backslash \mathbb{R}$ and $I_j(z_i)\neq 0$, the elliptic function solutions are modulationally unstable; if $\Gamma_{i}/\Lambda_{i}\in \mathbb{C}\backslash \mathbb{R}$ with $I_j(z_i)\rightarrow 0$, $\Omega_j(z_i)\rightarrow 0$, $z_i\in \mathcal{B}$, the elliptic function solutions are baseband modulationally unstable;
	\item[(2)] For any $z\in S_b$, if $\Omega_j(z)\in \mathbb{R}$, $I_j(z)\neq 0$, the elliptic function solutions are modulationally stable; if $\Gamma_{i}/\Lambda_{i}\in \mathbb{R}$ with $I_j(z_i)\rightarrow 0$, $\Omega_j(z_i)\rightarrow 0$, $z_i\in \mathcal{B}$, the elliptic function solutions are baseband modulationally stable.
\end{itemize}

\subsection{The stability of INS equations}\label{sec:stability-INS-MI}
Before delving into the stability of elliptic function solutions for the INS equations, we are going to calculate some parameters that are useful in the stability analysis.
For $z_i\in \mathcal{B}$, $i=1,2,3,4$, parameters $\hat{y}_i=y'(z_i)/(2\ii K)$ defined in Eq. \eqref{eq:lambda-y} could be expressed by $d_{1,2,3}$ as
\begin{equation}\label{eq:define-y-hat}
	\hat{y}_{1,2}=\alpha^2\left(d_1^2 \pm d_2d_3\right)\left(d_3\mp d_2\right)\pm\ii  \alpha^2d_1\left(d_3\mp d_2\right)^2,
\end{equation}
and $\hat{y}_{3,4}=-\hat{y}^*_{1,2}$;
parameters $\lambda_i=\lambda(z_i)$ defined in Eq. \eqref{eq:lambda-y} are 
\begin{equation}\label{eq:define-lambda-hat}
	\lambda_{1,2}=\pm \alpha d_1/2+\ii \alpha(d_3 \mp d_2)/2,
\end{equation}
$\lambda_{3,4}=-\lambda^*_{1,2}$;
parameters $\Lambda_i$ defined in Eq. \eqref{eq:define-Gamma-Lambda} could be written as
\begin{equation}\label{eq:define-Lambda}
	\Lambda_{1,2}=-\alpha d_1(d_2\mp d_3) -\ii \alpha(d_1^2\pm d_2d_3-E/K),
\end{equation}  
and $\Lambda_{3,4}=-\Lambda^*_{1,2}$.
The detailed calculation process of equations \eqref{eq:define-y-hat}-\eqref{eq:define-Lambda} is provided in Appendix \ref{appendix:Lemma}. Based on them, we are going to study the stability of three representative INS equations: the \eqref{eq:NLS-equation} equation, the \eqref{eq:mKdV-equation} equation, and the \eqref{eq:SG-equation} equation.

For the \eqref{eq:NLS-equation} equation, the perturbed functions \eqref{eq:eigenvectors} of elliptic function solutions \eqref{eq:solution-seed} have the form $p(\xi,\eta)$ in Eq. \eqref{eq:perturbed-solutions} and the linearized NLS equation is converted into
\begin{equation}\label{eq:NLSE-linearized}
	\mathcal{L}\mathbf{p}
	\!=\!\begin{bmatrix}
		\frac{1}{2}\partial_{\xi}^2+2|q|^2-\kappa & q^2\\
		-\left(q^*\right)^2 & -\frac{1}{2}\partial_{\xi}^2+2|q|^2-\kappa
	\end{bmatrix}\!\mathbf{p}
	=\Omega \mathbf{p}.
\end{equation}
By the stationary zero curvature equation \eqref{eq:zero-curvature-equation-L}, the Lax pair \eqref{eq:NLS-Lax-pair}, and the coordinate transformation \eqref{eq:coordinate-transformation}, we get
\begin{equation}\nonumber
	\ii \sigma_3 \mathbf{L}_{\eta}^{\mathrm{off}}+\frac{1}{2}\mathbf{L}_{\xi\xi}^{\mathrm{off}}+2\mathbf{Q}^2\mathbf{L}^{\mathrm{off}}-\mathbf{Q}\mathbf{L}^{\mathrm{off}}\mathbf{Q}-\kappa \mathbf{L}^{\mathrm{off}}=0, 
\end{equation}
which implies that $L_{12}$ and $L_{21}$ satisfy the linearized NLS equation \eqref{eq:NLSE-linearized}. 
Due to the su(2)-symmetry \eqref{eq:symmetry-su} of $\mathbf{U}(\lambda;\mathbf{Q})$ and $\mathbf{V}(\lambda;\mathbf{Q})$, the matrix function $\mathbf{L}$ could be expressed by matrix functions $\Phi$: $\mathbf{L}=\Phi\sigma_3\left(\ii \sigma_2\right)\Phi^{\top}\left(\ii \sigma_2\right)/2$ with $L_{12}=\Phi_{11}\Phi_{12}$ and $L_{21}=-\Phi_{21}\Phi_{22}$. 
Thus, two linearly independent solutions $\mathbf{p}_{1,2}$ are expressed in Eq. \eqref{eq:eigenvectors} and the related eigenvalues are $\Omega_{1,2}(z)=\pm 2y(z)$.
Since $y(z_i)=0$, $z_i \in \mathcal{B}$, we obtain $\Omega_{1,2}(z_i)=0$.
Subsequently, we need to study parameters $\Gamma_i$ and $\Lambda_i$ defined in Eq. \eqref{eq:define-Gamma-Lambda}. 
By the definition of functions $\Omega_{1,2}(z)$ in Eq. \eqref{eq:E1-E2-W}, we obtain $\Omega'_{1,2}(z_i)=4\ii K \hat{y}_i$. 
Since $\Gamma_i=\Omega'_{1}(z_i)/(4\ii K)$, we get
\begin{equation}\label{eq:define-Gamma-NLS}
	\Gamma_{1,2}\!=\!\hat{y}_{1,2}\!=\!\alpha^2 (d_3\mp d_2)( (d_1^2 \pm d_2d_3)-\ii d_1(d_2\mp d_3)),
\end{equation}
$\Gamma_{3,4}=-\Gamma_{1,2}^*$. By parameters $\Lambda_i$ \eqref{eq:define-Lambda}, we get $\Gamma_{i} /\Lambda_{i}\in \mathbb{C}\backslash\mathbb{R}$, $l\in \left[0,-\ii \tau/4\right]$. 
Thus, elliptic function solutions of the NLS equation are baseband modulationally unstable.

The linearized \eqref{eq:mKdV-equation} equation could be expressed as
\begin{equation}\label{eq:mKdV-linearized}
	\mathcal{L}
	\mathbf{p}=\begin{bmatrix}
		\mathcal{L}_{11} & -6uu_{\xi}\\
		-6uu_{\xi} & 	\mathcal{L}_{22}
	\end{bmatrix}\mathbf{p}=\Omega
	\mathbf{p},
\end{equation}
with $\mathcal{L}_{22}=\mathcal{L}_{11}=-\partial_{\xi}^3+\left(s-6u^2\right)-6uu_{\xi}$. 
Together with the stationary zero curvature equation \eqref{eq:zero-curvature-equation-L}, the Lax pair \eqref{eq:mKdV-Lax-pair}, and the coordinate transformation \eqref{eq:coordinate-transformation}, we get
\begin{equation}\nonumber
	\mathbf{L}_{\eta}^{\mathrm{off}}+\mathbf{L}_{\xi\xi\xi}^{\mathrm{off}}+6\mathbf{Q}^2\mathbf{L}^{\mathrm{off}}-6\mathbf{Q}\mathbf{L}^{\mathrm{off}}\mathbf{Q}_{\xi}-s\mathbf{L}^{\mathrm{off}}=0.
\end{equation}
Then, we obtain eigenvectors $\mathbf{p}_{1,2}$ \eqref{eq:eigenvectors} of the operator $\mathcal{L}$ \eqref{eq:mKdV-linearized} and the corresponding eigenvalues $\Omega_{1,2}(z)=\pm 8 \lambda(z) y(z)$ of the operator $\mathcal{L}$. Since $\Omega_{1,2}(z_i)=0$, $z_i\in \mathcal{B}$, $i=1,2,3,4$, we consider the limitation \eqref{eq:limitation}. By Eqs. \eqref{eq:define-y-hat} and \eqref{eq:define-lambda-hat}, we gain
\begin{equation}\label{eq:define-Gamma-mKdV}
	\begin{split}
		\Gamma_{1,2}=& \pm2\alpha^3d_1d_3(d_1\pm \ii d_3)^2 \in \mathbb{C}\backslash(\ii \mathbb{R}\cup\mathbb{R}), \,\,\, l=0,\\
		\Gamma_{1,2}=&\pm 2 \ii\alpha^3(d_3\mp d_2)^2d_2d_3 \in \ii \mathbb{R}, \quad l=-\ii \tau/4,
	\end{split}
\end{equation}
$\Gamma_{3,4}=-\Gamma_{1,2}^*$ and
\begin{equation}\label{eq:define-Lambda-mKdV}
	\begin{split}
		\!\!\!\!
		\Lambda_{1,2}=&\pm \alpha d_1 d_3-\ii \alpha(d_1^2-E/K) \!\in \!\mathbb{C}\backslash(\ii \mathbb{R}\cup\mathbb{R}),l=0, \!\!\!\!\\
		\!\!\!\!
		\Lambda_{1,2}=& \ii \alpha(E/K\mp d_2d_3) \in \ii \mathbb{R},\qquad l=-\ii \tau/4,
	\end{split}
\end{equation}
$\Lambda_{3,4}=-\Lambda_{1,2}^*$ by Eq. \eqref{eq:define-Lambda}, which implies $\Gamma_{i}/\Lambda_{i}\in \mathbb{C}\backslash\mathbb{R}$ when $l=0$ and $ \Gamma_{i}/\Lambda_{i}\in \mathbb{R}$ when $l=-\ii\tau/4$. Thus, $\cn$-type solutions ($l=0$) correspond to baseband MI and $\dn$-type solutions ($l=-\ii \tau/4$) are baseband modulationally stable.

The linearized \eqref{eq:SG-equation} equation could be expressed as
\begin{equation}\label{eq:SG-linearized}
	\mathcal{L}\mathbf{p}=\begin{bmatrix}
		\partial_{\xi}^2+\cos(v) & 0 \\
		0 & 	\partial_{\xi}^2+\cos(v)
	\end{bmatrix}
	\mathbf{p}=\Omega \partial_{\xi}
	\mathbf{p}.
\end{equation}
By Eqs. \eqref{eq:SG-Lax-pair}, \eqref{eq:zero-curvature-equation-L}, and \eqref{eq:coordinate-transformation}, we could get
\begin{equation}\nonumber
	\mathbf{L}_{\xi \eta}^{\mathrm{off}}-\ii \left(\left[\mathbf{Q}_{\eta},\mathbf{L}\right]\right)^{\mathrm{off}}+\frac{\ii}{2}\sigma_3\left(\mathbf{V}_{\eta}(\lambda;\mathbf{Q})\mathbf{L}\right)^{\mathrm{off}}-\frac{\ii}{2}\sigma_3\mathbf{L}^{\mathrm{off}}=0.
\end{equation}
Then, the eigenvectors of the linearized SG equation \eqref{eq:SG-linearized} are $\mathbf{p}_i$ \eqref{eq:eigenvectors} and the corresponding eigenvalues are $\Omega_{1,2}(z)=\pm 2s y(z)/ \lambda(z)$, which also satisfy $\Omega_{1,2}(z_i)=0$, $z_i\in \mathcal{B}$. 
To parameters $\Gamma_i$ and $\Lambda_i$, 
by Eqs. \eqref{eq:define-y-hat}, \eqref{eq:define-lambda-hat} and the definition of parameters $\Gamma_i$ defined in Eq. \eqref{eq:define-Gamma-Lambda}, we obtain
\begin{equation}\nonumber
	\Gamma_{1,2}=2s\alpha\frac{\left(d_1^2 \pm d_2d_3\right)\left(d_3\mp d_2\right)\pm\ii  d_1\left(d_3\mp d_2\right)^2}{\pm  d_1+\ii (d_3 \mp d_2)},
\end{equation}
$\Gamma_{3,4}=-\Gamma_{1,2}^*$.
When $l=0$ ($d_2=0$) or $l=-\ii\tau/4$ ($d_1=0$), parameters $\Gamma_i$ are simplified as 
\begin{equation}\label{eq:define-Gamma-SG}
	\begin{split}
		\Gamma_{1,2}=&\pm 2\alpha s d_1 d_3 \in \mathbb{R}, \quad l=0,\\
		\Gamma_{1,2}=&\mp 2 \ii\alpha s d_2d_3 \in \ii \mathbb{R}, \quad l=-\ii \tau/4,
	\end{split}
\end{equation}
$\Gamma_{3,4}=-\Gamma_{1,2}^*$ and
parameters $\Lambda_i$ are expressed in Eq. \eqref{eq:define-Lambda-mKdV}.
It can be readily deduced that if $l=0$, then $\Gamma_{j}/\Lambda_{j}\in \mathbb{C}\backslash\mathbb{R}$; 
if $l=-\ii \tau/4$, then $\Gamma_{j}/\Lambda_{j}\in \mathbb{R}$. 
Therefore, rotational wave solutions ($l=-\ii \tau/4$) of the SG equation are baseband modulationally stable; while librational wave solutions ($l=0$) of the SG equation are baseband modulationally unstable.

\section{Rational-elliptic-localized wave solutions}\label{section:eRWs}

It is well known that RWs on plane wave backgrounds could be expressed in the rational form  with respect to temporal and spatial variables. 
In this context, our objective is to construct eRWs expressed in a rational form, akin to RWs.
Firstly, in view of the Darboux-B\"{a}cklund transformation, multi-elliptic-localized wave solutions of the INS equations are expressed in theta functions. 
Secondly, building upon the symmetric of solutions $\Phi$ at branch points $z_i\in \mathcal{B}$, we rewrite the multi-elliptic-localized wave solutions.
Taking into account the limitations of these solutions as $\epsilon\rightarrow 0$, we proceed to construct the higher-order elliptic-localized wave solutions for INS equations.
Finally, we illustrate elliptic-localized wave solutions of the NLS equation, the mKdV equation, and the SG equation in Fig. \ref{fig:order-1}.
These solutions not only encompass eRW solutions but also include elliptic-soliton solutions.

\subsection{Elliptic-localized wave solutions}
As is known to all, the Darboux transformation allows us to  generate a new Lax pair:
\begin{equation}\label{eq:Phi-1}
	\Phi_{ \xi}^{[1]}=\mathbf{U}\left(\lambda;\mathbf{Q}^{[1]}\right)\Phi^{[1]}, \qquad 
	\Phi^{[1]}:=\mathbf{T}^{[1]}\Phi,
\end{equation} 
where $\mathbf{T}^{[1]}=\mathbf{T}^{[1]}(\lambda;\xi,\eta)$.
Based on the su(2)-symmetry \eqref{eq:symmetry-su} and twist-symmetry \eqref{eq:symmetry-twist} of matrices $\mathbf{U}(\lambda;\mathbf{Q})$ and $\mathbf{V}(\lambda;\mathbf{Q})$, the matrix $\Phi(\xi,\eta;\lambda)$ satisfies equations $\Phi(\xi,\eta;\lambda)\Phi^{\dagger}(\xi,\eta;\lambda^*)=\mathbb{I}_2$ and $\Phi(\xi,\eta;\lambda)\Phi^{\top}(\xi,\eta;-\lambda)=\mathbb{I}_2$. 
In accordance with Eq. \eqref{eq:Phi-1}, the Darboux matrix $\mathbf{T}^{[1]}(\lambda;\xi,\eta)$ satisfies
\begin{equation}\label{eq:T-sym}\nonumber
	\begin{split}
		&\mathbf{T}^{[1]}(\lambda;\xi,\eta)\left(\mathbf{T}^{[1]}(\lambda^*;\xi,\eta)\right)^{\dagger}=\mathbb{I}_2,\\ &\mathbf{T}^{[1]}(\lambda;\xi,\eta)\left(\mathbf{T}^{[1]}(-\lambda;\xi,\eta)\right)^{\top}=\mathbb{I}_2.
	\end{split}
\end{equation} 
Subsequently, the Darboux matrix is divided into the following two cases. 
If matrices $\mathbf{U}(\lambda;\mathbf{Q})$ and $\mathbf{V}(\lambda;\mathbf{Q})$ just satisfy the su(2)-symmetry, the Darboux matrix could be expressed as 
\begin{equation}\label{eq:T-1}\nonumber
	\mathbf{T}_1=\mathbb{I}_2-\frac{\lambda_1-\lambda_1^*}{\lambda-\lambda_1^*}\frac{\psi_1\psi_1^{\dagger}}{\psi_1^{\dagger}\psi_1}, \quad 
	\psi_1:=\Phi( \xi,\eta;\lambda_1)\mathbf{c}_1,
\end{equation}	
where $\mathbf{c}_i
=\begin{bmatrix}
	c_{i1} & c_{i2}
\end{bmatrix}^{\top}$, $\lambda_1\equiv\lambda(z_1)\in  \mathbb{C} \backslash \mathbb{R}$, $\mathbb{I}_2$ is the $2\times 2$ identity matrix and $\Phi$ is defined in Eq. \eqref{eq:Phi-solution}.
If matrices $\mathbf{U}(\lambda;\mathbf{Q})$ and $\mathbf{V}(\lambda;\mathbf{Q})$ satisfy both the su(2)-symmetry and twist-symmetry, the Darboux matrix would be divided into two cases:
\begin{equation}\nonumber
	\mathbf{T}_1^{\mathbf{P}}= \mathbb{I}_2-\frac{\lambda_1-\lambda_1^*}{\lambda-\lambda_1^*}\frac{\psi_1\psi_1^{\dagger}}{\psi_1^{\dagger}\psi_1}, \,\, \lambda_1\in \ii \mathbb{R}, \,\,
	\psi_1\psi_1^{\dagger}=(\psi_1\psi_1^{\dagger})^{\top},
\end{equation}
and
\begin{equation}\nonumber
	\begin{split}
		\mathbf{T}_1^{\mathbf{C}}
		= & \mathbb{I}_2-
		\begin{bmatrix}
			\psi_1 & \psi_1^*
		\end{bmatrix}
		\mathbf{M}^{-1}_2(\lambda \mathbb{I}_2-\mathbf{D}_2)^{-1}\begin{bmatrix}
			\psi_1^{\dagger}\\[4pt]
			\psi_1^{\top}\\
		\end{bmatrix},\\
		\mathbf{M}_2=&\begin{bmatrix}
			\frac{\psi_1^{\dagger}\psi_1}{\lambda_1-\lambda_1^*}&\frac{\psi_1^{\dagger}\psi_1^*}{-\lambda_1^*-\lambda_1^*} \\[4pt]
			\frac{\psi_1^{\top}\psi_1}{\lambda_1+\lambda_1}& \frac{\psi_1^{\top}\psi_1^*}{-\lambda_1^*+\lambda_1} \end{bmatrix},
	\end{split}
\end{equation}
where $\lambda_1\in \mathbb{C}\backslash \left(\ii \mathbb{R}\cup \mathbb{R}\right)$, $\mathbf{D}_2=\diag\left(\lambda_1^*, -\lambda_1\right)$.

Based on the elementary Darboux transformation $\mathbf{T}_1$, $\mathbf{T}_1^{\mathbf{P}}$, and $\mathbf{T}_1^{\mathbf{C}}$, we can iterate them to derive multi-order transformations, collectively referred to as the multi-fold Darboux matrix:
\begin{equation}\label{eq:higher-order-T}
	\begin{split}
		\mathbf{T}^{[n]}
		=&\mathbf{T}_{n}^{\mathbf{J}}\mathbf{T}_{n-1}^{\mathbf{J}}\cdots\mathbf{T}_{2}^{\mathbf{J}}\mathbf{T}_{1}^{\mathbf{J}}\\
		=&\mathbb{I}_2-\mathbf{X}_m\mathbf{M}_m^{-1}\left(\lambda\mathbb{I}_m-\mathbf{D}_m\right)^{-1}\mathbf{X}^{\dagger}_m,
	\end{split}
\end{equation}
where $\mathbf{J}=\mathbf{P}, \mathbf{C}, \emptyset,$ $n\le m \le 2n$, and
\begin{equation}\label{eq:define-X-D-M}
	\begin{split}
		&\mathbf{X}_m\!=\!\begin{bmatrix}
			\psi_1 & \psi_2 & \cdots & \psi_m
		\end{bmatrix},	\mathbf{D}_m\!=\!\diag\left(\lambda_1^*,\cdots,\lambda_m^*\right),\\
		&\mathbf{M}_m=\left( 	\frac{\psi_i^{\dagger}\psi_j}{\lambda_j-\lambda_i^*} \right)_{1\le i,j\le m}.
	\end{split}
\end{equation} 
For the NLS equation, we choose $\mathbf{J}=\emptyset$; whereas, for the mKdV equation and the SG equation, we select $\mathbf{J}=\mathbf{P}$ or $\mathbf{C}$. 
Upon reviewing the Darboux matrix $\mathbf{T}^{[n]}$ as defined in Eq. \eqref{eq:higher-order-T} and combining it with the B\"{a}cklund transformation, we can obtain an n-order elliptic-localized wave solution after an iteration.  
Let the numbers of matrices $\mathbf{T}_i^{\mathbf{P}}$ and $\mathbf{T}_i^{\mathbf{C}}$ provided in Eq. \eqref{eq:higher-order-T} be denoted as $m_1$ and $m_2$. 
The values of $n=m_1+m_2$ and $m=m_1+2m_2$ hinge on the selection of the number of Darboux matrices $\mathbf{T}_1^{\mathbf{P}}$ and $\mathbf{T}_1^{\mathbf{C}}$.
If $\mathbf{T}^{[n]}$ is obtained by multiplying $\mathbf{T}_i$, the values of $n$ and $m$ satisfy the condition $n=m$.

Based on the definition of the Darboux matrix, we deduce the matrix
\begin{equation}\nonumber
	\mathbf{U}\left(\lambda;\mathbf{Q}^{[n]}\right)
	=\mathbf{T}_{\xi}^{[n]}\left(\mathbf{T}^{[n]}\right)^{-1}
	+\mathbf{T}^{[n]}\mathbf{U}\left(\lambda;\mathbf{Q}\right)\left(\mathbf{T}^{[n]}\right)^{-1}.
\end{equation}
By the Sherman-Morrison-Woodbury-type matrix identity, 
the new solution could be written as
\begin{equation}\label{eq:solution-higher-order}
	\mathbf{Q}^{[n]}_{12}
	=\frac{\det(\mathbf{Q}_{12}\mathbf{M}_m-2\mathbf{X}_{m,2}^{\dagger}\mathbf{X}_{m,1})}{\mathbf{Q}_{12}^{m-1}\det\left(\mathbf{M}_m\right)},
\end{equation}
where $\mathbf{Q}_{12}$ is the $(1,2)$-element of $\mathbf{Q}$ and $\mathbf{X}_{m,i}$ is the $i$-row of $\mathbf{X}_m$ defined in Eq. \eqref{eq:define-X-D-M}.
Combined with Eqs. \eqref{eq:solution-seed}, \eqref{eq:Phi-solution}, and \eqref{eq:solution-higher-order}, 
the multi-elliptic-localized wave solutions of INS equations could be written in theta functions as 
\begin{equation}\label{eq:solution-localized}
	\mathbf{Q}_{12}^{[n]}=\gamma\left(\frac{\vartheta_4(\hat{\alpha}\xi)}{\vartheta_2(\hat{\alpha}\xi+2\ii l)}\right)^{m-1}\frac{\det(\hat{\mathcal{P}})}{\det(\hat{\mathcal{H}})}\ee^{\ii (\omega \xi +\kappa \eta)},
\end{equation}
where matrices $\hat{\mathcal{P}}$ and $\hat{\mathcal{H}}$ are both $m\times m$ matrices, whose $(i,j)$-elements are given by $\hat{\mathcal{P}}_{ij}=\hat{\mathcal{P}}(z_i^*,z_j)$, $\hat{\mathcal{H}}_{ij}=\hat{\mathcal{H}}(z_i^*,z_j)$, and
\begin{equation}\label{eq:define-P-H-hat}
	\begin{split}
		\hat{\mathcal{P}}(z^*\!,z)\!=&\mathbf{c}^{\dagger}\mathbf{E}^{\dagger}\!\!
		\begin{bmatrix}
			\frac{\vartheta_2(\ii(z^*-z+2l)+\hat{\alpha}\xi)r}{-\vartheta_1(\ii(z^*-z))r^*} \!& \!
			\frac{\vartheta_4(\ii(z^*+z+2l)+\hat{\alpha}\xi)}{-\vartheta_3(\ii(z^*+z))rr^*} \\
			\frac{\vartheta_4(\hat{\alpha}\xi-\ii(z^*z-2l))r r^*}{\vartheta_3(\ii(-z^*-z))} \!& \!
			\frac{\vartheta_2(\ii(z-z^*+2l)+\hat{\alpha}\xi)r^*}{\vartheta_1(\ii(z-z^*))r}
		\end{bmatrix}\!\!\mathbf{E}\mathbf{c},\\ \hat{\mathcal{H}}(z^*\!,z)\!=&\mathbf{c}^{\dagger}\mathbf{E}^{\dagger}
		\begin{bmatrix}
			-\frac{\vartheta_4(\ii(z^*-z)+\hat{\alpha}\xi)}{\vartheta_1(\ii(z^*-z))} & 
			\frac{\vartheta_2(\ii(z^*+z)+\hat{\alpha}\xi)}{\vartheta_3(\ii(z^*+z))} \\
			\frac{\vartheta_2(-\ii(z^*+z)+\hat{\alpha}\xi)}{\vartheta_3(-\ii(z^*+z))} & 
			\frac{\vartheta_4(\ii(z-z^*)+\hat{\alpha}\xi)}{\vartheta_1(\ii(z-z^*))}
		\end{bmatrix}
		\mathbf{E}\mathbf{c},
	\end{split}
\end{equation}
where $z_i, z_j\in S\backslash \mathcal{B}$, $\mathbf{c}=[c_1,c_2]^{\top}$, $c_{1,2}\in \mathbb{C}$; $r=r(z)=:\vartheta_2(\ii(z+l))/\vartheta_4(\ii(z-l))$;
and the matrix $\mathbf{E}$ is defined in Eq. \eqref{eq:Phi-solution}.

\subsection{Rational-elliptic-localized wave solutions}
The solution \eqref{eq:solution-localized} with different parameters, can manifest diverse dynamic behaviors, including multi-elliptic-breathers, multi-elliptic-solitons, and multi-elliptic-soliton-breathers, as vividly illustrated in \cite{FengLT-20,LingS-21-mKdV-stability,LingS-22-mKdV,LingS-22-SG}. 
Therefore, a detailed presentation of these variations will not be reiterated here. 
It is worth noting that, precisely at the branch points, the elliptic-localized wave solutions of INS equations cannot be directly obtained using Eq. \eqref{eq:solution-localized}.

At the branch point $z_i\in \mathcal{B}$, functions $\lambda(z)$ and $y(z)$ satisfy the symmetric 
\begin{equation}\label{eq:lambda-y-zi}
	\lambda(z_i+\epsilon)=\lambda(z_i-\epsilon), \,\,\,
	y(z_i+\epsilon)=-y(z_i-\epsilon), \,\,\, \epsilon\in \mathbb{C}.
\end{equation}
Building upon the aforementioned symmetry of functions $y(z)$ and $\lambda(z)$, at the neighborhood of branch points, functions $\hat{\mathcal{P}}(z^*,z)$ and $\hat{\mathcal{H}}(z^*,z)$ could be expressed as the quadratic form:
\begin{equation}\label{eq:P-H-expand}
	\begin{split}
		\!\!\!\!\!&\hat{\mathcal{H}}(\hat{z}_i^*,\hat{z}_j) \!= \! 4\Sigma_i^{\dagger} \mathbf{H}(z_i^*,z_j) \Sigma_j,  \hat{\mathcal{P}}(\hat{z}_i^*,\hat{z}_j)\!= \!
		4\Sigma_i^{\dagger} \mathbf{P}(z_i^*,z_j) \Sigma_j,\! \!\!\!\!\!\!\!\!\!\!\!\! \\
		\!\!\!\!\!&\mathbf{H}(z_i^*,z_j)
		=\begin{bmatrix}
			\mathcal{H}^{[1,1]}(z_i^*,z_j) & \mathcal{H}^{[1,3]}(z_i^*,z_j) & \cdots \\
			\mathcal{H}^{[3,1]}(z_i^*,z_j) & \mathcal{H}^{[3,3]}(z_i^*,z_j) & \cdots \\
			\vdots & \vdots & \ddots \\
		\end{bmatrix}, \\
		\!\!\!\!\!&\mathbf{P}(z_i^*,z_j)
		=\begin{bmatrix}
			\mathcal{P}^{[1,1]}(z_i^*,z_j) & \mathcal{P}^{[1,3]}(z_i^*,z_j) & \cdots \\
			\mathcal{P}^{[3,1]}(z_i^*,z_j) & \mathcal{P}^{[3,3]}(z_i^*,z_j) & \cdots \\
			\vdots & \vdots & \ddots \\
		\end{bmatrix},
	\end{split}	
\end{equation}	
where $\Sigma_j=\begin{bmatrix} \epsilon_j & \epsilon_j^3 & \epsilon_j^5 & \cdots \end{bmatrix}^{\top}$, $\hat{z}_i=z_i+\epsilon_i$, $\epsilon_{i,j}$ are in the neighborhood of origin and
\begin{equation}\label{eq:define-P-H-2j-1}
	\begin{split}
	\mathcal{P}(z^*,z)=& -|E_1(z)|^2\frac{\vartheta_2(\ii(z^*-z+2l)+\hat{\alpha}\xi)r(z)}{\vartheta_1(\ii(z^*-z))r(z^*)},\\
	\mathcal{H}(z^*,z)=&-|E_1(z)|^2	\frac{\vartheta_4(\ii(z^*-z)+\hat{\alpha}\xi)}{\vartheta_1(\ii(z^*-z))},
	\end{split}
\end{equation}
with $E_1(z)=\exp(\ii W_1(z)\xi+\ii V_1(z)\eta+h(z))$. 
The function $h(z)$ is a polynomial function about $z$.
The function $r$ defined in Eq. \eqref{eq:define-P-H-hat} could be expressed as 
\begin{equation}\label{eq:R-hat-1234}
	r(z)=r_i+2\ii K r_iR_i(z-z_i)+\mathcal{O}\left((z-z_i)^2\right),
\end{equation}
where $R_{1,2}=-(d_3\mp d_2)-\ii (\omega/\alpha \mp d_1)$, $R_{3,4}=-R_{1,2}^*$, and $r_i=r(z_i)$ with $z_i\in \mathcal{B}$.
Appendix \ref{appendix:Lemma} provides details of the relevant formulas.

Combining with Eqs. \eqref{eq:solution-localized} and \eqref{eq:P-H-expand}, we could rewrite the elliptic-localized wave solutions as 
\begin{equation}\label{eq:localized-1}\nonumber
	\mathbf{Q}_{12}^{[1]}=\gamma\frac{4\Sigma_i^{\dagger}\mathbf{P}(z_i^*,z_i)\Sigma_i}{4\Sigma_i^{\dagger}\mathbf{H}(z_i^*,z_i)\Sigma_i}\ee^{\ii ( \omega \xi+ \kappa \eta)}, \quad z_i\in \mathcal{B},
\end{equation}
where $\epsilon_i\in \mathbb{C}$ is a small parameter; $\Sigma_i=[\epsilon_i, \epsilon_i^3,\cdots]^{\top}$;
\begin{equation}\label{eq:define-H-bf}
	\begin{split}
		&\mathbf{H}(z_i^*,z_j)
		=\begin{bmatrix}
			\mathcal{H}^{[1,1]}(z_i^*,z_j) & \mathcal{H}^{[1,3]}(z_i^*,z_j) & \cdots\\
			\mathcal{H}^{[3,1]}(z_i^*,z_j) & \mathcal{H}^{[3,3]}(z_i^*,z_j) & \cdots\\
			\vdots & \vdots & \ddots\\
		\end{bmatrix},\\
		&\mathcal{H}^{[N_i,N_j]}(z_i^*,z_j)	=\left.\frac{\dd^{N_i+N_j}\mathcal{H}(z^*,z)}{N_i!N_j!\dd z^{*N_i}\dd z^{N_j}}\right|_{z^*=z_i^*,z=z_j};
	\end{split}
\end{equation}
the definition of $\mathbf{P}(z_i^*,z_i)$ is similar to $\mathbf{H}(z_i^*,z_i)$; $\mathcal{H}(z^*,z)$ and $E_1(z)$ are defined in Eq. \eqref{eq:define-P-H-2j-1};
$r(z)=\vartheta_2(\ii(z+l))/\vartheta_4(\ii(z-l))$.
The rational-elliptic-localized wave solution could be expressed as
\begin{equation}\label{eq:rogue-wave-lim}
\hat{\mathbf{Q}}_{i}^{(1)}\!= \! \lim_{\epsilon_i\rightarrow 0} \! \mathbf{Q}_{12}^{[1]}
=\gamma\frac{\mathcal{P}^{[1,1]}(z_i^*,z_i)}{\mathcal{H}^{[1,1]}(z_i^*,z_i)}\ee^{\ii (\omega \xi+\kappa \eta)},\ z_i\in \mathcal{B}.
\end{equation}
Together with Eqs. \eqref{eq:define-P-H-2j-1} and \eqref{eq:define-H-bf}, a rational-elliptic-localized wave solution of INS equations could be written as 
\begin{equation}\label{eq:rogue-wave}
	\hat{\mathbf{Q}}_{i}^{(1)}=\ii \gamma\frac{\vartheta_2(2\ii l+\hat{\alpha}\xi-1/2)}{\vartheta_4(\hat{\alpha}\xi-1/2)}\left(1+
	\frac{\mathcal{P}_{i}}{\mathcal{H}_{i}}\right)\ee^{\ii (\omega\xi+\kappa \eta)},
\end{equation}
where $i=1,2$, $\mathcal{P}_{i}=2\hat{E}_{i}^{\mathbf{i}}(f_{1}-\ii R_{i}^{\mathbf{i}}-g_1)-2\ii R_{i}^{\mathbf{r}}\hat{E}_{i}^{\mathbf{r}}+f_{2}-g_2-(R_{i}^{\mathbf{r}})^2-g_1^2+(f_{1}-\ii R_{i}^{\mathbf{i}})^2$, $\mathcal{H}_i=(\hat{E}_i^{\mathbf{r}})^2+(\hat{E}_i^{\mathbf{i}}+g_1)^2+g_2$,
$g_1=Z(K+\alpha \xi)$, $g_2=\dn^2(K+\alpha\xi)$, $f_1=Z(4\ii l K+\alpha \xi+\ii K')-Z(K+\ii K')$, $f_2=\dn^2(4\ii l K+\alpha\xi+\ii K')$, $\hat{E}_{i}^{\mathbf{i}}=\Lambda_{i}^{\mathbf{i}}\xi+\Gamma_{i}^{\mathbf{i}}\eta$, $\hat{E}_{i}^{\mathbf{r}}=\Lambda_{i}^{\mathbf{r}}\xi+\Gamma_{i}^{\mathbf{r}}\eta$; parameters $R_i$ are defined in Eq. \eqref{eq:R-hat-1234}; $\Lambda_i$ and $\Gamma_{i}$ are defined in Eq. \eqref{eq:define-Gamma-Lambda}; superscripts $\mathbf{r}$ and $\mathbf{i}$ represent the real and imaginary parts of parameters respectively. 
Since $R_{3,4}=-R_{1,2}^*$, $\Lambda_{3,4}=-\Lambda_{1,2}^*$, and $\Gamma_{3,4}=-\Gamma_{1,2}^*$, we get $	\hat{\mathbf{Q}}_{3,4}^{(1)}=\hat{\mathbf{Q}}_{1,2}^{(1)}$.
The detailed calculation process is provided in Appendix \ref{Appendix:localized-wave}. 

\begin{figure}[ht]
	\centering
	\includegraphics[width=1\linewidth]{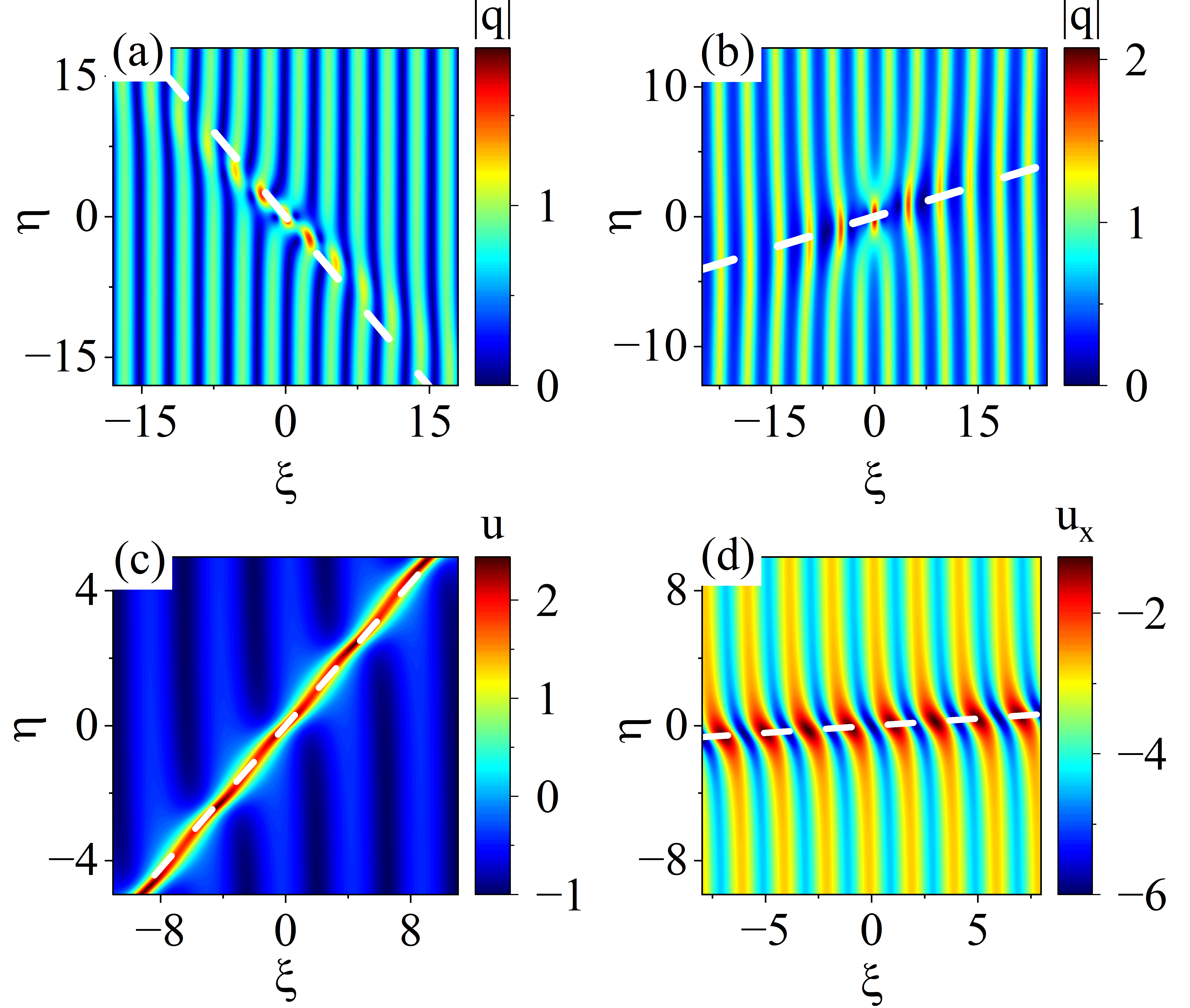}
	\caption{(a): The eRW of the NLS equation with the maximum $1.875$: $l=0$, $z_1=-\ii(1+\tau)/4$, $\alpha=5/4$, $k=3/4$, $h(z_1)=0$; (b): the eRW of the NLS equation with the maximum $2.113$: $z_1=-\ii(1-\tau)/4$, $l=-7\tau\ii/32$, $\alpha=5/4$, $k=19/20$, $h(z_1)=0$; (c): the elliptic-soliton of the mKdV equation: $k=9/10$, $\alpha=1$, $z_1=-\ii(1+\tau)/4$, $l=-\ii \tau/4$, $h(z_1)=0$; (d): the elliptic-soliton of the SG equation: $k=4/5$, $\alpha=2$, $z_1=-\ii(1-\tau)/4$, $l=-\ii \tau/4$, $h(z_1)=0$. }
	\label{fig:order-1}
\end{figure}

The above elliptic-localized wave solutions \eqref{eq:rogue-wave} exhibit two different dynamic behaviors. 
One of them corresponds to the eRW solution, and the other manifests as the elliptic-soliton solution, as depicted in Fig. \ref{fig:order-1}.
The above rational-eRW solutions are consistent with solutions provided in articles \cite{Chen-19-mKdV,PelinovskyW-20-SG,ChenP-21,ChenPW-20,ChenPW-19,FengLT-20,LiG-20-SG} with different forms.
The elliptic-localized wave solutions we provided are expressed in the quadratic polynomial form \eqref{eq:rogue-wave} with respect to $\xi$ and $\eta$ variables, which is greatly similar to the form of rational-RWs on plane wave backgrounds. 
Drawing upon this analogy, we can explore additional properties of these solutions by extending methods from plane wave backgrounds to elliptic function backgrounds.
As $|\xi|+|\eta|\rightarrow \infty$, both solutions $\hat{\mathbf{Q}}_{1}^{(1)}$ and $\hat{\mathbf{Q}}_{2}^{(1)}$ would degenerate into 
\begin{equation}\nonumber
	\hat{\mathbf{Q}}_{1,2}^{(1)}\rightarrow \ii \gamma\frac{\vartheta_2(2\ii l+\hat{\alpha}\xi-1/2)}{\vartheta_4(\hat{\alpha}\xi-1/2)}
	\ee^{\ii( \omega \xi+\kappa \eta)},
\end{equation}
which could be seen as a shift $\xi \rightarrow \xi-1/(2\hat{\alpha})$ and a phase transformation $\ee^{\ii \omega\pi/(2\hat{\alpha})}$ on the elliptic function solution \eqref{eq:solution-seed}. 
Solutions $\hat{\mathbf{Q}}_{1,2}^{(1)}$ attain their maximum value at the zero point with the maximum value being $|\mathbf{Q}_{12}(0,0)|+2\lambda_{1,2}^{\mathbf{i}}=|-\ii \alpha k\cn(4\ii l K)/\dn(4\ii l K)|+\alpha (d_3\mp d_2)=\alpha (2d_3\mp d_2)$.

Considering the \eqref{eq:NLS-equation} equation, on the $\cn$-type backgrounds ($l=0$), we get $\omega=0$, $d_1=-k'=-\sqrt{1-k^2}$, $d_2=0$, $d_3=k$. By Eq. \eqref{eq:transformation-elliptic-theta}, the rational-eRW solution could be simplified as 
\begin{equation}\nonumber
	\hat{\mathbf{Q}}_{i}^{(1)}
	=\ii \alpha k\cn(\alpha \xi -K) \! \left(1+
	\frac{\mathcal{P}_i}{\mathcal{H}_i}\right) \! \ee^{\ii\kappa \eta}, \quad i=1,2,
\end{equation}
where $\mathcal{P}_i$ and $\mathcal{H}_i$ are defined in Eq. \eqref{eq:rogue-wave} with parameters $l=\omega=d_2=0$, $d_1=-k'=-\sqrt{1-k^2}$, and $d_3=k$.
As $|\xi|+|\eta|\rightarrow \infty $, solutions $\hat{\mathbf{Q}}_{1,2}^{(1)}$ both degenerate into the $\cn$-type solution: $\ii \alpha k\cn(\alpha\xi -K)\ee^{\ii \kappa \eta}$. 
As $k\rightarrow 0$, solutions degenerate into zero.
On $\dn$-type backgrounds ($l=-\ii \tau/4$) with $\omega= \hat{\alpha} \pi$, $d_1=0$, $d_2=k'$, $d_3=1$,  combined with Eqs. \eqref{eq:transformation-thetas} and \eqref{eq:transformation-elliptic-theta}, the rational-eRW solution \eqref{eq:rogue-wave} is simplified as 
\begin{equation}\nonumber
	\hat{\mathbf{Q}}_{i}^{(1)}
	=\ii \alpha\dn(\alpha\xi-K)
	\!\left(1+\frac{\mathcal{P}_i}{\mathcal{H}_i}\right)\!\ee^{\ii \kappa\eta}, \quad i=1,2,
\end{equation}
where $\mathcal{P}_i$ and $\mathcal{H}_i$ are defined in Eq. \eqref{eq:rogue-wave} with parameters $l=-\ii \tau/4$, $\omega= \hat{\alpha} \pi$, $d_1=0$, $d_2=k'$, and $d_3=1$.
As $k\rightarrow 0$, combined with the definition of complete elliptic integrals in Def. \ref{define:complete-elliptic-integrals}, the limitations of $K$ and $E$ satisfy $\lim_{k\rightarrow 0}E/K=1$, $\lim_{k\rightarrow 0}\omega/\alpha= \lim_{k\rightarrow 0}\pi/(2K)=1$, $\lim_{k\rightarrow 0}f_2=\lim_{k\rightarrow 0}\dn^2(\alpha \xi)=1$, $\lim_{k\rightarrow 0}g_2=1$, $\lim_{k\rightarrow 0}g_1=\lim_{k\rightarrow 0}Z(K+\alpha \xi)=0$, and $\lim_{k\rightarrow 0}f_1=\lim_{k\rightarrow 0}Z(\alpha \xi)-\ii \pi/K+\ii \pi /(2K)=-\ii$ by Eqs. \eqref{eq:approximate-elliptic}, \eqref{eq:addition-zeta}, and \eqref{eq:limitation-zeta}.
Then, the above two solutions degenerate into
\begin{equation}\nonumber
	\begin{split}
		&\lim_{k\rightarrow 0}\hat{\mathbf{Q}}_{1}^{(1)}\xlongequal[\eqref{eq:limitation-zeta}]{\eqref{eq:approximate-elliptic}}\ii \alpha\ee^{\ii \alpha^2\eta},\\
		&\lim_{k\rightarrow 0}\hat{\mathbf{Q}}_{2}^{(1)}\xlongequal[\eqref{eq:limitation-zeta}]{\eqref{eq:approximate-elliptic}}\ii \alpha \left(1-\frac{8\ii\alpha^2\eta+4}{(2\alpha^2\eta)^2+1+(2\alpha\xi)^2}\right)\ee^{\ii \alpha^2\eta}.
	\end{split}
\end{equation}
Thus, as $k\rightarrow 0$, rational-eRWs on $\dn$-type backgrounds could degenerate into rational-RWs or plane waves.

Consider the rational-elliptic-localized wave solutions of the \eqref{eq:mKdV-equation} equation.
Plugging $l=-\ii \tau/4$ into Eq. \eqref{eq:rogue-wave}, 
the solution could be simplified as 
\begin{equation}\label{eq:asy-k-0-mKdV}
	\!\!\hat{\mathbf{Q}}_{i}^{(1)}
	\xlongequal{\eqref{eq:transformation-elliptic-theta}}\ii \alpha \dn(\alpha\xi -K)\left(1+\frac{\mathcal{P}_i}{(\hat{E}_{i}^{\mathbf{i}}+g_1)^2+g_2}\right)\!,
\end{equation}
where $i=1,2$,  $\mathcal{P}_i=2\hat{E}_{i}^{\mathbf{i}}(f_{1}+\ii \omega/\alpha-g_1)+f_{2}-g_2-(1\mp k')^2-g_1^2+(f_{1}+\ii \omega/\alpha)^2$,
$\hat{E}_{1,2}^{\mathbf{i}}=\alpha(E/K\mp k')\xi-2\alpha^3k'(1\mp k')^2\eta$, $\hat{E}_{1,2}^{\mathbf{r}}=0$, and $f_{1,2},g_{1,2}$ are defined in Eq. \eqref{eq:rogue-wave}.
It is easy to get that along trajectories $\hat{E}_{1,2}^{\mathbf{i}}=\mathrm{const}$ the solution evolves periodically with period $2K/\alpha$. 
As $|\xi|\rightarrow \infty$, solutions $\hat{\mathbf{Q}}_{1,2}^{(1)}$ would degenerate into $\dn$-type solutions $\ii \alpha \dn(\alpha \xi-K)$.
For the \eqref{eq:SG-equation} equation with $l=-\ii \tau/4$, 
the solution \eqref{eq:rogue-wave} could also be simplified in Eq. \eqref{eq:asy-k-0-mKdV} with  $\hat{E}_{1,2}^{\mathbf{i}}=\alpha(E/K\mp k') \xi\mp 2 \alpha s k'\eta$.
It is easy to find that along lines $\hat{E}_{1,2}^{\mathbf{i}}=\mathrm{const}$, the solution evolves periodically. 
As $|\xi|\rightarrow \infty$, we show rotational wave solutions.

Then, we provide some examples to illustrate the dynamic behaviors of these rational-elliptic-localized wave solutions and depict them in Fig. \ref{fig:order-1}.
Figures \ref{fig:order-1}(a) and \ref{fig:order-1}(b) show two solutions of the NLS equation on elliptic function backgrounds. 
Figure \ref{fig:order-1}(a) exhibits the localization of the RW on the $\cn$-type background, called the rational-eRW solution.
Figure \ref{fig:order-1}(b) illustrates the localization of the RW on the non-trivial phase solution background, also referred to as the rational-eRW solution.
Figures \ref{fig:order-1}(c) and \ref{fig:order-1}(d) show the solutions of the mKdV equation and the SG equation, respectively.
Figure \ref{fig:order-1}(c) exhibits the soliton localization on the $\dn$-type background, denoted as the rational-elliptic-soliton solution.
Figure \ref{fig:order-1}(d) portrays the soliton localization on the rotational wave background, also termed the rational-elliptic-soliton solution.

\section{MI-eRW correspondence}\label{section:correspondence}

On the plane wave background, the MI and baseband MI are crucial mechanisms in studying the exciting conditions of RWs and the existence of RWs \cite{BaronioCDLOW-14,BaronioCGWC-15,ZhaoL-16,LingZY-17,LingZ-19}.
We aim to extend this mechanism to the elliptic function background and quantitatively elucidate the relationship between the dynamic behaviors of these solutions and the baseband MI/MS of elliptic function solutions \cite{BaronioCDLOW-14,BaronioCGWC-15}.
In this section, we illustrate a pair of parameters $(\Lambda_{i},\Gamma_{i})$, $i=1,2$, that establishes a quantitative connection between the baseband MI/MS analysis and dynamic behaviors of elliptic-localized wave solutions.

The pair of parameters $(\Lambda_i,\Gamma_{i})$ serves a dual purpose in both investigating the stability of elliptic function solutions for INS equations and the derivation of dynamic behaviors for elliptic-localized wave solutions. 
As discussed in Sec. \ref{section:MI}, the ratio of $\Gamma_{i}$ to $\Lambda_{i}$ offers insights into the stability of the system.
To be precise, if $\Gamma_i/\Lambda_{i}\in \mathbb{C}\backslash \mathbb{R}$  elliptic function solutions of INS equations are the baseband modulationally unstable; and if $\Gamma_{i}/\Lambda_i\in \mathbb{R}$, elliptic function solutions of INS equations are the baseband modulationally stable.

Additionally, the pair of parameters $(\Lambda_{i},\Gamma_{i})$ plays a significant role in deriving the polynomials $\hat{E}_i^{\mathbf{i}}=\Lambda_{i}^{\mathbf{i}}\xi +\Gamma_{i}^{\mathbf{i}}\eta$ and $\hat{E}_i^{\mathbf{r}}=\Lambda_{i}^{\mathbf{r}}\xi +\Gamma_{i}^{\mathbf{r}}\eta$ in the rational form of elliptic-localized wave solutions \eqref{eq:rogue-wave}, which are closely connected to eRW solutions. When $\Lambda_{i}=\Gamma_{i}=0$, the solution \eqref{eq:rogue-wave} becomes a periodic function since both polynomials are zero, indicating $\hat{E}_{i}^{\mathbf{i}}=\hat{E}_i^{\mathbf{r}}=0$. However, if this condition is not satisfied, let us assume, without loss of generality, that the parameter $\hat{E}_i^{\mathbf{i}}\neq 0$ with $(\xi,\eta )\in \mathbb{R}^2$. 
We will examine the dynamics of the solution \eqref{eq:rogue-wave} along trajectories where $\hat{E}_i^{\mathbf{i}}\equiv C$, expressed as $\hat{\mathbf{Q}}_{i}^{(1)}(\xi,\eta;\hat{E}_i^{\mathbf{i}}\equiv C)$, where $C$ represents a constant. Conversely, if $\hat{E}_i^{\mathbf{i}}\equiv 0$, we analyze the solution \eqref{eq:rogue-wave} along trajectories where $\hat{E}_i^{\mathbf{r}}\equiv C$. 

Functions $f_{1,2}$, $g_{1,2}$ and $\vartheta_{2,4}(\cdot)$ defined in Eq. \eqref{eq:rogue-wave} are all $2K$ periodic with respect to the variable $\xi$.
It can be inferred that these functions remain bounded for $\xi\in \mathbb{R}$. 
The excitation of eRW is contingent upon the polynomial form $\hat{E}_i^{\mathbf{r}}$ with $(\xi,\eta)\in \mathbb{R}^2$. 
The existence of the polynomial form occurs when either $\Lambda_{i}^{\mathbf{r}}\neq0$ or $\Gamma_{i}^{\mathbf{r}}\neq 0$. 
Consequently, elliptic-localized wave solutions \eqref{eq:rogue-wave} exhibit the localized wave as eRWs. 
In the special case where $\Lambda_{i}^{\mathbf{r}} =0$ and $\Gamma_{i}^{\mathbf{r}}=0$ (i.e., $\hat{E}_{i}^{\mathbf{r}}\equiv 0$), 
it is straightforward to observe that the solution $\hat{\mathbf{Q}}_{i}^{(1)}(\xi,\eta;\hat{E}_{i}^{\mathbf{i}}\equiv C)$ becomes periodic along all trajectories $\hat{E}_{i}^{\mathbf{i}}\equiv C$. 
Therefore, when solutions \eqref{eq:rogue-wave} are baseband modulationally stable, they can be characterized as elliptic-soliton solutions. 

In summary, when $\Gamma_i/\Lambda_{i}\in \mathbb{R}$, we derive $\hat{E}_i^{\mathbf{r}}=0$ or $\hat{E}_i^{\mathbf{i}}=0$, considering the linear independence of $\Lambda_i$ and $\Gamma_{i}$. 
On the other hand, when $\Gamma_{i}/\Lambda_i\in \mathbb{C}\backslash\mathbb{R}$, we deduce $\hat{E}_i^{\mathbf{r}}\neq 0$ and $\hat{E}_i^{\mathbf{i}}\neq 0$. 
Building upon the aforementioned investigations, the pair $(\Lambda_{i}, \Gamma_i)$ establishes a quantitative relationship between baseband MI/MS and the dynamic behaviors of eRWs/elliptic-solitons.

Indeed, the baseband MS not only corresponds to elliptic-soliton solutions but also maintains a significant connection with elliptic-breather solutions. 
However, the explicit expression of the elliptic-localized wave solution \eqref{eq:rogue-wave} does not manifest elliptic-breathers. 
Only when the following two conditions are met can elliptic-breather solutions be achieved. 
Firstly, the spectral parameter $\lambda_{i}$ employed in constructing solutions must be complex numbers, that is, $\lambda_i\equiv\lambda(z_i)\in \mathbb{C}\backslash(\ii \mathbb{R}\cup \mathbb{R})$, $z_i\in \mathcal{B}$. 
Secondly, the related elliptic function solutions of INS equations need to be baseband MS. 
Unfortunately, rational-elliptic-breather solutions for the NLS equation, the mKdV equation, and the SG equation were not obtained.  
This restriction arises from the fact that when $\lambda_i \in \mathbb{C}\backslash(\ii \mathbb{R}\cup \mathbb{R})$, elliptic function solutions on the associated backgrounds yield baseband MI solutions that fail to satisfy the second condition. 
In summary, the concealed correspondence between elliptic-localized wave solutions and the MS/MI analysis results can be summarized as follows:
\begin{itemize}
	\item[(1)] {\bf [MI and eRWs correspondence]}
	If the elliptic function solutions of the INS equations correspond to baseband MI, it implies that there exists a pair of parameters $(\Lambda_{i},\Gamma_{i})$ satisfying $\Gamma_{i}/\Lambda_{i}\in \mathbb{C}\backslash \mathbb{R}$, indicating that $\hat{E}_i^{\mathbf{r}}\neq 0$ and $\hat{E}_i^{\mathbf{i}}\neq 0$, $(\xi,\eta)\in \mathbb{R}^2$. 
	Therefore, the solution \eqref{eq:rogue-wave} expressed in a rational form represents the presence of eRWs. 
	Conversely, if the solution \eqref{eq:rogue-wave} exhibits eRWs, it must be expressed in terms of polynomials, i.e., $\hat{E}_{i}^{\mathbf{r}}\neq 0$ and $\hat{E}_{i}^{\mathbf{i}}\neq 0$, which further implies that $\Gamma_{i}/\Lambda_{i}\in \mathbb{C}\backslash\mathbb{R}$ due to the linear independence of the parameters $\Gamma_{i}$ and $\Lambda_{i}$. 
	Hence, it becomes evident that the related elliptic function solutions correspond to baseband MI.
	
	\item[(2)] {\bf [MS and elliptic-solitons correspondence]} 
	When the elliptic function solutions of the INS equations are baseband modulationally stable (meaning $\Gamma_{i}/\Lambda_{i}\in \mathbb{R}$, $i=1,2$), equations $\hat{E}_{i}^{\mathbf{r}}\equiv0$ or $\hat{E}_{i}^{\mathbf{i}}\equiv 0$ are satisfied due to the linear independence of $\Gamma_{i}$ and $\Lambda_{i}$. 
	Along the trajectories where $\hat{E}_{i}^{\mathbf{i}}\equiv C$ or $\hat{E}_{i}^{\mathbf{r}}\equiv C$, the solution evolves periodically, signifying that the elliptic-localized wave solutions \eqref{eq:rogue-wave} represent elliptic solitons. 
	Conversely, if the solutions represent elliptic solitons, it indicates their correspondence to baseband MS.

	\item[(3)] {\bf [MS and elliptic-breathers correspondence]} 
	When the elliptic function solutions of the INS equations correspond to baseband MS, the pair of parameters $(\Lambda_{i},\Gamma_{i})$ satisfy $\Gamma_{i}/\Lambda_{i}\in \mathbb{R}$. 
	Due to the linear independence of $\Gamma_{i}$ and $\Lambda_{i}$, we can deduce $\hat{E}_{i}^{\mathbf{r}}\equiv0$ or $\hat{E}_{i}^{\mathbf{i}}\equiv 0$. 
	Under these conditions, the elliptic-localized wave solutions with parameters $\lambda_i\equiv\lambda(z_i)\in \mathbb{C}\backslash(\ii \mathbb{R}\cup \mathbb{R})$, $z_i\in \mathcal{B}$, exhibit elliptic-breathers. 
	The reverse is also true.
\end{itemize}

Applying the aforementioned results to the INS equations, 
we elucidate the dynamic behaviors of elliptic-localized wave solutions under different cases. 
For the NLS equation, parameters $(\Lambda_{i},\Gamma_i)$ presented in Eqs. \eqref{eq:define-Lambda} and \eqref{eq:define-Gamma-NLS} fulfill the condition $\Gamma_{i}/\Lambda_{i}\in \mathbb{C}\backslash \mathbb{R}$ for any $l\in [0,-\ii \tau/4]$. 
Some examples in Fig. \ref{fig:order-1} vividly exhibit the aforementioned correspondence.
The evolution directions of these elliptic-localized waves align with trajectories $\hat{E}_{i}^{\mathbf{i}}\equiv 0$ depicted in Fig. \ref{fig:order-1}, indicated by white dashed lines.
The baseband MI of elliptic function solutions ($\Lambda_{i}/\Gamma_{i}\in \mathbb{C}\backslash \mathbb{R}$ in Eq. \eqref{eq:define-Gamma-NLS}) indicates that solutions \eqref{eq:rogue-wave} correspond to rational-eRWs, as illustrated in Figs. \ref{fig:order-1}(a) and \ref{fig:order-1}(b).
For the mKdV equation, the parameters $(\Lambda_{i},\Gamma_i)$ discussed in Eqs. \eqref{eq:define-Gamma-mKdV} and \eqref{eq:define-Lambda-mKdV} satisfy the condition $\Gamma_{i}/\Lambda_{i}\in \mathbb{C}\backslash \mathbb{R}$ for $l=0$ and $\Gamma_{i}/\Lambda_{i}\in \mathbb{R}$ for $l=-\ii \tau/4$. Consequently, on $\dn$-type backgrounds, the solutions correspond to elliptic-soliton solutions, as depicted in Fig. \ref{fig:order-1}(c). 
On the other hand, on $\cn$-type backgrounds, the solutions represent eRW solutions, as shown in Fig. \ref{fig:rogue-wave-mKdV-SG}(c).
Concerning the SG equation, the parameters $(\Lambda_{i},\Gamma_i)$ described in Eqs. \eqref{eq:define-Lambda-mKdV} and  \eqref{eq:define-Gamma-SG} meet the condition $\Gamma_{i}/\Lambda_{i}\in \mathbb{C}\backslash \mathbb{R}$ for $l=0$, and $\Gamma_{i}/\Lambda_{i}\in \mathbb{R}$ for $l=-\ii \tau/4$. As a result, on rotational wave backgrounds, the solutions correspond to elliptic-soliton solutions, as demonstrated in Fig. \ref{fig:order-1}(d). Conversely, on librational wave backgrounds, the solutions represent eRWs, as shown in Fig. \ref{fig:rogue-wave-mKdV-SG}(d). 
It is important to note that we were unable to obtain elliptic-breather solutions for the NLS equation, the mKdV equation, and the SG equation, since for any parameters $(\Gamma_{i},\Lambda_{i})$ satisfying $\Gamma_{i}/\Lambda_{i}\in \mathbb{R}$ the related spectral parameter $\lambda_i$ satisfies $\lambda_i\in \ii \mathbb{R}$, i.e., $\lambda_i\notin \mathbb{C}\backslash(\mathbb{R}\cup \ii \mathbb{R})$, which do not satisfy the MS and elliptic-breathers correspondence.

The aforementioned correspondence can also be extended to higher-order elliptic-localized wave solutions. 
In the upcoming section, we will construct higher-order elliptic-localized solutions in a rational form and utilize the established correspondence between the solutions and stability analysis.

\section{Multi-higher-order eRW solutions}\label{section:higher-order}
By extending the method of constructing the solution \eqref{eq:rogue-wave} to higher-order cases,
the multi-higher-order $($$(N_1,N_2)$-order$)$ elliptic-localized wave solutions of INS equations could also be expressed in the rational form as follows:
\begin{equation}\label{eq:higher-order-N}
	\begin{split}
	&\hat{\mathbf{Q}}^{(N_1,N_2)}\\=&\gamma\mathcal{J}^{N-1}
	\!\frac{\det\!
		\begin{pmatrix}
			\mathbf{P}_{N_1 \times N_1}(z_1^*,z_1) \!& \!
			\mathbf{P}_{N_1 \times N_2}(z_1^*,z_2)\\
			\mathbf{P}_{N_2 \times N_1}(z_2^*,z_1) \!&\! \mathbf{P}_{N_2 \times N_2}(z_2^*,z_2)
		\end{pmatrix}
	}{\det\!
		\begin{pmatrix}
			\mathbf{H}_{N_1 \times N_1}(z_1^*,z_1) \!& \!
			\mathbf{H}_{N_1 \times N_2}(z_1^*,z_2)\\
			\mathbf{H}_{N_2 \times N_1}(z_2^*,z_1) \!& \!
			\mathbf{H}_{N_2 \times N_2}(z_2^*,z_2)
	\end{pmatrix}}\ee^{\ii (\omega \xi +\kappa \eta)},
	\end{split}
\end{equation}	
where $\mathcal{J}=\vartheta_4(\hat{\alpha}\xi)/\vartheta_2(\hat{\alpha}\xi+2\ii l)$, $N=N_1+N_2$, $\mathbf{H}_{N_i \times N_j}(z_i^*,z_j)$ and $\mathbf{P}_{N_i \times N_j}(z_i^*,z_j)$ are the first $N_i$ row and $N_j$ column matrices $\mathbf{H}(z_i^*,z_j)$ and $\mathbf{P}(z_i^*,z_j)$ respectively;
parameters $\gamma$, $\kappa$ are defined in Eq. \eqref{eq:define-d-1-2-3-Kl} and the $(a,b)$-element of matrices $\mathbf{H}_{N_i \times N_j}(z_i^*,z_j)$ and $\mathbf{P}_{N_i \times N_j}(z_i^*,z_j)$ are $\mathcal{H}^{[N_a,N_b]}(z_i^*,z_j)$ and $\mathcal{P}^{[N_a,N_b]}(z_i^*,z_j)$ expressed as
\begin{equation}\nonumber
	\begin{split}
	\mathcal{H}^{[N_a,N_b]}(z_i^*,z_j)
	\!=\!&\left(\mathbf{E}_i^{[N_a]}\right)^{\dagger}\mathbf{G}_{i,j}^{[N_a,N_b]}\mathbf{E}_j^{[N_b]}, \\
	\mathcal{P}^{[N_a,N_b]}(z_i^*,z_j)\!=\!&\left(\mathbf{E}_i^{[N_a]}\right)^{\dagger}\!\!\left(\mathbf{R}^{[N_a]\dagger}_{i}\right)^{-1}\!\!\mathbf{F}_{i,j}^{[N_a,N_b]}\mathbf{R}_{j}^{[N_b]}\mathbf{E}_j^{[N_b]}\!;
	\end{split}		
\end{equation}
and matrices $\mathbf{F}_{i,j}^{[N_a,N_b]}$, $\mathbf{G}_{i,j}^{[N_a,N_b]}$, $\mathbf{R}_{j}^{[N_b]}$, and $\mathbf{E}_j^{[N_b]}$ are defined in Eq. \eqref{eq:define-F-R-G-E}. The detailed calculation process is provided in Appendix \ref{Appendix:higher-order}.
 
\begin{figure}[ht]
	\centering
	\includegraphics[width=1\linewidth]{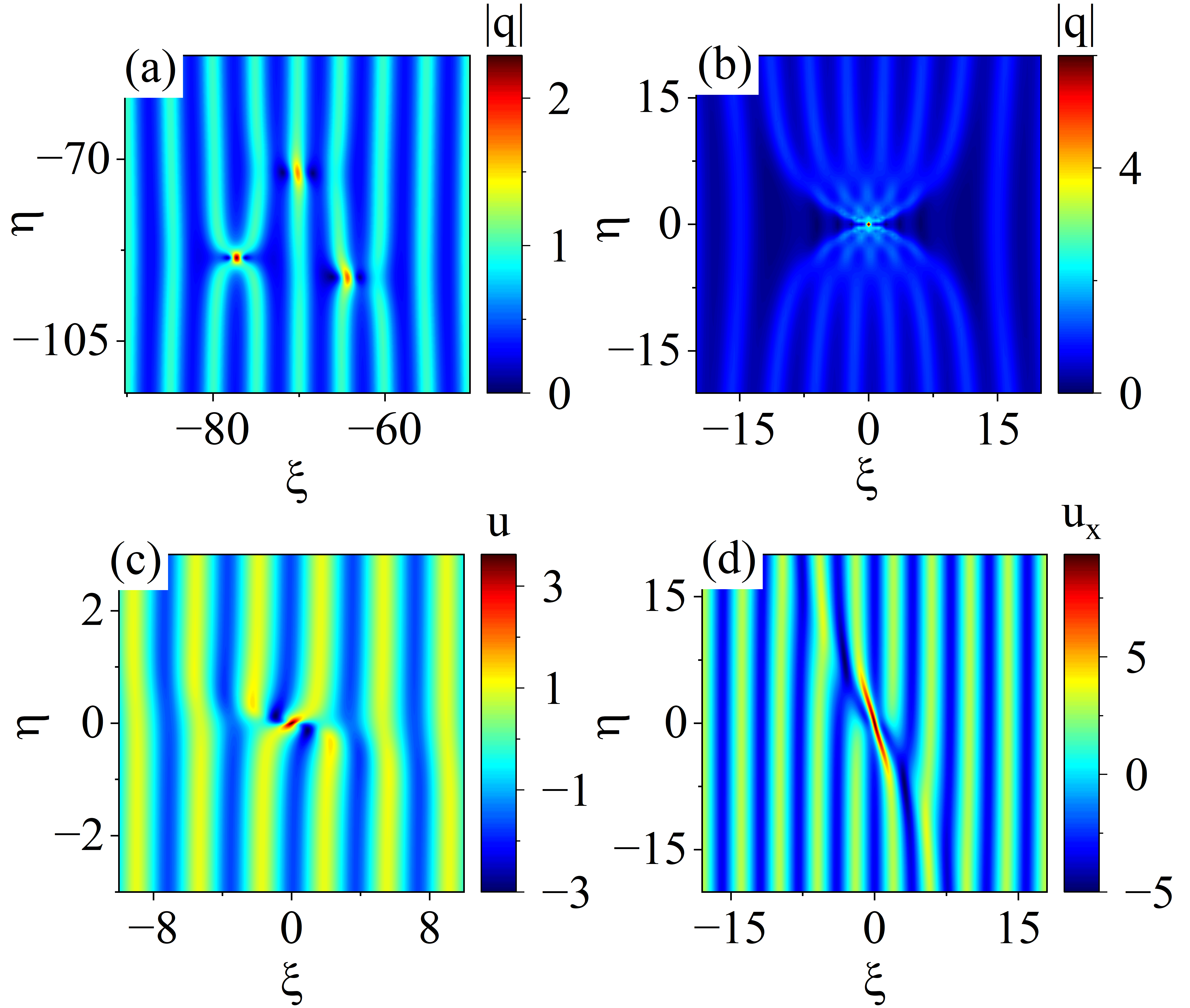}
	\caption{(a) and (b): second-order (separated-type) and fifth-order (polymeric-type) eRWs solutions of the NLS equation with parameters $\alpha=1$, $k=0.94$, $z_{1}=-\ii(1-\tau)/4$, $l=-9\ii \tau/40$, $c_{12}=0$, $h(z_1)=300(1+\ii \sqrt{3})z_1^2$ for (a) and $h(z_1)=0$ for (b). (c): the eRW of the mKdV equation with $k=2/3$, $\alpha=2$, $z_{1,2}=-\ii(1\pm \tau)/4$; (d): the eRW of the SG equation with $k=4/5$, $\alpha=2$, $z_{1,2}=-\ii(1\pm \tau)/4$, $h(z_1)=0$.}
	\label{fig:rogue-wave-mKdV-SG}
\end{figure}

Since the function $E_1(z)$ represents an exponential function, its differentiation with respect to the parameter $z$ generates a polynomial form involving variables $\xi$ and $\eta$.
The differentials of functions $F(z^*,z)$ and $G(z^*,z)$ exhibit elliptic functions with respect to $\xi$. 
The function $r(z)$ remains independent of $\xi$ and $\eta$. 
In summary, the solution $\hat{\mathbf{Q}}^{(N_1,N_2)}$ is expressed in a rational form in terms of $\xi$ and $\eta$. 
Consequently, we provide the multi-higher-order rational-elliptic-localized wave solutions \eqref{eq:higher-order-N} expressed in higher-order polynomials with respect to $\xi$ and $\eta$ for the INS equations.
These higher-order solutions have never been provided before.
Some multi-higher-order eRW solutions of INS equations are shown in Fig. \ref{fig:rogue-wave-mKdV-SG}.

Figure \ref{fig:rogue-wave-mKdV-SG}(a) illustrates a second-order eRW of the separated type composed of three eRWs, analogous to RWs on plane wave backgrounds.
In Figure \ref{fig:rogue-wave-mKdV-SG}(b), a fifth-order eRW of the polymeric type is depicted, reaching a maximum value of $7.674$ at the origin.
The $(1,1)$-eRW solutions of the mKdV equation on $\cn$-type backgrounds and the SG equation on librational wave backgrounds are presented in Figs. \ref{fig:rogue-wave-mKdV-SG}(c) and \ref{fig:rogue-wave-mKdV-SG}(d) respectively, both satisfying the baseband MI mechanism.

\section{Conclusions}\label{section:conclution}
We report the rational-elliptic-localized wave solutions of the INS equations in the polynomial form using a systematic method.
These solutions are expressed as polynomials involving temporal and spatial variables, establishing a close relationship between eRWs and RWs in INS equations.
Combining this method with the modified squared wave function approach, we unveil a quantitative correspondence between MI and eRWs through the parameters $(\Lambda_{i},\Gamma_{i})$, $i=1,2$.
We also demonstrate that the MI of elliptic function solutions leads to rational-form solutions exhibiting eRWs, while the MS of elliptic function solutions results in rational-form solutions displaying elliptic-solitons or elliptic-breathers. 
Furthermore, the higher-order eRW solutions are also expressed in the rational form with respect to variables $\xi$ and $\eta$, which have never been reported on elliptic function backgrounds.

This methodology is applicable for generating eRWs and perform stability analyses for other integrable equations \cite{BertolaJT-23,Smirnov1994finite}. 
The study unveils an intrinsic connection between MI and eRW solutions, as well as MS and solutions resembling elliptic-solitons/breathers. 
This theoretical foundation supports the creation of eRWs on elliptical function backgrounds.
Moreover, our research enhances the comprehension of eRW phenomena within intricate dynamics \cite{HeSM-22,Dudley2019rogue,WalczakRS-15}, offering valuable insights for the observation of eRWs in physical experiments \cite{ZhangLXSS-2023experimental,Mao-hoefer-2023,TikanBRETDC-22,DucrozetANT-21,XuCPK-20}.

\section*{Acknowledgments}
The research of L. Ling is supported by the National Natural Science Foundation of China (Grant No. 12122105).

\appendix

\titleformat{\section}[display]
{\centering \bfseries}{ }{11pt}{ }

\setcounter{section}{0}

\renewcommand{\appendixname}{Appendix \ \Alph{section} }

\section{\appendixname. The definitions and properties of elliptic functions}\label{appendix:Elliptic functions}

\setcounter{equation}{0}
\renewcommand\theequation{\Alph{section}.\arabic{equation}}

\renewcommand\thedefine{\Alph{section}.\arabic{define}}

\begin{define}[Jacobi elliptic functions {\cite[p.18]{ByrdF-54}}]\label{define:elliptic-function}
	The inverse function of 
	\begin{equation}\nonumber
		u(y,k)=\int_{0}^{y}\frac{\dd t}{\sqrt{(1-t^2)(1-k^2t^2)}},
	\end{equation}
	is defined by $y=\sn(u,k)$ or briefly $y=\sn(u)$. Functions $\cn(u,k)$ and $\dn(u,k)$ are defined by $\cn(u,k)=\sqrt{1-y^2}$ and $\dn(u,k)=\sqrt{1-k^2y^2}$ with $\cn(0,k)=1$ and $\dn(0,k)=1$. These three functions are called Jacobi elliptic functions and $k$ is the modulus. The complementary modulus is denoted by $k'=\sqrt{1-k^2}$.
\end{define}
It is worth noting that we can confine our consideration to the modulus $k$ within the range $(0,1)$.
This limitation arises from the fact that elliptic functions with $k$ in the interval $(1,\infty)$ can be transformed into the case $(0,1)$ through reciprocal modulus transformation formulas \cite[p.38]{ByrdF-54}.
Specifically, when $k=0$, elliptic functions $\sn(u)$, $\cn(u)$, and $\dn(u)$ degenerate into trigonometric functions $\sin(u)$, $\cos(u)$, and $1$, respectively. 
Similarly, when $k=1$, elliptic functions degenerate into hyperbolic functions $\tanh(u)$, $\mathrm{sech}(u)$, and $\mathrm{sech}(u)$, respectively.

\begin{define}[Complete elliptic integrals {\cite[p.9]{ByrdF-54}}]\label{define:complete-elliptic-integrals}
	The elliptic integrals
	\begin{equation}\label{eq:complete-elliptic-integrals}
		\begin{split}
		&K\equiv K(k)=\int_{0}^{\frac{\pi}{2}}\frac{\dd \theta}{\sqrt{1-k^2\sin^2\theta}}, \\
		&E\equiv E(k)=\int_{0}^{\frac{\pi}{2}}\sqrt{1-k^2\sin^2\theta}\ \dd \theta,
		\end{split}
	\end{equation}	
	are called the first and second complete elliptic integrals respectively. 
	The parameter $K'$ is defined as $K'\equiv K(k')$. 
	The normal elliptic integral of the second kind is $	E(u)=\int_{0}^{u}\dn^2(v)\dd v$. 
	When $u=K$, the normal elliptic integral would convert into the complete elliptic integral.
\end{define}

Hence, the aforementioned elliptic functions are not independent, giving rise to various formulas. 
Here are some useful formulas:
\begin{itemize}
	\item The shift formulas of elliptic functions \cite[p.22]{ByrdF-54}:
	\begin{equation}\label{eq:shift-elliptic}
		\begin{split}
			&\sn(u+2 m K+2\ii n K')= (-1)^m\sn(u),\\
			&\cn(u+2 m K+2\ii n K')= (-1)^{m+n}\cn(u), \\
			&\dn(u+2 m K+2\ii n K')= (-1)^{n}\dn(u),\\
			&\dn(u+ K+\ii K')=  \ii k'\sn(u)/\cn(u),\\
			&\sn(u+\ii K')=1/(k\sn(u)),\\
			&\cn(u+\ii K')=-\ii \dn(u)/(k\sn(u)),\\
			&\dn(u+\ii K')=-\ii \cn(u)/\sn(u);
		\end{split}	
	\end{equation}
	
	\item The addition formulas of elliptic functions \cite[p.23]{ByrdF-54}:
	\begin{equation}\label{eq:addition-elliptic}
		\begin{split}
			&\sn(u-v)=\frac{\sn(u)\cn(v)\dn(v)-\sn(v)\cn(u)\dn(u)}{1-k^2\sn^2(u)\sn^2(v)},\\
			&\sn(u+v)\sn(u-v)=\frac{\sn^2(u)-\sn^2(v)}{1-k^2\sn^2(u)\sn^2(v)};
		\end{split}
	\end{equation}

	\item  The approximation formulas of elliptic functions \cite[p.24]{ByrdF-54}:
	\begin{equation}\label{eq:approximate-elliptic}
		\begin{split}
			&\cn(u,k)\approx \cos(u)\!+ \! k^2\sin(u)(u-\sin(u)\cos(u))/4,\\
			&\dn(u,k)\approx 1-k^2\sin^2(u)/2, \qquad k \ll 1;
		\end{split}
	\end{equation}

\item The derivative formulas of elliptic functions \cite[p.284-285]{ByrdF-54}:
\begin{equation}\label{eq:derivative-elliptic}
	\begin{split}
		\sn'(z)=&\cn(z)\dn(z), \quad 
		\cn'(z)=-\sn(z)\dn(z), \\
		\dn'(z)=&-k^2\cn(z)\sn(z);
	\end{split}	
\end{equation}

	\item The double and half arguments of elliptic functions \cite[p.24]{ByrdF-54}:
	\begin{equation}\label{eq:argument-elliptic}
		\begin{split}
			\sn\left(2u\right)=& \frac{2\sn(u)\cn(u)\dn(u)}{1-k^2\sn^4(u)},\\
			\cn\left(2u\right)=&\frac{\cn^2(u)-\sn^2(u)\dn^2(u)}{1-k^2\sn^4(u)}, \\
			\dn\left(2u\right)=&\frac{\dn^2(u)-k^2\sn^2(u)\cn^2(u)}{1-k^2\sn^4(u)},\\
			\sn^2\left(u\right)=& \frac{1-\cn(2u)}{1+\dn(2u)},\\
			\cn^2\left(u\right)=&\frac{\dn(2u)+\cn(2u)}{1+\dn(2u)},\\
			\dn^2\left(u\right)=&\frac{\dn(2u)+\cn(2u)}{1+\cn(2u)}.
		\end{split}
	\end{equation} 

\end{itemize}
Combining with Eq. \eqref{eq:argument-elliptic}, we obtain
\begin{equation}\label{eq:add-1}
	\begin{split}
	&\frac{1-\cn(2u)}{\sn(2u)}\\
	\xlongequal{\eqref{eq:argument-elliptic}}&
	\frac{1-k^2\sn^4(u)-\cn^2(u)+\sn^2(u)\dn^2(u)}{2\sn(u)\cn(u)\dn(u)}\\
	=&\frac{\sn(u)\dn(u)}{\cn(u)},
	\end{split}
\end{equation}
by which we get
\begin{equation}\label{eq:add-2}
	\begin{split}
		\scd (u)
		\xlongequal[\eqref{eq:add-1}]{\eqref{eq:argument-elliptic}}&\frac{(1-\cn(2u))(\dn(2u)+\cn(2u))}{\sn(2u)(1+\dn(2u))}.
	\end{split}
\end{equation}

\begin{define}[Theta functions {\cite[p.302]{Farkas-92-Riemann}}]\label{define:theta}
	The theta functions are defined as:
	\begin{equation}\nonumber
		\begin{split}
		\vartheta_1(u,\ee^{\ii \pi \tau}):=\Theta\begin{bmatrix}
			1 \\ 1
		\end{bmatrix}(u,\tau), \,\,\,
	\vartheta_2(u,\ee^{\ii \pi \tau}):=\Theta\begin{bmatrix}
			1 \\ 0
		\end{bmatrix}(u,\tau), \\ 
	\vartheta_3(u,\ee^{\ii \pi \tau}):=\Theta\begin{bmatrix}
			0 \\ 0
		\end{bmatrix}(u,\tau), \,\,\, 
	\vartheta_4(u,\ee^{\ii \pi \tau}):=\Theta\begin{bmatrix}
			0 \\ 1
		\end{bmatrix}(u,\tau),
		\end{split}
	\end{equation}
	where $\tau=\ii K'/K$, parameters $K$ and $K'$ are defined in Definition \ref{define:complete-elliptic-integrals}, and
	\begin{equation}\nonumber
		\begin{split}
			&\Theta\begin{bmatrix}
				\epsilon\\
				\epsilon'
			\end{bmatrix}(u,\tau)\\
			=&\sum_{n=-\infty}^{+\infty}\exp\left\{2\pi \ii \left[\frac{1}{2}\left(n+\frac{\epsilon}{2}\right)^2\tau+\left(n+\frac{\epsilon}{2}\right)\left(z+\frac{\epsilon'}{2}\right)\right]\right\}.
		\end{split}
	\end{equation}
\end{define}
Typically, we omit the parameter $\ee^{\ii \pi \tau}$ and express theta functions as $\vartheta_i(u)\equiv\vartheta_i(u,\ee^{\ii \pi \tau})$. 
The mentioned theta functions are not independent.
Below are transformation formulas \cite[p.86]{ArmitageE-06-elliptic-function} that are useful in the subsequent discussion:
\begin{equation}\label{eq:transformation-thetas}
	\begin{split}
		\vartheta_3(z)=&\vartheta_1(z+(1+\tau)/2)\ee^{\ii \tau \pi/4+\ii z \pi},\\ \vartheta_2(z)=&\vartheta_4(z+(1+\tau)/2)\ee^{\ii \tau \pi/4+\ii z \pi},\\
		\vartheta_4(z)=&\ii\vartheta_2(z+(1+\tau)/2)\ee^{\ii \tau \pi/4+\ii z \pi}, \\ \vartheta_1(z)=&-\ii\vartheta_3(z+(1+\tau)/2)\ee^{\ii \tau \pi/4+\ii z \pi},\\
		\vartheta_2(z)=&\vartheta_3(z+\tau/2)\ee^{\ii \tau \pi/4+\ii z \pi}, \\
		\vartheta_1(z)=&-\ii\vartheta_4(z+\tau/2)\ee^{\ii \tau \pi/4+\ii z \pi},\\
		\vartheta_3(z)=&\vartheta_2(z+\tau/2)\ee^{\ii \tau \pi/4+\ii z \pi}.
	\end{split}
\end{equation}
The transformation formulas connecting elliptic function solutions and theta functions are
\begin{equation}\label{eq:transformation-elliptic-theta}
	\!\!\!\!\cn(2uK)\!=\!\frac{\vartheta_4\vartheta_2(u)}{\vartheta_2\vartheta_4(u)},\,\,
	\dn(2uK)\!=\!\frac{\vartheta_4\vartheta_3(u)}{\vartheta_3\vartheta_4(u)}, \,\,
	k\!=\!\frac{\vartheta_2^2}{\vartheta_3^2},\!\!\!\!
\end{equation}
where $\vartheta_{i}\equiv \vartheta_{i}(0)$, $i=1,2,3,4$.

\begin{define}[Zeta function {\cite[p.33]{ByrdF-54}}]\label{define:Zeta}
	The Zeta function $Z(u)\equiv Z(u,k)$ is defined by
	\begin{equation}\label{eq:derivative-zeta}
		\begin{split}
		\!\!\!\! \!\!\!	&Z(u)=\int_{0}^{u}\left(\dn^2(v)-E/K\right) \dd v=E(u)-Eu/K, \! \!\!\!\\
		\!\!\!\! \!\!\!	&Z(u)=\vartheta_4'(u/(2K))/\vartheta_4(u/(2K)),
		\end{split}
	\end{equation}	
	where functions $K$ and $E$ are the first and second complete elliptic integrals defined in Eq. \eqref{eq:complete-elliptic-integrals}; $E(u)$ is the normal elliptic integral of the second kind defined in Def. \ref{define:complete-elliptic-integrals}. It is noteworthy that the two equations mentioned above are equivalent.
\end{define}
Here are some shift and transformation formulas for the Zeta function:
\begin{itemize}
	\item The addition formulas of the Zeta function \cite[p.33-34]{ByrdF-54}:
	\begin{equation}\label{eq:addition-zeta}
		\begin{split}
	\!\!\!\!\!\! &	Z(u+\ii K')  \!=\! Z(u)+\cn(u)\dn(u)/\sn(u)-\ii \pi /(2K),\!\!\! \!\!\! \\
	\!\!\!\!\!\! &	Z(u+2\ii K')\!=\! Z(u)-\ii \pi /K,\!\!\!\\
	\!\!\!\!\!\! &	Z(u\pm v)\!=\!  Z(u)\pm Z(v)\mp k^2\sn(u)\sn(v)\sn(u\pm v);\!\!\!\!\!\! 
		\end{split}
	\end{equation}
	\item The derivative formulas of theta functions are derived by Eqs. \eqref{eq:transformation-thetas} and \eqref{eq:derivative-zeta}:
	\begin{equation}\label{eq:derivative-theta}
		\begin{split}
			\ln(\vartheta_1(z))_z
			\!=& 2K\!\left(Z(2zK-K)+\frac{\cn(2zK)}{\sn(2zK)\dn(2zK)}\right),\\
			\ln(\vartheta_4(z))_z\!=&2KZ(2zK),\,\, \ln(\vartheta_3(z))_z\!=2KZ(2zK+K),\\
			\ln(\vartheta_2(z))_z
			\!=& 2K\!\left(Z(2zK)-\frac{\sn(2zK)\dn(2zK)}{\cn(2zK)}\right);
		\end{split}
	\end{equation}
	\item As $k\rightarrow 0$, the limitation of the Zeta function satisfies 
	\begin{equation}\label{eq:limitation-zeta}
		\lim_{k\rightarrow 0}Z(u,k)=\lim_{k\rightarrow 0}E(u,k)-Eu/K=0,
	\end{equation}
	based on the definition of function $E(u)$ in Def. \ref{define:complete-elliptic-integrals}.
\end{itemize}
By studying zeros and poles of elliptic functions, we could obtain the following formulas \cite{LingS-21-mKdV-stability}:
\begin{equation}\label{eq:trans-elliptic-theta}
	\begin{split}
		&k^2\left(\sn^2(2uK)-\sn^2(2vK)\right)
		\! =\! \frac{\vartheta_2^2\vartheta_4^2\vartheta_1(u-v)\vartheta_1(u+v)}{\vartheta_3^2\vartheta_4^2(u)\vartheta_4^2(v)},\\
		&\int_{0}^{x} \! \!\frac{\scd(2vK) \dd (2uK)}{\sn^2(2uK)-\sn^2(2vK)}
		\!=\!\frac{1}{2}\ln \! \left(\frac{\vartheta_1(v-x)}{\vartheta_1(v+x)}\right) \! \!+ \! xZ(2vK).
	\end{split}
\end{equation}
where $\scd(u)$ is defined in Eq. \eqref{eq:lambda-y}.

\begin{widetext}
\section{\appendixname. The proof of Eqs. \eqref{eq:W-1-2}, \eqref{eq:define-y-hat}-\eqref{eq:define-Lambda}, \eqref{eq:lambda-y-zi}-\eqref{eq:P-H-expand} and \eqref{eq:R-hat-1234}}
\label{appendix:Lemma}

\textbf{The proof of the Eq. \eqref{eq:W-1-2}:}
Combined with Eq. \eqref{eq:addition-zeta}, the function $W_2(z)$ satisfies
\begin{equation}\label{eq:W-2}
	\begin{split}
		W_2(z)-\ii \alpha Z(K_l)
		\xlongequal{\eqref{eq:addition-zeta}}&\lambda+\hat{\alpha}\pi-\ii \alpha Z(\ii K'-z_l)-\ii \alpha k^2\sn(\ii K'-z_l)\sn(\ii K'- K_l-z_l)\sn(K_l)\\
		\xlongequal[\eqref{eq:shift-elliptic}]{\eqref{eq:addition-zeta}}&\lambda+\ii \alpha Z(z_l)+\ii \alpha \cn(z_l)\dn(z_l)/\sn(z_l)-\ii \alpha\sn(K_l)/(\sn(z_l)\sn(K_l+z_l)).
	\end{split}
\end{equation}
And then, we get
\begin{equation}\nonumber
	\begin{split}
		W_2(z)-\ii \alpha Z(K_l)+W_1(z)
		\xlongequal[\eqref{eq:W-2}]{\eqref{eq:E1-E2-W}}&2\lambda+\ii \alpha \frac{\cn(z_l)\dn(z_l)}{\sn(z_l)}-\ii \alpha\frac{\sn(K_l)}{\sn(z_l)\sn(K_l+z_l)}\\
		\xlongequal{\eqref{eq:lambda-y}}&\ii \alpha\left(\frac{\scd(z_l)-\scd(K_l)}{\sn^2(K_l)-\sn^2(z_l)}+\frac{\cn(z_l)\dn(z_l)}{\sn(z_l)}-\frac{\sn(K_l)}{\sn(z_l)\sn(K_l+z_l)}\right)\\
		=&\ii \alpha\left(\frac{\sn(K_l)(\cn(z_l)\dn(z_l)\sn(K_l)-\cn(K_l)\dn(K_l)\sn(z_l))}{(\sn^2(K_l)-\sn^2(z_l))\sn(z_l)}-\frac{\sn(K_l)}{\sn(z_l)\sn(K_l+z_l)}\right)\\
		\xlongequal{\eqref{eq:addition-elliptic}}&\ii \alpha\left(\frac{\sn(K_l)\sn(K_l-z_l)}{\sn(K_l-z_l)\sn(K_l+z_l)\sn(z_l)}-\frac{\sn(K_l)}{\sn(z_l)\sn(K_l+z_l)}\right)=0.
	\end{split}
\end{equation}
Thus, we obtain $-W_1(z)=W_2(z)-\ii \alpha Z(K_l)=W_2(z)-\omega$, which implies $W'_2(z)=-W_1'(z)$.

\textbf{The proof of the Eq. \eqref{eq:define-y-hat}:} Differentiating $y(z)$ in Eq. \eqref{eq:y_mu_lambda} with respect to $z$ gives
	\begin{equation}\label{eq:y-d}
		\begin{split}
			y'(z)
			&\!\xlongequal{\eqref{eq:derivative-elliptic}}\!
			\ii \alpha^2 k^2 K\left(
			\scd(z_l)-\scd(z_l+K_l+\ii K')\right)\\
			&\!\xlongequal[\eqref{eq:shift-elliptic}]{\eqref{eq:add-2}}\!
			\ii \alpha^2 k^2 K\!\Big(\frac{(1-\cn(2z_l))(\dn(2z_l)+\cn(2z_l))}{\sn(2z_l)(1+\dn(2z_l))}
			\!+\!\frac{(1+\cn(2z_l+2K_l))(\cn(2z_l+2K_l)+\dn(2z_l+2K_l))}{\sn(2z_l+2K_l)(1-\dn(2z_l+2K_l))}\Big).
		\end{split}
	\end{equation}
	Substituting $z=z_1$ into $z_l=2\ii (z-l)K$, we get $2z_{l_1}=4\ii (z_1-l)K=-K_l+2K+\ii K'$ and $2z_{l_1}+2K_l=K_l+2K+\ii K'$. 
	Then, plugging $z_1$ into Eq. \eqref{eq:y-d}, we get
	\begin{equation}\label{eq:derivative-y}
			\hat{y}_{1}=\frac{y'(z_1)}{2\ii K}
			\xlongequal{\eqref{eq:shift-elliptic}}
			\alpha^2 k^2 \left(
			\frac{(k\sn(K_l)+\ii\dn(K_l))(\ii\dn(K_l)-\ii k\cn(K_l) )}{(k\sn(K_l)+\ii k\cn(K_l))}\right)
			\xlongequal{\eqref{eq:define-d-1-2-3-Kl}}
			\alpha^2 (d_3-\ii d_1)(d_2 +\ii d_1)(d_3-d_2).
	\end{equation}
	Similarly, we get $\hat{y}_{i}$, $i=2,3,4$. Thus, we obtain the Eq. \eqref{eq:define-y-hat}.
	
	\textbf{The proof of the Eq. \eqref{eq:define-lambda-hat}:} Combined with Eqs. \eqref{eq:define-d-1-2-3-Kl}, \eqref{eq:y-d}, \eqref{eq:derivative-y}, \eqref{eq:shift-elliptic}, and \eqref{eq:argument-elliptic}, the following equations hold:
	\begin{equation}\label{eq:scd-sn}
		\begin{split}
			\scd(z_{l_1})\xlongequal[\eqref{eq:derivative-y}]{\eqref{eq:y-d}}&
			\frac{(1-\cn(K_l+\ii K'))(\cn(K_l+\ii K')-\dn(K_l+\ii K'))}{k^2\sn(K_l+\ii K')(1-\dn(K_l+\ii K'))}
			\xlongequal{\eqref{eq:derivative-y}}
			\frac{(d_3-\ii d_1)(d_2 +\ii d_1)(d_3-d_2)}{k^2},\\
			\sn^2(z_{l_1})
			\xlongequal{\eqref{eq:argument-elliptic}}&\frac{1-\cn(z_{l_1})}{1+\dn(z_{l_1})}
			\xlongequal{\eqref{eq:shift-elliptic}}\frac{k\sn(K_l)+\ii\dn(K_l)}{k\sn(K_l)+\ii k\cn(K_l)}
			\xlongequal[\eqref{eq:define-d-1-2-3-Kl}]{\eqref{eq:shift-elliptic}}\frac{(d_3-\ii d_1)(d_3-d_2)}{k^2},
		\end{split}
	\end{equation}
	where $z_{l_1}=2\ii (z_1-l)K$.
	Plugging $z=z_1$ into Eq. \eqref{eq:lambda-y} and combining with Eq. \eqref{eq:scd-sn}, we obtain 
	\begin{equation}\nonumber
		\lambda_1=\lambda(z_1)
		\xlongequal[\eqref{eq:scd-sn}]{\eqref{eq:lambda-y}}\frac{\ii \alpha}{2}\frac{(d_3-\ii d_1)(d_2 +\ii d_1)(d_3-d_2)-\ii d_1d_2d_3}{d_3^2-(d_3-\ii d_1)(d_3-d_2)}
		=\alpha(d_1+\ii (d_3-d_2))/2.
	\end{equation}
	Similarly, we obtain parameters $\lambda_{2,3,4}$ and get the Eq. \eqref{eq:define-lambda-hat}.

	\textbf{The proof of the Eq. \eqref{eq:define-Lambda}:} By Eqs. \eqref{eq:argument-elliptic}, \eqref{eq:derivative-zeta}, and \eqref{eq:E1-E2-W}, the following equations hold
	\begin{equation}\nonumber
		I'_1(z)\xlongequal[\eqref{eq:argument-elliptic}]{\eqref{eq:derivative-zeta}}
		-4 K \alpha\left(\frac{\cn(2z_l)+\dn(2z_l)}{1+\cn(2z_l)}-\frac{E}{K}\right). 
	\end{equation}
	Substituting $z=z_i$, $i=1,2,3,4$, into the above equations, we obtain
	\begin{equation}\nonumber
		\begin{split}
			\Lambda_1=\frac{I'_1(z_1)}{4\ii K}
			\xlongequal{\eqref{eq:shift-elliptic}}\ii \alpha \left(\frac{-\ii \dn(K_l)+\ii k\cn(K_l)}{k\sn(K_l)-\ii\dn(K_l)}-\frac{E}{K}\right)
			=\ii \alpha \left((\ii d_1+d_2)(d_3-\ii d_1)-\frac{E}{K}\right).
		\end{split}
	\end{equation}
	As above, we could obtain parameters $\Lambda_i$, $i=2,3,4$ expressed in Eq. \eqref{eq:define-Lambda}.

\textbf{The proof of the Eq. \eqref{eq:lambda-y-zi}:} Plugging $z=z_1+\epsilon=-\ii(1+\tau)/4+\epsilon$ into Eq. \eqref{eq:lambda-y}, we get
	\begin{equation}\nonumber
		\begin{split}
			y(z_1+\epsilon)
			=&\alpha^2 k^2(\sn^2(2\ii (z_1+\epsilon)K-2\ii lK)
			-\sn^2(-2\ii (z_1+\epsilon)K-2\ii lK+K+\ii K'))/4\\
			=& \alpha^2 k^2(\sn^2((K+\ii K')/2-2\ii lK+2\ii \epsilon K)
			-\sn^2((K+\ii K')/2-2\ii lK-2\ii\epsilon K))/4,
		\end{split}
	\end{equation}
	which implies $y(z_1+\epsilon)=-y(z_1-\epsilon)$. 	
	Then, we consider the symmetry of $\lambda(z)$. 
	By \cite{FengLT-20}, when $l\in [0,-\ii \tau/4)$, the spectral parameter $\lambda(z)$ with $z=z_1+\epsilon$ could be rewritten as 
	\begin{equation}\nonumber
		\begin{split}
			\lambda(z_1+\epsilon)
			=&\ii \alpha(\ii k^2\sn(K_l)\cn(K_l )
			-\dn(2\ii(z_1+\epsilon-l)K)
			\dn(\ii K'+K-2\ii (z_1+\epsilon+l)K))/(2\dn(K_l))\\
			=&\ii \alpha(\ii k^2\sn(K_l )\cn(K_l )-\dn((K+\ii K')/2-2\ii lK+2\ii\epsilon K)
			\dn((K+\ii K')/2-2\ii lK-2\ii \epsilon K))/(2\dn(K_l)),
		\end{split}
	\end{equation}
	which means $\lambda(z_1+\epsilon)=\lambda(z_1-\epsilon)$.
	When $l=-\ii \tau/4$, the function $\lambda(z)$ satisfies
	\begin{equation}\nonumber
		\begin{split}
			\lambda(z_1+\epsilon)
			=&\ \frac{\ii \alpha \sn(2\ii(z_1+\epsilon-l)K)\cn(2\ii(z_1+\epsilon-l)K)}{2\dn(2\ii(z_1+\epsilon-l)K)}
			= \frac{\ii \alpha \sn(K/2+2\ii\epsilon K)\cn(K/2+2\ii \epsilon K)}{2\dn(K/2+2\ii \epsilon K)}\\
			\xlongequal{\eqref{eq:shift-elliptic}}
			&-\frac{\ii \alpha \sn(-K/2+2\ii\epsilon K)\cn(-K/2+2\ii \epsilon K)}{2\dn(-K/2+2\ii \epsilon K)}
			=\lambda(z_1-\epsilon).
		\end{split}	
	\end{equation}
	In summary, we obtain that symmetries of functions $\lambda(z)$ and $y(z)$ hold at the branch point $z_1=-\ii (1+\tau)/4$.
	In the same way, we could obtain Eq. \eqref{eq:lambda-y-zi} at branch points $z_{2,3,4}$.

\textbf{The proof of the Eq. \eqref{eq:P-H-expand}:} Based on the conformal map $\lambda(z)$, the function $y(z)$, and symmetries \eqref{eq:lambda-y-zi}, the fundamental solution $\Phi$ in Eq. \eqref{eq:Phi-elliptic} could be rewritten as
\begin{equation}\label{eq:define-Phi}
	\Phi(\xi,\eta;\lambda(\hat{z}_i))
	\xlongequal{\eqref{eq:lambda-y-zi}}\Phi(\xi,\eta;\lambda(z_i+\epsilon_i))
	=\begin{bmatrix}
		e_1^{+}(z_i) & e_1^{-}(z_i) \\
		b_1^{+}(z_i)e_1^{+}(z_i) & 
		b_1^{-}(z_i)e_1^{-}(z_i) \\
	\end{bmatrix}
	\begin{bmatrix}
		\ee^{\theta_1^{+}(z_i)} & 0 \\ 0 & \ee^{\theta_1^{-}(z_i)}
	\end{bmatrix},
\end{equation}
where $\hat{z}_i=z_i+\epsilon_i\in S$, $z_i\in\mathcal{B}$, $\epsilon_i\in \mathbb{C}$; the function $\theta_1(z)$ is defined in Eq. \eqref{eq:define-theta-1-2}; and $e_1^{\pm}(z_i):=e_1(z_i\pm \epsilon_i)$, $b_1^{\pm}(z_i):=b_1(z_i\pm \epsilon_i)$, $\theta_1^{\pm}(z_i):=\theta_1(z_i\pm \epsilon_i)$, $y^{\pm}(z_i):=y(z_i\pm \epsilon_i)$, $\lambda^{\pm}(z_i):=\lambda(z_i\pm \epsilon_i)$.
Functions $b_1(z)$, $e_1(z)$ and $\beta_{1,2}(z)$ are defined as
\begin{equation}\nonumber
	b_{1}(z)=\sqrt{\frac{(\varrho(\xi)-\beta_2(z))(\lambda(z)+\mu)}{(\varrho(\xi)-\beta_1(z))(\lambda(z)-\mu)}},
\end{equation}
$e_{1}(z)=\sqrt{\varrho(\xi)-\beta_1(z)}$, $\beta_{1,2}(z)=2\lambda^2(z)+\alpha^2(d_3^2+d_2^2-d_1^2)/2\mp2y(z)$ and functions $\varrho(\xi)$ and $\mu$ are defined in Eqs. \eqref{eq:Phi-elliptic} and \eqref{eq:elements-L}.
Based on the definition of functions $\lambda(z)$ and $y(z)$, it is easy to obtain 	
\begin{equation}\nonumber
	\lambda^*(z)=\lambda(z^*), \qquad  y^*(z)=y(z^*),\qquad 
	\beta_{1,2}^*(z)=\beta_{1,2}(z^*),
\end{equation}
where $d_{1,2,3}\in \mathbb{R}$ are defined in Eq. \eqref{eq:define-d-1-2-3-Kl}.
Furthermore, we obtain 
\begin{equation}\nonumber
	\varrho^*(\xi)=k^2\alpha^2\left(\sn^2(K-4\ii lK)-\sn^2(\alpha\xi)\right)
	=k^2\alpha^2\left(\sn^2(-K-4\ii lK)-\sn^2(\alpha\xi)\right)=\varrho(\xi),
\end{equation}
	which implies 
	\begin{equation}\nonumber
		e_1^*(z)=\sqrt{\varrho^*(\xi)-\beta_1^*(z)}=\sqrt{\varrho(\xi)-\beta_1(z^*)}=e_1(z^*),\qquad 
		b_2(z^*)=b_1^*(z)=\sqrt{\frac{(\varrho(\xi)-\beta_2(z^*))(\lambda(z^*)-\mu)}{(\varrho(\xi)-\beta_1(z^*))(\lambda(z^*)+\mu)}},
	\end{equation}
	and $\theta_{1}^*(z)=-\theta_1(z^*)$ with $\theta_1(z)$ defined in Eq. \eqref{eq:define-theta-1-2}.
	Then, we get 
	\begin{equation}\label{eq:define-Psi-Phi-dagger}
		\Psi(\xi,\eta;\lambda(z_i^*+\epsilon_i^*))
		=\left(\Phi(\xi,\eta;\lambda(z_i+\epsilon_i))\right)^{\dagger}
		=\begin{bmatrix}
			\ee^{-\theta_1^+(z_i^*)} & 0 \\ 0 & \ee^{-\theta_1^-(z_i^*)}
		\end{bmatrix}
		\begin{bmatrix}
			e_1^+(z_i^*) & b_2^+(z_i^*)e_1^+(z_i^*) \\
			e_1^-(z_i^*) & 	b_2^-(z_i^*)e_1^-(z_i^*) \\
		\end{bmatrix},
	\end{equation}
	where $e_1^{\pm}(z_i^*):=e_1(z_i^*\pm \epsilon_i^*)$, $b_2^{\pm}(z_i^*):=b_2(z_i^*\pm \epsilon_i^*)$, and $\theta_1^{\pm}(z_i^*):=\theta_1(z_i^*\pm \epsilon_i^*)$.
	Since parameters $c_{i1}$ and $c_{i2}$ are arbitrary, we assume that $c_{i1}=\exp(h^+(z_i))$ and $c_{i2}=-\exp(h^-(z_i))$ satisfy $c_{i1}^*=\exp(h^+(z_i^*))$ and $c_{i2}^*=-\exp(h^-(z_i^*))$ with $h^{\pm}(z_i)=h(z_i\pm \epsilon)$, where $h(z)$ is a polynomial function about variable $z$.
	Combined with the Darboux-B\"{a}cklund transformation, $\hat{\mathcal{H}}(z^*,z)$ could be expressed as
	\begin{equation}\nonumber
		\begin{split}
			\hat{\mathcal{H}}(\hat{z}_i^*,\hat{z}_j)
			=&\frac{\vartheta_4(\hat{\alpha}\xi)}{2\ii \gamma \vartheta_3(2\ii l)}
			\frac{\mathbf{c}_i^{\dagger}(\Phi(\xi,\eta;\lambda(\hat{z}_i)))^{\dagger}\Phi(\xi,\eta;\lambda(\hat{z}_j))\mathbf{c}_j}{\lambda(\hat{z}_j)-\lambda(\hat{z}_i^*)}
			\xlongequal{\eqref{eq:define-Psi-Phi-dagger}}\frac{\vartheta_4(\hat{\alpha}\xi)}{2\ii \gamma \vartheta_3(2\ii l)}
			\frac{\mathbf{c}_i^{\dagger}\Psi(\xi,\eta;\lambda(\hat{z}_i^*))\Phi(\xi,\eta;\lambda(\hat{z}_j))\mathbf{c}_j}{\lambda(\hat{z}_j)-\lambda(\hat{z}_i^*)}\\
			\xlongequal[\eqref{eq:define-Phi}]{\eqref{eq:define-Psi-Phi-dagger}}&\frac{\vartheta_4(\hat{\alpha}\xi)}{2\ii \gamma \vartheta_3(2\ii l)}\left(\sum_{a,b=1}^{2}(-1)^{a+b}\hat{\mathcal{H}}_{ab}(\hat{z}_i^*,\hat{z}_j)\right),
		\end{split}		
	\end{equation}
	where $\hat{z}_j=z_j+\epsilon_j$, $\hat{z}_i^*=z_i^*+\epsilon_i^*$ and 
	\begin{equation}\nonumber
		\begin{split}
			\hat{\mathcal{H}}_{1m}(\hat{z}^*_i,\hat{z}_j)=&\ee^{-\theta_1^+(z^*_i)+h^{+}(z_i^*)+\theta_1^{\pm}(z_j)+h^{\pm}(z_j)}
			\frac{e_1^{+}(z^*_i)e_1^{\pm}(z_j)(1+b_3^{+}(z^*_i)b_1^{\pm}(z_j))}{\lambda^{\pm}(z_j)-\lambda^{+}(z_i^*)}, \\
			\hat{\mathcal{H}}_{2m}(\hat{z}^*_i,\hat{z}_j)=&\ee^{-\theta_1^-(z^*_i)+h^{-}(z_i^*)+\theta_1^{\pm}(z_j)+h^{\pm}(z_j)}
			\frac{e_1^{-}(z^*_i)e_1^{\pm}(z_j)(1+b_3^{-}(z^*_i)b_1^{\pm}(z_j))}{\lambda^{\pm}(z_j)-\lambda^{-}(z_i^*)},
		\end{split}
	\end{equation} 
	$m=1,2$ ($m=1$ and $m=2$ correspond to ``$+$" and ``$-$" respectively).
	As $\epsilon_j\rightarrow 0$, we get $e_1^{+}(z_j)=e_1^{-}(z_j)$, $b_{1,2}^{+}(z_j)=b_{1,2}^{-}(z_j)$, $\theta_1^{+}(z_j)=\theta_1^{-}(z_j)$, $h^{+}(z_j)=h^{-}(z_j)$ , $z_j\in \mathcal{B}$. 
	As $\epsilon_i^*\rightarrow 0$, we could obtain similar conclusions.
	Thus, we know $\lim_{\epsilon_i^*,\epsilon_j\rightarrow 0}\hat{\mathcal{H}}(\hat{z}^*_i,\hat{z}_j)=\hat{\mathcal{H}}(z^*_i,z_j)=0$. 
	The same as above, the function $\hat{\mathcal{P}}(z^*,z)$ is expressed as 
	\begin{equation}\nonumber
		\begin{split}
			\hat{\mathcal{P}}(\hat{z}^*_i,\hat{z}_j)=&\frac{\vartheta_4(\hat{\alpha}\xi)}{2\ii \gamma \vartheta_3(2\ii l)}\left(\frac{q\mathbf{c}_i^{\dagger}(\Phi(\xi,\eta;\lambda(\hat{z}_i)))^{\dagger}\Phi(\xi,\eta;\lambda(\hat{z}_j))\mathbf{c}_j}{2(\lambda(\hat{z}_j)-\lambda(\hat{z}_i^*))}
			-\mathbf{c}_i^{\dagger}(\Phi(\xi,\eta;\lambda(\hat{z}_i)))^{\dagger}
			\begin{bmatrix}
				0 & 0 \\ 1 & 0
			\end{bmatrix}
			\Phi(\xi,\eta;\lambda(\hat{z}_j))\mathbf{c}_j \right)\\
			=&\frac{\vartheta_4(\hat{\alpha}\xi)}{2\ii \gamma \vartheta_3(2\ii l)}\left(\frac{q\mathbf{c}_i^{\dagger}\Psi(\xi,\eta;\lambda(\hat{z}_i^*))\Phi(\xi,\eta;\lambda(\hat{z}_j))\mathbf{c}_j}{2(\lambda(\hat{z}_j)-\lambda(\hat{z}_i^*))}
			-\mathbf{c}_i^{\dagger}\Psi(\xi,\eta;\lambda(\hat{z}_i^*))
			\begin{bmatrix}
				0 & 0 \\ 1 & 0
			\end{bmatrix}
			\Phi(\xi,\eta;\lambda(\hat{z}_j))\mathbf{c}_j \right),
		\end{split}
	\end{equation}
	and also satisfies $\lim_{\epsilon_i^*,\epsilon_j\rightarrow 0}\hat{\mathcal{P}}(\hat{z}^*_i,\hat{z}_j)=\hat{\mathcal{P}}(z^*_i,z_j)=0$, $z_i^*\in \mathcal{B}$, $z_j\in \mathcal{B}$.
	Consider Taylor expansions of functions $\hat{\mathcal{H}}(\hat{z}^*_i,\hat{z}_j)$ and $\hat{\mathcal{P}}(\hat{z}^*_i,\hat{z}_j)$ at branch points $z_{i,j}$ as $\epsilon_i$ and $\epsilon_j$ at the neighborhood of origin.
	Based on the definition of functions $\Psi(\xi,\eta;\lambda(z^*))$ in Eq. \eqref{eq:define-Psi-Phi-dagger}, the Taylor expansion of the function $\Psi(\xi,\eta;\lambda(z^*))$ has similar properties of $\Phi(\xi,\eta;\lambda(z))$, since $\Psi(\xi,\eta;\lambda(z^*))=\sum_{j=0}^{\infty}\Psi^{[j]}(\xi,\eta;\lambda(z_i^*))(z^*-z_i^*)^j$ and $(\Phi(\xi,\eta;\lambda(z)))^{\dagger}=\sum_{j=0}^{\infty}(\Phi^{[j]}(\xi,\eta;\lambda(z_i)))^{\dagger}(z^*-z_i^*)^j$ with $(\Phi^{[j]}(\xi,\eta;\lambda(z_i)))^{\dagger}=\Psi^{[j]}(\xi,\eta;\lambda(z_i^*))$.
	Combined with the definition of functions $\hat{\mathcal{H}}_{ab}(z^*_i,z_j)$, $a,b=1,2$, it is easy to obtain
	\begin{equation}\nonumber
		\hat{\mathcal{H}}(\hat{z}^*_i,\hat{z}_j)
		=\sum_{m,n=0}^{\infty}\sum_{a,b=1}^{2}\frac{\vartheta_4(\hat{\alpha}\xi)\hat{\mathcal{H}}_{ab}^{[m,n]}(z^*_i,z_j)\epsilon_j^n(\epsilon^*_i)^m}{(-1)^{a+b}2\ii \gamma \vartheta_3(2\ii l)}, \qquad 
		\hat{\mathcal{H}}_{ab}^{[m,n]}(z^*_i,z_j)
		=\left.\frac{\dd^{m+n}\hat{\mathcal{H}}_{ab}(z^*,z)}{m!n!\dd z^{*m}\dd z^n}\right|_{z^*=z^*_i,z=z_j}.
	\end{equation}
	We obtain $\hat{\mathcal{H}}_{11}^{[m,n]}(z^*_i,z_j)
	=(-1)^{n+m}\hat{\mathcal{H}}_{22}^{[m,n]}(z^*_i,z_j)
	=(-1)^{n}\hat{\mathcal{H}}_{12}^{[m,n]}(z^*_i,z_j)
	=(-1)^{m}\hat{\mathcal{H}}_{21}^{[m,n]}(z^*_i,z_j)
	$.
	Therefore, only when $n$ and $m$ are both odd numbers, $\hat{\mathcal{H}}_{11}^{[m,n]}(z^*_i,z_j)=-\hat{\mathcal{H}}_{12}^{[m,n]}(z^*_i,z_j)=-\hat{\mathcal{H}}_{21}^{[m,n]}(z^*_i,z_j)=\hat{\mathcal{H}}_{22}^{[m,n]}(z^*_i,z_j)\neq 0$.
	As $\epsilon_{i,j}\rightarrow 0$, we get 
	\begin{equation}\label{eq:H-hat-limitation}\nonumber
		\hat{\mathcal{H}}(z^*_i+\epsilon_i^*,z_j+\epsilon_j)
		=\frac{\vartheta_4(\hat{\alpha}\xi)}{2\ii \gamma \vartheta_3(2\ii l)}\sum_{m,n=0}^{\infty}4\hat{\mathcal{H}}_{11}^{[2m-1,2n-1]}(z_i^*,z_j)\epsilon_j^n(\epsilon_i^*)^m.
	\end{equation}
	Together with the definition of the function $\hat{\mathcal{H}}(z^*,z)$, we get $\hat{\mathcal{H}}(\hat{z}_i^*,\hat{z}_j)=4\mathcal{H}(\hat{z}_i^*,\hat{z}_j)$ in Eq. \eqref{eq:P-H-expand}. 
	The function $\hat{\mathcal{P}}(z^*,z)$ could also be expressed in the same way.

\textbf{The proof of the Eq. \eqref{eq:R-hat-1234}:}
By the derivative formulas of theta functions \eqref{eq:derivative-theta} and the addition formulas of elliptic functions \eqref{eq:add-1} and \eqref{eq:addition-zeta}, we get
	\begin{equation}\nonumber
		\begin{split}
			\ln\left(r(z)\right)_z
			\xlongequal[]{\eqref{eq:derivative-theta}}&\ 2\ii K \Big(Z(z_l+4\ii l K)-Z(z_l)
			-\frac{\sn(z_l+4\ii l K)\dn(z_l+4\ii l K)}{\cn(z_l+4\ii l K)}\Big)\\
			\xlongequal[\eqref{eq:add-1}]{\eqref{eq:addition-zeta}}&\ \ii K \Big(Z(2z_l+8\ii l K)-Z(2z_l)
			-\frac{(1-\cn(2z_l+8\ii l K))(1+\dn(2z_l+8\ii l K))}{\sn(2z_l+8\ii l K)}
			-k^2\frac{1-\cn(2z_l)}{1+\dn(2z_l)}\sn(2z_l)\Big).
		\end{split}
	\end{equation}
	Plugging $z=z_i$, $i=1,2,3,4$, into the above equation and combining with addition formulas of the Zeta function \eqref{eq:addition-zeta} and shift formulas of elliptic functions \eqref{eq:shift-elliptic}, we get
	\begin{equation}\nonumber
		\begin{split}
			\left.\ln\left(r(z)\right)_z\right|_{z=z_1}
			\xlongequal[\eqref{eq:addition-zeta}]{\eqref{eq:shift-elliptic}}
			&\ii K \Big(Z(\ii K'+K_l)-Z(\ii K'-K_l)
			-2\frac{(1-\cn(\ii K'+K_l))(1+\dn(\ii K'+K_l))}{\sn(\ii K'+K_l)}\Big)\\
			\xlongequal[]{\eqref{eq:shift-elliptic}}
			&\ii K \Big(2Z(K_l)+2\frac{\cn(K_l)\dn(K_l)}{\sn(K_l)}
			-\frac{(\sn(K_l)-\ii \dn(K_l))(\sn(K_l)-\ii \cn(K_l))}{\sn(K_l)}\Big)\\
			\xlongequal{\eqref{eq:define-d-1-2-3-Kl}}
			&2\ii K(d_2-d_3+\ii (d_1-\omega)).
		\end{split}
	\end{equation}
	Then, we get $r'(z_1)=2\ii K r_1R_1$, where $R_1=d_2-d_3+\ii (d_1-\omega)$.
	Similarly, we could obtain $R_{i}$, $i=2,3,4$.

\section{\appendixname. Elliptic-localized wave solution of INS equations (the proof of Eq. \eqref{eq:rogue-wave})}\label{Appendix:localized-wave}
For ease of representation, we introduce the following functions to represent elements of matrices provided in $\mathcal{P}(z^*,z)$ and $\mathcal{H}(z^*,z)$.
	Define functions $G(z^*,z)$ and $F(z^*,z)$ as follows:
	\begin{equation}\label{eq:define-G-F}
		G(z^*,z)=-\frac{\vartheta_4(\ii (z^*-z)+\hat{\alpha}\xi)}{\vartheta_1(\ii (z^*-z))},\qquad  F(z^*,z)=-\frac{\vartheta_2(\ii (z^*-z+2l)+\hat{\alpha}\xi)}{\vartheta_1(\ii (z^*+z))}.
	\end{equation}
	By the derivative formulas of theta functions \eqref{eq:derivative-theta} and the Zeta function \eqref{eq:derivative-zeta}, differentiating $F(z^*,z)$ with respect to $z$, we get
	\begin{equation}\label{eq:derivative-G}
		\begin{split}
			\hat{G}_1(z^*,z)
			:=&-\left(\ln\left(G(z^*,z)\right)\right)_z
			=\left(\ln\left(G(z^*,z)\right)\right)_{z^*}
			\xlongequal[ ]{\eqref{eq:derivative-theta}}
			2\ii K\! (Z(\hat{z}+\alpha \xi)-Z(\hat{z}-K)
			-\cn(\hat{z})/(\sn(\hat{z})\dn(\hat{z}))),\\
			\hat{G}_2(z^*,z)
			:=&\left(\ln\left(G(z^*,z)\right)\right)_{zz}
			\xlongequal[\eqref{eq:derivative-zeta}]{\eqref{eq:derivative-elliptic}}
			(2\ii K)^2 \Big(\dn^2(\hat{z}+\hat{\alpha}\xi)-\dn^2(\hat{z}-K)
			-\frac{k^2\sn^2(\hat{z})\cn^2(\hat{z})-\dn^2(\hat{z})}{\sn^2(\hat{z})\dn^2(\hat{z})}\Big)\\
			\xlongequal[]{\eqref{eq:shift-elliptic}}&
			\left(2\ii K\right)^2\Big(\dn^2(\hat{z}+\hat{\alpha}\xi)+k'^2\frac{\sn^2(-\hat{z}-\ii K')}{\cn^2(-\hat{z}-\ii K')}
			-\frac{\dn^2(\hat{z}+\ii K')}{\cn^2(\hat{z}+\ii K')}+k^2\sn^2(\hat{z}+\ii K')\Big)\\
			=
			&\left(2\ii K\right)^2\left(\dn^2(\hat{z}+\hat{\alpha}\xi)-\dn^2(\hat{z}+\ii K')\right),
		\end{split}
	\end{equation}
	where $\hat{z}=2\ii(z^*-z)K$. 
	Similarly, we obtain
	\begin{equation}\label{eq:derivative-F}
		\begin{split}
			\hat{F}_1(z^*,z):=&-\left(\ln\left(F(z^*,z)\right)\right)_z
			=\left(\ln\left(F(z^*,z)\right)\right)_{z^*}
			\xlongequal[ ]{\eqref{eq:derivative-theta}}
			2\ii K (Z(\hat{z}+K_l+\alpha \xi+\ii K')
			-Z(\hat{z}+\ii K')),\\
			\hat{F}_2(z^*,z):=&\left(\ln\left(F(z^*,z)\right)\right)_{zz}
			\xlongequal[ ]{\eqref{eq:derivative-zeta}}
			(2\ii K)^2 (\dn^2(\hat{z}+K_l+\alpha \xi+\ii K')
			-\dn^2(\hat{z}+\ii K')).
		\end{split}
	\end{equation}
	
By Eq. \eqref{eq:rogue-wave-lim}, to obtain the exact expression of solution $\hat{\mathbf{Q}}_{i}^{(1)}$, we need to get the expression of $\mathcal{P}^{[1,1]}(z_i^*,z_i)$ and $\mathcal{H}^{[1,1]}(z_i^*,z_i)$.
Together with Eqs. \eqref{eq:E1-E2-W}, \eqref{eq:define-Gamma-Lambda}, \eqref{eq:R-hat-1234}, \eqref{eq:derivative-G}, \eqref{eq:derivative-F}, and $h(z)=0$, functions $\mathcal{H}^{[1,1]}(z_i^*,z_i)$ and $\mathcal{P}^{[1,1]}(z_i^*,z_i)$ are rewritten as
		\begin{equation}\label{eq:define-H-11}\nonumber
	\begin{split}
		\!\!\!\!\mathcal{H}^{[1,1]}(z_i^*,z_i)
		=&\begin{bmatrix}
			E_1^{[1]*}(z_i) & 	E_1^{[0]*}(z_i)
		\end{bmatrix}
		\begin{bmatrix}
			G(z_i^*,z_i) &
			G^{[0,1]}(z_i^*,z_i) \\
			G^{[1,0]}(z_i^*,z_i) &
			G^{[1,1]}(z_i^*,z_i)
		\end{bmatrix}
		\begin{bmatrix}
			E_1^{[1]}(z_i)\\[3pt] 	E_1^{[0]}(z_i)
		\end{bmatrix}\\
		\xlongequal[\eqref{eq:derivative-G}]{\eqref{eq:E1-E2-W},\eqref{eq:define-Gamma-Lambda}}&|E_1(z_i)|^2  G(z_i^*,z_i)
		\begin{bmatrix}
			2\ii K(\ii \hat{E}_i^*)& 	1
		\end{bmatrix}
		\begin{bmatrix}
			1 &
			-\hat{G}_1(z_i^*,z_i) \\
			\hat{G}_1(z_i^*,z_i) &
			-(\hat{G}_1^2(z_i^*,z_i)+\hat{G}_2(z_i^*,z_i))
		\end{bmatrix}
		\begin{bmatrix}
			2\ii K(\ii \hat{E}_i)\\ 	1
		\end{bmatrix}\\
		=&-(2\ii K)^2|E_1(z_i)|^2  G(z_i^*,z_i)
		\begin{bmatrix}
			-\ii \hat{E}_i^*& 	1
		\end{bmatrix}
		\begin{bmatrix}
			1 &
			-g_1(z_i^*,z_i) \\
			-g_1(z_i^*,z_i) &
			g_1^2(z_i^*,z_i)+g_2(z_i^*,z_i)
		\end{bmatrix}
		\begin{bmatrix}
			\ii \hat{E}_i\\ 	1
		\end{bmatrix},
	\end{split}
\end{equation}
\begin{equation}\label{eq:define-H-11}\nonumber
		\begin{split}
			\!\!\!\!\!\!	
			\mathcal{P}^{[1,1]}(z_i^*,z_i)
			=&\begin{bmatrix}
				E_1^{[1]*}(z_i) & 	E_1^{[0]*}(z_i)
			\end{bmatrix}
			\begin{bmatrix}
				((r_i^*)^{-1})^{[0]}	&  0\\[3pt]
				((r_i^*)^{-1})^{[1]} & ((r_i^*)^{-1})^{[0]}
			\end{bmatrix}
			\begin{bmatrix}
				F(z_i^*,z_i) &
				F^{[0,1]}(z_i^*,z_i) \\
				F^{[1,0]}(z_i^*,z_i) &
				F^{[1,1]}(z_i^*,z_i)
			\end{bmatrix}
			\begin{bmatrix}
				r_i^{[0]}	&  r_i^{[1]}\\[3pt]
				0 & r_i^{[0]}
			\end{bmatrix}
			\begin{bmatrix}
				E_1^{[1]}(z_i)\\[3pt] 	E_1^{[0]}(z_i)
			\end{bmatrix}\!\!\!\!\!\!\!\!\\	
			\xlongequal[\eqref{eq:R-hat-1234}]{\eqref{eq:E1-E2-W},\eqref{eq:define-Gamma-Lambda}}&
			\frac{|E_1(z_i)|^2r_i}{r_i^*}\begin{bmatrix}
				2\ii K (\ii \hat{E}_i^*) & 	1
			\end{bmatrix}
			\begin{bmatrix}
				1	&  0\\[3pt]
				2\ii K R_i^* & 1
			\end{bmatrix}
			\begin{bmatrix}
				F(z_i^*,z_i) &
				F^{[0,1]}(z_i^*,z_i) \\
				F^{[1,0]}(z_i^*,z_i) &
				F^{[1,1]}(z_i^*,z_i)
			\end{bmatrix}
			\begin{bmatrix}
				1	&  2\ii K  R_i\\[3pt]
				0 & 1
			\end{bmatrix}
			\begin{bmatrix}
				2\ii K (\ii \hat{E}_i)\\[3pt] 	1
			\end{bmatrix}\!\!\!\!\!\!\\
			\xlongequal[]{\eqref{eq:derivative-F}}&
			\frac{(2\ii K)^2|E_1(z_i)|^2r(z_i)F(z_i^*,z_i)}{-r(z_i^*)}
			\begin{bmatrix}
				-\ii \hat{E}_i^* & 	1
			\end{bmatrix}\!
			\begin{bmatrix}
				1	&  0\\
				-R_i^* & 1
			\end{bmatrix}\!
			\begin{bmatrix}
				1 &
				-f_1(z_i^*,z_i) \\
				-f_1(z_i^*,z_i) &
				f_1^2(z_i^*,z_i)+f_2(z_i^*,z_i)
			\end{bmatrix}\!
			\begin{bmatrix}
				1	&   R_i\\
				0 & 1
			\end{bmatrix}\!
			\begin{bmatrix}
				\ii \hat{E}_i\\ 	1
			\end{bmatrix}\!,\!\!\!\!\!\!\!\!
			\end{split}
		\end{equation}
where $\hat{E}_i=\Lambda_i\xi +\Gamma_i\eta$, $g_1(z_i^*,z_i)=\hat{G}_1(z_i^*,z_i)/(2\ii K)$, $g_2(z_i^*,z_i)=\hat{G}_2(z_i^*,z_i)/(2\ii K)^2$, $f_1(z_i^*,z_i)=\hat{F}_1(z_i^*,z_i)/(2\ii K)$, $f_2(z_i^*,z_i)=\hat{F}_2(z_i^*,z_i)/(2\ii K)^2$, and
	\begin{equation}\label{eq:define-different-F-G-E-r}\nonumber
		\begin{split}
			F^{[p,q]}(z_i^*,z_i)=&\left.\frac{\dd^{q+p} F(z^*,z)}{p!q!\dd z^q \dd (z^*)^p}\right|_{z=z_j,z^*=z_i^*}, \qquad G^{[p,q]}(z_i^*,z_i)=\left.\frac{\dd^{q+p} G(z^*,z)}{p!q!\dd z^q \dd (z^*)^p}\right|_{z=z_j,z^*=z_i^*},\\
			E_n^{[q]}(z_j)=&\left.\frac{\dd^q E_{n}(z)}{q!\dd z^q}\right|_{z=z_j},\quad\quad
			r_j^{[q]}=\left.\frac{\dd^q r(z)}{q!\dd z^q}\right|_{z=z_j},\quad\quad
			\Big(\frac{1}{r^*_j}\Big)^{[q]}=\left.\frac{\dd^q (r^*(z))^{-1}}{q!\dd z^q}\right|_{z=z_j}.
		\end{split}
	\end{equation} 
	By Eq. \eqref{eq:define-branch}, it is easy to obtain $\hat{z}_{1,2}=2\ii (z_{1,2}^*-z_{1,2})=K$ and $\hat{z}_{3,4}=2\ii (z_{3,4}^*-z_{3,4})=-K$. Since functions $g_{1,2}(z_i^*,z_i)$ and $f_{1,2}(z_i^*,z_i)$ are $2K$-periodic functions, we obtain 
	\begin{equation}\label{eq:define-g-f-1-2}\nonumber
		\begin{split}
			g_1:=&\ g_1(z_i^*,z_i)=Z(K+\alpha\xi), \qquad f_1:=f_1(z_i^*,z_i)
			=Z(\alpha\xi +\ii K'+4\ii l K)-Z(K+\ii K'),\\
			g_2:=&\ g_2(z_i^*,z_i)=\dn^2(K+\alpha\xi), \qquad
			f_2:=f_2(z_i^*,z_i)
			=\dn^2(4\ii l K+\ii K'+\alpha\xi)-\dn^2(K+\ii K'),
		\end{split}
	\end{equation}
	$i=1,2,3,4$.
	Then, the solution $\hat{\mathbf{Q}}_i^{(1)}$ in Eq. \eqref{eq:rogue-wave-lim} is rewritten as the rational form:
		\begin{equation}\nonumber
			\begin{split}
				\hat{\mathbf{Q}}_{i}^{(1)}
				=&\gamma\frac{r_iF(z_i^*,z_i)}{r_i^*G(z_i^*,z_i)}
				\frac{\begin{bmatrix}
						-\ii\hat{E}_i^* & 	1
					\end{bmatrix}
					\begin{bmatrix}
						1	&  0\\
						-R_i^{*} & 1
					\end{bmatrix}
					\begin{bmatrix}
						1 &
						-f_1 \\
						-f_1 &
						f_1^2+f_2
					\end{bmatrix}
					\begin{bmatrix}
						1	&  R_i\\
						0 & 1
					\end{bmatrix}
					\begin{bmatrix}
						\ii\hat{E}_i\\	1
				\end{bmatrix}}{\begin{bmatrix}
						-\ii\hat{E}_i^* & 	1
					\end{bmatrix}
					\begin{bmatrix}
						1 &
						-g_1 \\
						-g_1 &
						g_1^2+g_2
					\end{bmatrix}
					\begin{bmatrix}
						\ii\hat{E}_i\\	1
				\end{bmatrix}}\ee^{\ii( \omega
					\xi+ \kappa \eta)}\\
				=&\ii \gamma\frac{\vartheta_2(2\ii l+\hat{\alpha}\xi-1/2)}{\vartheta_4(\hat{\alpha}\xi-1/2)}
				\frac{(\hat{E}_{i}^{\mathbf{r}}-\ii R_{i}^{\mathbf{r}})^2+f_{2}+(f_{1}+\hat{E}_{i}^{\mathbf{i}}-\ii R_{i}^{\mathbf{i}})^2}{(\hat{E}_{i}^{\mathbf{r}})^2+g_2+(g_1+\hat{E}_{i}^{\mathbf{i}})^2}\ee^{\ii (\omega\xi+\kappa \eta)}\\
				=&\ii \gamma\frac{\vartheta_2(2\ii l+\hat{\alpha}\xi-1/2)}{\vartheta_4(\hat{\alpha}\xi-1/2)}\left(1+
				\frac{2(f_{1}-\ii R_{i}^{\mathbf{i}}-g_1)\hat{E}_{i}^{\mathbf{i}}-2\ii R_{i}^{\mathbf{r}}\hat{E}_{i}^{\mathbf{r}} -(R_{i}^{\mathbf{r}})^2 -g_2+\left(f_{1}-\ii R_{i}^{\mathbf{i}}\right)^2-g_1^2+f_{2}}{(\hat{E}_{i}^{\mathbf{r}})^2+(\hat{E}_{i}^{\mathbf{i}}+g_1)^2+g_2}\right)\ee^{\ii (\omega\xi+\kappa \eta)},
			\end{split}
		\end{equation}
	where $\hat{E}_{i}^{\mathbf{r}}=\Lambda_{i}^{\mathbf{r}}\xi+\Gamma_{i}^{\mathbf{r}}\eta$, $\hat{E}_{i}^{\mathbf{i}}=\Lambda_{i}^{\mathbf{i}}\xi+\Gamma_{i}^{\mathbf{i}}\eta$, $z_i\in \mathcal{B}$, superscripts $^\mathbf{r}$ and $^\mathbf{i}$ represent the real and imaginary parts of parameters respectively.
	Since $\Lambda_{3,4}=-\Lambda_{1,2}^*$, $\Gamma_{3,4}=-\Gamma_{1,2}^*$ and $R_{3,4}=-R_{1,2}^*$, we obtain $\hat{\mathbf{Q}}_{1,2}^{(1)}=\hat{\mathbf{Q}}_{3,4}^{(1)}$. 
Then, we obtain exact expressions of elliptic-localized wave solutions \eqref{eq:rogue-wave} for INS equations.

\section{\appendixname. Higher-order elliptic-localized wave solutions (the proof of Eq. \eqref{eq:higher-order-N})}\label{Appendix:higher-order}

Together with Eqs. \eqref{eq:define-P-H-hat} and \eqref{eq:P-H-expand}, the higher-order elliptic-localized wave solutions \eqref{eq:solution-localized} of INS equations could be rewritten as follows
\begin{equation}\nonumber	
	\begin{split}
		\mathbf{Q}_{12}^{[N]}\!
		=\!\gamma \mathcal{J}^{N-1}
		\frac{\det
			\begin{pmatrix}
				\hat{\mathcal{P}}
			\end{pmatrix}
		}{\det
			\begin{pmatrix}
				\hat{\mathcal{H}} 
		\end{pmatrix}}\ee^{\ii (\omega \xi +\kappa \eta)}
		\!=\!\gamma\mathcal{J}^{N-1}
		\frac{\det
			\begin{pmatrix}
				\hat{\mathbf{P}}
			\end{pmatrix}
		}{\det
			\begin{pmatrix}
				\hat{\mathbf{H}} 
		\end{pmatrix}}\ee^{\ii (\omega \xi +\kappa \eta)},
	\end{split}
\end{equation}
where $\mathcal{J}=\vartheta_4(\hat{\alpha}\xi)/\vartheta_2(\hat{\alpha}\xi+2\ii l)$; 
$\hat{\mathcal{P}}$ and $\hat{\mathcal{H}}$ are both $N\times N$ matrices; matrices $\hat{\mathbf{P}}$ and $\hat{\mathbf{H}}$ are defined as
\begin{equation}\nonumber
	\begin{split}
		\hat{\mathbf{P}}=&\begin{pmatrix}
			\mathbf{\Sigma}_1^{\dagger}\mathbf{P}(z_1^*,z_1)\mathbf{\Sigma}_1 & 
			\mathbf{\Sigma}_1^{\dagger}\mathbf{P}(z_1^*,z_2)\mathbf{\Sigma}_2\\
			\mathbf{\Sigma}_2^{\dagger}\mathbf{P}(z_2^*,z_1)\mathbf{\Sigma}_1 & \mathbf{\Sigma}_2^{\dagger}\mathbf{P}(z_2^*,z_2)\mathbf{\Sigma}_2
		\end{pmatrix},\qquad
		\hat{\mathbf{H}}=\begin{pmatrix}
			\mathbf{\Sigma}_1^{\dagger}\mathbf{H}(z_1^*,z_1)\mathbf{\Sigma}_1 & 
			\mathbf{\Sigma}_1^{\dagger}\mathbf{H}(z_1^*,z_2)\mathbf{\Sigma}_2\\
			\mathbf{\Sigma}_2^{\dagger}\mathbf{H}(z_2^*,z_1)\mathbf{\Sigma}_1 & \mathbf{\Sigma}_2^{\dagger}\mathbf{H}(z_2^*,z_2)\mathbf{\Sigma}_2
		\end{pmatrix};
	\end{split}
\end{equation}
the $(i,j)$-elements of them are $\hat{\mathcal{P}}_{i,j}$ and $\hat{\mathcal{H}}_{i,j}$ defined in Eq. \eqref{eq:define-P-H-hat}; 
matrices $\mathbf{P}(z_i^*,z_j)$ and $\mathbf{H}(z_i^*,z_j)$ are defined in Eq. \eqref{eq:P-H-expand};
$\mathbf{\Sigma}_i
=\begin{bmatrix}
	\Sigma_{i1} & \Sigma_{i2} & \cdots & \Sigma_{iN_i} 
\end{bmatrix}$ with $\Sigma_{ij}=\begin{bmatrix}
	\epsilon_{ij} & \epsilon_{ij}^3 & \cdots  
\end{bmatrix}$. 
Matrices $\mathbf{H}(z_i^*,z_j)$ and $\mathbf{\Sigma}_j$ could be expressed by block matrices:
\begin{equation}\nonumber
	\begin{split}
		\mathbf{H}(z_i^*,z_j)
		=&\begin{bmatrix}
			\mathbf{H}_{N_i \times N_j}(z_i^*,z_j) & \mathbf{H}_{N_i \times \infty}(z_i^*,z_j)\\
			\mathbf{H}_{\infty\times N_j}(z_i^*,z_j) & \mathbf{H}_{\infty\times \infty}(z_i^*,z_j)\\
		\end{bmatrix}, \qquad
		\mathbf{\Sigma}_j
		=\begin{bmatrix}
			\Sigma_{N_j\times N_j} \\ \Sigma_{\infty \times N_j}
		\end{bmatrix},
	\end{split}
\end{equation}
where the subscript $N_i\times N_j$ represents the dimension of related matrices.
It is easy to obtain
\begin{equation}\label{eq:Sigme-H-Sigma}
	\begin{split}
		\mathbf{\Sigma}_i^{\dagger}\mathbf{H}(z_i^*,z_j)\mathbf{\Sigma}_j
		=&\begin{bmatrix}
			\Sigma_{N_i\times N_i}^{\dagger} & \Sigma_{\infty\times N_i}^{\dagger}
		\end{bmatrix}
		\begin{bmatrix}
			\mathbf{H}_{N_i\times N_j}(z_i^*,z_j) & \mathbf{H}_{N_i\times \infty}(z_i^*,z_j)\\
			\mathbf{H}_{\infty\times N_j}(z_i^*,z_j) & \mathbf{H}_{\infty\times \infty}(z_i^*,z_j)\\
		\end{bmatrix}
		\begin{bmatrix}
			\Sigma_{N_j\times N_j} \\ \Sigma_{\infty\times N_j}
		\end{bmatrix}\\
		=&\Sigma^{\dagger}_{N_i\times N_i}\mathbf{H}_{N_i\times N_j}(z_i^*,z_j)\Sigma_{N_j\times N_j}
		+\Sigma^{\dagger}_{N_i\times N_i}\mathbf{H}_{N_i\times \infty}(z_i^*,z_j)\Sigma_{\infty\times N_j}\\
		&+\Sigma^{\dagger}_{\infty\times N_i}\mathbf{H}_{\infty\times N_j}(z_i^*,z_j)\Sigma_{N_j\times N_j}
		+\Sigma^{\dagger}_{\infty\times N_i}\mathbf{H}_{\infty\times \infty}(z_i^*,z_j)\Sigma_{\infty\times N_j}\\
		=&\Sigma^{\dagger}_{N_i\times N_i}\mathbf{H}_{N_i\times N_j}(z_i^*,z_j)\Sigma_{N_j\times N_j}
		+o\left((\epsilon^{*}_{im})^{2N_i},\epsilon^{2N_j}_{jn}\right).
	\end{split}
\end{equation}
Then, by Eq. \eqref{eq:Sigme-H-Sigma}, we get 
	\begin{equation}\nonumber
		\begin{split}
			\det
			\!\begin{pmatrix}
				\hat{\mathbf{H}}
			\end{pmatrix}
			=&\det\!\begin{pmatrix}
				\Sigma^{\dagger}_{N_1\times N_1}\mathbf{H}_{N_1\times N_1}(z_1^*,z_1)\Sigma_{N_1\times N_1}
				+o((\epsilon^{*}_{1m})^{2N_1},\epsilon^{2N_1}_{1m}) &
				\Sigma^{\dagger}_{N_1\times N_1}\mathbf{H}_{N_1\times N_2}(z_1^*,z_2)\Sigma_{N_2\times N_2}
				+o((\epsilon^{*}_{1m})^{2N_1},\epsilon^{2N_2}_{2n})\\
				\Sigma^{\dagger}_{N_2\times N_2}\mathbf{H}_{N_2\times N_1}(z_2^*,z_1)\Sigma_{N_1\times N_1}
				+o((\epsilon^{*}_{2n})^{2N_2},\epsilon^{2N_1}_{1m}) & \Sigma^{\dagger}_{N_2\times N_2}\mathbf{H}_{N_2\times N_2}(z_2^*,z_2)\Sigma_{N_2\times N_2}
				+o((\epsilon^{*}_{2n})^{2N_2},\epsilon^{2N_2}_{2n})
			\end{pmatrix}\\
			=&
			\det\begin{pmatrix}
				\mathbf{H}_{N_1\times N_1}(z_1^*,z_1)
				+o(\epsilon^{*}_{1m},\epsilon_{1m})  &
				\mathbf{H}_{N_1\times N_2}(z_1^*,z_2)
				+o(\epsilon^{*}_{1m},\epsilon_{2n})\\
				\mathbf{H}_{N_2\times N_1}(z_2^*,z_1)
				+o(\epsilon^{*}_{2n},\epsilon_{1m}) & \mathbf{H}_{N_2\times N_2}(z_2^*,z_2)
				+o(\epsilon^{*}_{2n},\epsilon_{2n})
			\end{pmatrix}\sum_{i=1}^{2}
			\det(\Sigma_{N_i\times N_i})\det(\Sigma^{\dagger}_{N_i\times N_i}).
		\end{split}
	\end{equation}
	Combined with the above equations, the solution $\mathbf{Q}^{[N]}_{12}(\xi,\eta)$ at branch points could be expressed as 
	\begin{equation}\nonumber
		\hat{\mathbf{Q}}^{(N_1,N_2)}\!
		=\!\lim_{\epsilon_{1m},\epsilon_{2n}\rightarrow 0}
		\mathbf{Q}_{12}^{[N]}
		\!=\!\lim_{\epsilon_{1m},\epsilon_{2n}\rightarrow 0}\gamma\mathcal{J}^{N-1}
		\frac{\det
			\begin{pmatrix}
				\hat{\mathbf{P}}
			\end{pmatrix}
		}{\det
			\begin{pmatrix}
				\hat{\mathbf{H}}
		\end{pmatrix}}\ee^{\ii (\omega \xi +\kappa \eta)}
		\!=\!\gamma\mathcal{J}^{N-1}
		\frac{
			\det\begin{pmatrix}
				\mathbf{P}_{N_1\times N_1}(z_1^*,z_1)  & 
				\mathbf{P}_{N_1\times N_2}(z_1^*,z_2)  \\
				\mathbf{P}_{N_2\times N_1}(z_2^*,z_1)  & \mathbf{P}_{N_2\times N_2}(z_2^*,z_2)
			\end{pmatrix}
		}{\det\begin{pmatrix}
				\mathbf{H}_{N_1\times N_1}(z_1^*,z_1)  &
				\mathbf{H}_{N_1\times N_2}(z_1^*,z_2) \\
				\mathbf{H}_{N_2\times N_1}(z_2^*,z_1)  & \mathbf{H}_{N_2\times N_2}(z_2^*,z_2)
		\end{pmatrix}}\ee^{\ii (\omega \xi +\kappa \eta)}.
	\end{equation}
	Then, we are going to express them in the rational form with respect to $\xi$ and $\eta$. Based on the definition of $\mathcal{P}(z^*,z)$, $\mathcal{H}(z^*,z)$ in Eq. \eqref{eq:define-P-H-2j-1} and functions $F(z^*,z)$, $G(z^*,z)$ , in Eq. \eqref{eq:define-G-F} the following equations hold
	\begin{equation}\nonumber
		\begin{split}
			\mathcal{P}^{[N_i,N_j]}(z_i^*,z_j)
			\xlongequal[\eqref{eq:define-G-F}]{\eqref{eq:define-P-H-2j-1}}&\left.\frac{\dd^{N_i+N_j}}{N_i!N_j!\dd z^{N_j} \dd z^{*N_i}}
			\frac{E^{*}_1(z)F(z^*,z)r(z)E_1(z)}{r(z^*)}\right|_{z^*=z_i^*,z=z_j}\\
			=&\sum_{m=0}^{N_i}\sum_{n=0}^{N_j}\sum_{a=0}^{m}\sum_{b=0}^{n}E^{[N_i-i-m]*}_1(z_i)\left(r^{-1}(z_i^*)\right)^{[i]}F^{[m,n]}_{i,j}(z_i^*,z_j)r^{[j]}(z_j)E_1^{[N_j-j-n]}(z_j)\\
			=&\mathbf{E}_i^{[N_i]\dagger}\left(\mathbf{R}_i^{[N_i]\dagger}\right)^{-1}\mathbf{F}_{i,j}^{[N_i,N_j]}\mathbf{R}_j^{[N_j]}\mathbf{E}_j^{[N_j]},\\
			\mathcal{H}^{[N_a,N_b]}(z_i^*,z_j)
			\xlongequal[\eqref{eq:define-G-F}]{\eqref{eq:define-P-H-2j-1}}&\left(\mathbf{E}_i^{[N_a]}\right)^{\dagger}\mathbf{G}_{i,j}^{[N_a,N_b]}\mathbf{E}_j^{[N_b]},
		\end{split}
	\end{equation}
	where $\mathbf{R}_j^{[N_j]}$, $\mathbf{E}_j^{[N_j]}$, and $\mathbf{F}_{i,j}^{[N_i,N_j]}$ are defined as
	\begin{equation}\label{eq:define-F-R-G-E}
		\begin{split}
			\mathbf{F}_{i,j}^{[N_a,N_b]} \!=& \!
			\begin{bmatrix}
				F^{[0,0]}(z^*_i,z_j) & F^{[0,1]}(z^*_i,z_j) &  \cdots & F^{[0,N_b]}(z^*_i,z_j) \\[3pt]
				F^{[1,0]}(z^*_i,z_j) & F^{[1,1]}(z^*_i,z_j) & \cdots & F^{[1,N_b]}(z^*_i,z_j) \\[3pt]
				\vdots & \vdots &  \ddots & \vdots \\[3pt]
				F^{[N_a,0]}(z^*_i,z_j) & F^{[N_a,1]}(z^*_i,z_j) & \cdots & F^{[N_a,N_b]}(z^*_i,z_j)
			\end{bmatrix} \!\!, \,
			\mathbf{R}_{j}^{[N_b]} \!= \! \!
			\begin{bmatrix}
				r^{[0]}(z_j) & r^{[1]}(z_j) & \cdots & r^{[N_b]}(z_j)\\[3pt]
				0 & r^{[0]}(z_j) & \cdots & r^{[N_b-1]}(z_j)\\[3pt]
				\vdots & \vdots & \ddots & \vdots \\[3pt]
				0 & 0 & \cdots & r^{[0]}(z_j)
			\end{bmatrix} \! \! ,\\
			\mathbf{G}_{i,j}^{[N_a,N_b]} \!=& \!
			\begin{bmatrix}
				G^{[0,0]}(z^*_i,z_j) & G^{[0,1]}(z^*_i,z_j) &  \cdots & G^{[0,N_b]}(z^*_i,z_j) \\[3pt]
				G^{[1,0]}(z^*_i,z_j) & G^{[1,1]}(z^*_i,z_j) & \cdots & G^{[1,N_b]}(z^*_i,z_j) \\[3pt]
				\vdots & \vdots &  \ddots & \vdots \\[3pt]
				G^{[N_a,0]}(z^*_i,z_j) & G^{[N_a,1]}(z^*_i,z_j) & \cdots & G^{[N_a,N_b]}(z^*_i,z_j)
			\end{bmatrix} \! \!,  
			\mathbf{E}_j^{[N_b]} \!=\! \!
			\begin{bmatrix}
				E_1^{[1]}(z_j) & E_1^{[3]}(z_j) & \cdots & E_1^{[N_b]}(z_j)\\[3pt]
				E_1^{[0]}(z_j) & E_1^{[2]}(z_j) & \cdots & E_1^{[N_b-1]}(z_j)\\[3pt]
				\vdots & \vdots & \ddots & \vdots \\[3pt]
				0 & 0 & \cdots & E_1^{[0]}(z_j) 
			\end{bmatrix} \! \! .
		\end{split}
	\end{equation}
\end{widetext}

\bibliographystyle{apsrev4-2}
\bibliography{reference}

\end{document}